\documentclass[12pt]{article}
\usepackage{authblk}
\usepackage{multirow} 
\usepackage{natbib}
\usepackage{booktabs}
\usepackage[colorlinks=true,linkcolor=blue,citecolor=blue,urlcolor=blue]{hyperref}
\usepackage[all]{xy}
\usepackage{subcaption} 
\usepackage{caption}    
\usepackage{graphicx}   
\usepackage{afterpage}  
\usepackage{amsmath}
\usepackage{graphicx,psfrag,epsf}
\usepackage{enumerate}
\usepackage{natbib}
\usepackage{hyperref} 
\usepackage{comment}    
\usepackage{pdfpages}
\usepackage{makecell}   
\usepackage{booktabs}   
\usepackage{longtable}  
\usepackage{float}      
\usepackage{caption}    
\usepackage{subcaption} 
\usepackage[font=scriptsize]{caption}
\usepackage{mwe}

\usepackage{amsthm}     
\usepackage{amsmath}    
\usepackage{amsmath,amssymb}
\newcommand{\blind}{0}

\begin{document}
\def\spacingset#1{\renewcommand{\baselinestretch}%
{#1}\small\normalsize} \spacingset{1}

\if0\blind
{
    \title{\bf Trade Dynamics of the Global Dry Bulk Shipping Network}   
    \author[1]{Yan Li\footnote{Corresponding author: yl702@sussex.ac.uk}}
    \author[1]{Carol Alexander}
    \author[1]{Michael Coulon}
    \author[2]{Istv\'an Kiss}

\affil[1]{\footnotesize Business School, University of Sussex, Falmer Brighton, BN1 9RH United Kingdom}
\affil[2]{Network Science Institute, Northeastern University London, New College of the Humanities, Devon House, 58 St Katharine’s Way, London, E1W 1LP, United Kingdom}

    \date{} 
    \maketitle
} \fi

\begin{abstract}
\thispagestyle{empty} 
This study investigates the inherently random structures of dry bulk shipping networks, often likened to a taxi service, and identifies the underlying trade dynamics that contribute to this randomness within individual cargo sub-networks. By analysing micro-level trade flow data from 2015 to 2023, we explore the evolution of dry commodity networks—including grain, coal, and iron ore—and suggest that the Giant Strongly Connected Components exhibit small-world phenomena, indicative of efficient bilateral trade. The significant heterogeneity of in-degree and out-degree within these sub-networks, primarily driven by importing ports, underscores the complexity of their dynamics. Our temporal analysis shows that while the Covid-19 pandemic profoundly impacted the coal network, the Ukraine conflict significantly altered the grain network, resulting in changes in community structures. Notably, grain sub-networks display periodic changes, suggesting distinct life cycles absent in coal and iron ore networks. These findings illustrate that the randomness in dry bulk shipping networks is a reflection of real-world trade dynamics, providing valuable insights for stakeholders in navigating and predicting network behaviours.
\end{abstract}

\spacingset{1}
\clearpage
\setcounter{page}{1}  

\section{Introduction}\label{sec:intro}
The dry bulk shipping industry, a segment of tramp shipping\footnote{Tramp shipping refers to a type of maritime transportation where vessels operate without fixed schedules or routes, in contrast to the operations of liner shipping, which typically follow published schedules. It encompasses dry bulk and oil tanker shipping sectors.}, accounts for 80\% of global seaborne trade. It serves as the backbone of the global economy by facilitating the movement of essential raw materials that sustain modern society and drive economic growth across nations. Three main bulk cargo types (coal, grain, and iron ore) have accounted for more than 60\% of the total dry bulk cargo volume, reinforcing the industry's critical role in global trade. As dry bulk shipping continues to support industrial development and economic transitions, it will remain integral to shaping global supply chains.

Compared to liner (or container) shipping, which operates with standardised cargo units, and tanker shipping, which primarily handles crude and refined oil, the dry bulk sector transports a diverse range of commodities, including grains, sugar, fertilizer, coal, iron ore, and other essential minerals. This diversity results in a highly segmented market, making the study of dry bulk shipping more complex.

Moreover, unlike liner shipping, dry bulk shipping faces challenges related to the supply and demand of sensitive commodities, particularly those tied to food and energy security. These commodities are often central to national strategies and geopolitical tensions. As global economic conditions evolve, demand cycles naturally fluctuate between peaks and troughs over the long term, continuing into the future. For instance, at the time of writing, in The Financial Times \cite{HookEtAl2025} report that the iron ore `supercycle' driven by China has come to an end following its peak in 2023. Thus, future demand is expected to shift toward resources essential for developing renewable energy and electricity infrastructure. This includes key minerals such as cobalt, copper, lithium, and nickel, potentially intensifying competition among a broader array of nations for these commodities. 

Such arguments are complemented by \cite{worldmaterial}, for example, who suggests that as environmental concerns for coal consumption increases, in the``short and medium term, pursuing environmental goals will require considerably more materials to build the electric cars, wind turbines, and solar panels needed to replace fossil fuels. The upshot is that in the coming decades, we are likely to extract more metals from the Earth's surface than ever before \citep{worldmaterial}[p.15]." Transporting raw materials from mining to processing and end users, is a fundamental component of the dry bulk shipping system. However, shifting demand patterns and potential changes in processing locations lead to ongoing network transformations.

These challenges, compounded by logistical risks such as volatile freight rates \citep{kingsman2017commodity}, create an unpredictable environment for stakeholders. As a result, the dry bulk transportation network is shaped by commercial trade patterns, geographical constraints, and profit-driven decisions, continuously adapting to global market dynamics. Despite its scale and significance, dry bulk shipping has received far less academic attention than liner shipping.

Our study approaches this gap from a trade flow shipping network perspective. Rather than inferring trade effects from a maritime transportation network angle, which is common in liner shipping, we analyse trade network dynamics directly. This approach emphasises our point that trade dynamics are the origin of all long-distance dry bulk vessel movements, or as \cite{maritimeeconomics} suggests, shipping is a `demand-derived' transportation industry. Hence, we argue that understanding the randomness in trade flow networks can clarify how the dry bulk transportation system evolves, particularly given its cargo-specific trades.

We use micro-level trade flow data from 2015 to 2023, focusing on grain, coal, and iron ore, as well as data from the full dry bulk sector. We offer insights into the unique feature of the inherent randomness of these trade flow networks, and how geopolitical events and shifting market conditions reshape global trade patterns. Here, we use examples of the disruptions caused by COVID-19 and the war in Ukraine to illustrate how temporal changes and community dynamics influence trade networks. 

Our methods combine elements from transportation analysis and trade economics, integrating network theory with detailed shipping trade flow data. These data align with maritime economics, which is often conflated with trade economics but differs in that it focuses on ``the physical quantity of cargo'' rather than the value of trade. Moreover, ``maritime trade analysis emphasises geographical regions over political states'' \citep{maritimeeconomics}. With these distinct features of the data sample, we emphasise that our research is not about identifying the important export and import regions, but to study the complex dry bulk shipping system from the angle of demand side using a multi-layered network approach.  

By applying network theory, we examine the randomness inherent in these networks, with particular attention to temporal network analysis. Temporality is vital for our intervention in shipping network scholarship as this approach not only addresses the granularity lacking in existing port call data studies, but also sheds light on significant structural changes triggered by geopolitical events and market conditions. This can further trigger trade relations to reform as a result. To identify these trade relations, we use the network community structure as a way to describe how networks are organised into clusters of nodes that are more densely connected internally than with the rest of the network. 

Given that our data spans a period which includes significant disruptions such as the COVID-19 pandemic and the Ukraine war, we highlight the need for shipping network studies to pay attention to micro-level dynamics and fluctuations in the shipping network. Conversely, this analysis also sheds light on mechanisms to mitigate disruptions from international geopolitical challenges and their impacts on essential trade. We aim to provide a foundation for understanding dry bulk shipping networks and offer a means to enhance their predictability for all stakeholders.

Accordingly, our research offers a granular, multi-layered analysis of dry bulk networks, demonstrating how market characteristics and global events reshape trade flows. This insight can help dry bulk fleet owners operating across multiple cargo segments worldwide optimise route selection, minimising ballast journeys.\footnote{Ballast journeys occur when a ship is empty and traveling to its next load location. In contrast, laden journeys refer to voyages where the ship carries cargo to the discharge port. Laden legs generate revenue for the shipowner by transporting cargo, whereas ballast legs are unprofitable since no freight revenue is earned during these trips.} Finally, since our analysis is based on trade flows, an aggregation of all voyages transporting bulk cargo from load ports to discharge ports, the results can be combined with freight rate data to enhance profitability, particularly for stakeholders with interests in the spot market. This interdisciplinary effort, combining approaches from network studies and maritime trade analysis, extends the applications of network theory to trade network evolution, challenging conventional methodologies and providing new perspectives on optimising global trade logistics for enhanced resilience and economic growth.

In the following, Section 2 situates our study within the context of relevant literature. Section 3 describes the dataset used in this study, highlighting its advantages and limitations. Section 4 outlines the network methods applied in this research, while Section 5 presents the results of the model. Finally, Section 6 concludes the paper.

\section{Literature review}
Recent advances in tracking and mapping technologies have enabled the analysis of ship trajectories, offering new perspectives on the spatial structure of maritime transport networks. Existing studies predominantly focus on liner shipping \citep{ducruet2010centrality, laxe2012maritime, tsiotas2015analyzing, wang2016study, calatayud2017connectivity, calatayud2017vulnerability, liu2018spatial, pan2019connectivity, cheung2020eigenvector, ducruet2012worldwide, bai2023data}, which is characterised by fixed schedules and routes. 

These studies use network methods to provide valuable insights into the hierarchical structure of global liner shipping port systems, with implications for trade patterns, connectivity, and efficiency \citep{ducruet2010centrality, laxe2012maritime, tsiotas2015analyzing, wang2016study, liu2018spatial, kanrak2019maritime, cheung2020eigenvector}. In doing so, they rely on port call data to study vessel movements and resultant maritime networks \citep{ducruet2010centrality, ducruet2010ports, ducruet2012worldwide, williams2014degree, tsiotas2015analyzing, calatayud2017vulnerability, ducruet2013network}. 

Despite significant advancements in network analysis, the existing literature, discussed above, largely overlooks the tramp shipping sector in favour of liner shipping. In contrast to liner shipping, tramp shipping, encompassing oil tankers and dry bulk carriers, operates with more variability, akin to a taxi service where routes are determined by immediate demand \citep{kaluza2010complex, brancaccio2020geography}. Yet, only two existing studies,  \cite{kaluza2010complex} and \cite{ducruet2013network}, examine dry bulk shipping networks. Of these, \cite{kaluza2010complex} identifies three primary ship types — container ships, dry bulk carriers, and oil tankers — each exhibiting distinct movement patterns. Here, the authors outline how dry bulk carriers tend to follow less predictable routes than container ships and oil tankers, reflecting their tramp shipping nature. 

\cite{ducruet2013network} focuses on coupled networks, defined as systems with multiple types of connections among the same set of ports. The authors utilise multi-layered graphs to analyse relationships between different maritime flows, such as dry bulk, liquid bulk, and container traffic. Their findings indicate that ports handling diverse commodities experience higher traffic volumes and centrality within maritime networks. 

However, neither study provides detailed characteristics of the dynamics of dry bulk and oil tanker shipping. While \cite{kaluza2010complex} notes randomness in dry bulk shipping, they do not explain the specifics causing these patterns, and thus offer limited predictive value for stakeholders. In addition, the authors focus on using aggregation of dry bulk vessel trajectories (port call data) to conduct static network analysis, which is unable to account for the complex dynamics of the segmented sector, for example, the impacts of current geopolitical events on international trade. Hence, they can only provide limited insights for stakeholders to adapt new strategies to counter concurrent events. 

Both of these studies suggest that future research should target dynamic networks and interdisciplinary applications to better understand and, consequently, optimise maritime dry bulk transport systems. Building on these assertions, this study proposes that the perceived randomness in transport systems often originates from international dry bulk trade (focused on main dry bulk commodities such as coal, grain, and iron ore), whereby trade flow data — which tracks laden voyages — is more suitable than port call data to understand the composition of tramp shipping networks. 

Our research interests lie in analysing dry bulk trade flow networks, and builds on the work of \cite{brancaccio2020geography}. While some studies have explored globalisation's impact on trade networks using a network approach \citep{iapadre2014emerging, calatayud2017connectivity}, they often rely on macro-level trade data, neglecting the specific dynamics of shipping networks informed by actual ship movements. In contrast, \cite{brancaccio2020geography} utilises micro-level ship movement data to analyse the interplay between transportation systems and international trade, especially in dry bulk. They propose that transportation costs are endogenous, influenced by the equilibrium behaviours of ships and exporters. This model reveals that shipping costs are affected by the attractiveness of both origins and destinations, which form the basis for the interconnectedness of trade routes. Additionally, their research illustrates how the transportation sector can mitigate disparities in comparative advantage by reallocating production from net exporters to net importers. Such a perspective suggests that the shipping sector has a major role to play in influencing global trade patterns and reducing trade imbalances.

In addition to the cargo type the intermediate locations within each cargo network are also important in understanding the complexity of shipping networks. \cite{brancaccio2020geography}'s approach aligns with \cite{rodrigue2006challenging}, who asserts that cargo movements are motivated by more integrated demand in addition to economic activity, which is driven by vertical integration in supply chain management. This perspective underscores the need for a more complex understanding of modern logistics and the vital importance of including cargo shipping hubs in this analysis, emphasising the critical role of intermediate locations  enhancing freight distribution efficiency. Their findings are important for understanding the constitution of the supply chain, including the role of maritime shipping within it. However, this supply chain-level perspective does not account for the important role of processing plants in the dry bulk sector. Hence, the intermediate locations have to be studied specifically for their roles in dry bulk shipping.

In light of the identified gaps in existing research and the limitations of dry bulk shipping network analysis, this study analyses data from 2015 to 2023 to identify some of the key dynamics of the full dry bulk shipping network as well as the three main bulk carrier sub-networks, including grain, coal, and iron ore, using the multi-layered network approach. Temporality is particularly vital for our intervention in shipping network scholarship as this approach demonstrates how market characteristics and global events reshape trade flows. By adopting an interdisciplinary combination of network studies and trade economics, we provide new perspectives on optimising global trade logistics for enhanced resilience and economic growth.

\section{Data }
\label{sec:data}
For our analysis of dry bulk shipping networks, we utilise data from Oceanbolt Maritime Market Intelligence, provided by Veson Nautical. This dataset is well suited for our study due to its completeness and direct derivation from AIS data, ensuring comprehensive coverage of shipping activities along with detailed attributes such as vessel identity, size, commodity types, timestamps, geographic coordinates for loading and discharging, and trade flow distances. This granularity allows for an in-depth analysis of trade flows, revealing temporal and spatial patterns crucial for understanding dry bulk shipping dynamics. Covering the period from 2015 to 2023, these data provide a robust temporal context for examining trends and changes in global maritime trade networks. A sample of the full dataset is provided in Supplemental Materials \ref{supplementA}.

To ensure the integrity and reliability of our analysis, we performed critical data cleaning steps on the raw data. First, we removed all trade flow entries with unknown load or discharge ports, as these incomplete records could skew the network analysis. Additionally, we filtered out flows categorized as ``Transit'' and``Yard'', which do not involve cargo offloading and do not contribute to international maritime trade networks. These steps are essential to creating a dataset that accurately represents trade activities. 

The final dataset includes only relevant data points, encompassing 16,121 ships and 1,676,143 trade flows over a nine-year period, as shown in Table \ref{tab:commodity_summary}. The complete dry bulk dataset comprises 21 identified commodities. However, some journeys for which the cargo content is unknown. Notably, coal, grain, and iron ore together account for 64\% of the total seaborne trade volume from 2015 to 2023, yet the total number of trade flows for these three commodities represents only 32\%. In the shipping industry, these three cargo types are referred to as the major bulks, while the remaining cargo types are classified as minor bulks, given that the main bulks collectively account for more than half of the total trade volume. Within all the vessels available in the market, more than half of the vessels transport, but not limited to, the main bulk cargo. A histogram shows that vessels can carry a wide variety of commodities.\footnote{For additional details, see Figure \ref{fig:his_vessel} in the Supplementary Material.} 

\begin{table}[ht!]
\tiny
\centering
\begin{tabular}{@{}p{4cm}p{4cm}p{4cm}p{4cm}@{}}
\toprule
\multicolumn{1}{l}{} & \multicolumn{1}{l}{\textbf{Proportion of vessels}} & \multicolumn{1}{l}{\textbf{Proportion of Trade   Flows}} & \multicolumn{1}{l}{\textbf{Proportion of Total Volume}} \\ 
\multicolumn{1}{l}{} & \multicolumn{1}{l}{(n = 16,121)} & \multicolumn{1}{l}{(n = 1,676,143)} & \multicolumn{1}{l}{($\Sigma = 53.9$ bn GT)} \\ 
\addlinespace
\multicolumn{1}{l}{\textbf{Full dry bulk}} & 100\% & 100\% & 100\% \\
\midrule
\multicolumn{1}{l}{\textbf{Grain}}     & 63\%   & 9\%       & 9\%            \\
\multicolumn{1}{l}{\textbf{Coal}}      & 78\%   & 16\%      & 28\%           \\
\multicolumn{1}{l}{\textbf{Iron Ore}}  & 53\%   & 7\%       & 27\%           \\ 
\multicolumn{1}{l}{\textbf{\textbf{Subtotal}}} &  & \textbf{32\%} &\textbf{64\%}\\ 
\bottomrule
\end{tabular}
\caption{\textbf{Data summary} This table presents Oceanbolt trade flow data for grain, coal, and iron ore cargo, highlighting the proportionate representation of each commodity within the overall dry bulk trade. It includes total vessel counts (n = 16,121), trade flows (n = 1,676,143), and an aggregate volume of 53.9 billion Gross Tonnage (GT). }
\label{tab:commodity_summary}
\end{table}

One limitation of the data is the relatively large number of trade flows categorised as unknown cargo types, as shown in Table \ref{tab:commodity_sum} in Supplemental Material A. However, incomplete data sets are a common feature of data collection in the maritime shipping industry. Despite this limitation, we are confident that the dataset includes sufficient observations to effectively reflect market trends. 

\subsection{Network construction}
The construction of our network model is crucial for analysing global dry bulk shipping and uncovering maritime trade patterns across commodities. The nodes in our network represent ports extracted from the Oceanbolt trade flow data. Here, directed edges are formed based on trade flows between load and discharge ports. We employ a weighted edge list, assigning frequency, deadweight tonnage, and cargo volume as edge weights.

To provide a foundational overview, we first adopt a static network configuration by aggregating trade flows over the study period into a single network. This captures on-going persistent patterns and essential structural properties of the dry bulk shipping networks, facilitating insights into long-term trade relationships. This static model serves as a foundation for further temporal analyses of dynamic changes over time.

Using a multi-layered network model, we analyse three primary commodities — grain, coal, and iron ore — each represented in distinct layers. This allows for a detailed examination of trade dynamics specific to each commodity. By focusing on each commodity, our research reveals the formulation of unique sub-network in the global shipping network and the relevance of external factors, such as pandemics and conflicts, on shipping operations. 

Drawing on \cite{kaluza2010complex}, our research focuses on the main dry bulk sub-networks: coal, iron ore, and grain, alongside analysis of the overall network. This layered approach allows us to examine static features across quarterly (static) sub-networks from 2015 to 2023. Table \ref{tab:travel_statistics} shows the summary statistics of the number of days ships travel from load to discharge ports. As the average trade flow journey length is approximately 35 days, more than a month, setting quarterly time windows ensures a well-connected network for meaningful structural analysis.

\begin{table}[ht!]
\tiny
\centering
\begin{tabular}{@{}p{1.5cm}p{1cm}p{1cm}p{1cm}p{1cm}p{1cm}p{1cm}p{1cm}p{1cm}@{}}
\toprule
\multicolumn{1}{l}{} &
  \multicolumn{1}{c}{\textbf{Count}} &
  \multicolumn{1}{c}{\textbf{Mean}} &
  \multicolumn{1}{c}{\textbf{Std}} &
  \multicolumn{1}{c}{\textbf{Min}} &
  \multicolumn{1}{c}{\textbf{25\%}} &
  \multicolumn{1}{c}{\textbf{50\%}} &
  \multicolumn{1}{c}{\textbf{75\%}} &
  \multicolumn{1}{c}{\textbf{Max}} \\
  \midrule
\textbf{Dry\_bulk} & 1,667,827 & 28 & 26 & 0 & 12 & 24 & 39 & 2,854 \\
\textbf{Grain}     & 156,608   & 42 & 24 & 1 & 26 & 39 & 56 & 1,382 \\
\textbf{Coal}      & 273,023   & 27 & 21 & 1 & 15 & 23 & 33 & 2,060 \\
\textbf{Iron Ore}  & 126,944   & 34 & 23 & 1 & 21 & 28 & 42 & 2,014\\
\bottomrule
\end{tabular}
\caption{\textbf{Descriptive statistics of dry bulk trade flow travel duration (number of days) per network.} This table summarises the descriptive statistics of travel durations in dry bulk trade flows across different commodity networks: dry bulk, grain, coal, and iron ore, aggregated from Oceanbolt data (2015–2023).  For each category, the table includes a calculation of trade flows, mean duration, standard deviation, minimum and maximum durations, as well as the 25th, 50th (median), and 75th percentiles. }
\label{tab:travel_statistics}
\end{table}

\subsection{Advantage of using trade flow dataset}
There are two common ways of organising ship movement data: port calls and trade flows. Port calls simply record all of the ports at which a ship stopped along its journey. Most existing literature \citep{kaluza2010complex, abouarghoub2018reconciling} relies on port calls or similar data derived from AIS data for maritime transportation network analyses. For our purposes, the trade flow data provide a number of benefits and has implications for the structure of the networks that are formed.

First, since our data are based on trade flows with laden legs only, the direction of the flows excludes the ballast legs. This is a key difference from the network analysis undertaken by researchers using port call data, such as \cite{brancaccio2020geography} and \cite{kaluza2010complex}. In our trade flow network, the direction of the edges only reflects the trade flow from the export port to an import port when cargo has been loaded. This focus on relevant data aligns with our research objective to accurately analyse maritime trade dynamics, as it avoids skewing the analysis with non-cargo-related movements.

Second, the edges can be directly associated with the freight rate data since, for each voyage, a ship owner or charterer is subject to receiving freight in response to paying for the ship's cost. Traditionally, the freight rates are named by the load and discharge region together with the cargo type information. Due to the availability of the trade flow network, researchers can directly build connections with the corresponding freight rates for more finance-related studies, where ballast legs are redundant.\footnote{Freight rates are typically applied only to the leg where the vessel carries cargo. While the ballast leg does incur operational costs (such as fuel, crew), it does not involve a freight transaction.} This approach lays the groundwork for future studies related to analysing the financial implications of trade flows. Yet, our research is unable to delve deeply into these economic aspects due to its limited scope.

\begin{figure}[ht!]
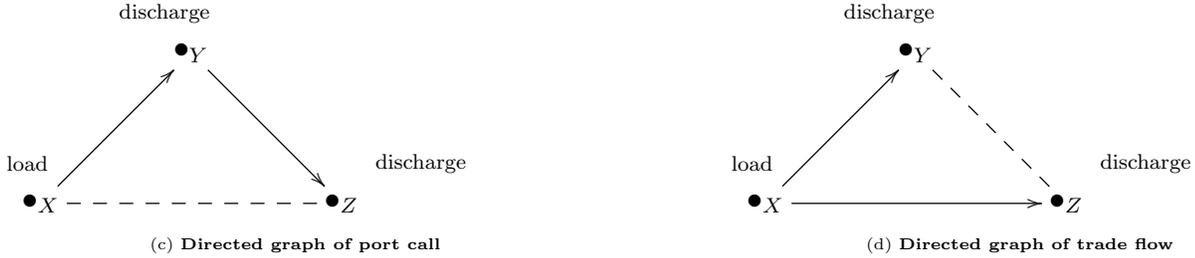

    \centering
    \text{(1) Multiple loads}
    \vspace{0.5em} 

    \begin{minipage}{0.45\textwidth}
        \centering
        \begin{subfigure}[b]{\textwidth}
            \centering
            \xygraph{  
            !{<0cm,0cm>;<1cm,0cm>:<0cm,1cm>::}  
            !{(-2,2) }*+[black]{\bullet_{A}}="A"   
            !{(0,4)}*+[black]{\bullet_{B}}="B"     
            !{(2,2) }*+[black]{\bullet_{C}}="C"    
            "A" *+!<5pt,-15pt>{\text{\scriptsize load}}    
            "B" *+!<10pt,-15pt>{\text{\scriptsize load}}   
            "C" *+!<-30pt,-15pt>{\text{\scriptsize discharge}} 
            "A":"B" 
            "B":"C"  
            "C" :@{--} "A"   
            }
            \caption{\textbf{Directed graph of port call}}
        \end{subfigure}
    \end{minipage}
    \hfill
    \begin{minipage}{0.45\textwidth}
        \centering
        \begin{subfigure}[b]{\textwidth}
            \centering
            \xygraph{  
            !{<0cm,0cm>;<1cm,0cm>:<0cm,1cm>::}  
            !{(-2,2) }*+[black]{\bullet_{A}}="A"   
            !{(0,4)}*+[black]{\bullet_{B}}="B"     
            !{(2,2) }*+[black]{\bullet_{C}}="C"    
            "A" *+!<5pt,-15pt>{\text{\scriptsize load}}    
            "B" *+!<10pt,-15pt>{\text{\scriptsize load}}   
            "C" *+!<-30pt,-15pt>{\text{\scriptsize discharge}} 
            "A":"C" 
            "B":"C"  
            "A" :@{--} "B"   
            }
            \caption{\textbf{Directed graph of trade flow}}
        \end{subfigure}
    \end{minipage}

    \vspace{1em} 

    \text{(2) Multiple discharges}
    \vspace{0.5em} 
    
    \begin{minipage}{0.45\textwidth}
        \centering
        \begin{subfigure}[b]{\textwidth}
            \centering
            \xygraph{  
            !{<0cm,0cm>;<1cm,0cm>:<0cm,1cm>::}  
            !{(-2,2) }*+[black]{\bullet_{X}}="X"   
            !{(0,4)}*+[black]{\bullet_{Y}}="Y"     
            !{(2,2) }*+[black]{\bullet_{Z}}="Z"    
            "X" *+!<5pt,-15pt>{\text{\scriptsize load}}    
            "Y" *+!<10pt,-15pt>{\text{\scriptsize discharge}}   
            "Z" *+!<-30pt,-15pt>{\text{\scriptsize discharge}} 
            "X":"Y" 
            "Y":"Z"  
            "Z" :@{--} "X"   
            }
            \caption{\textbf{Directed graph of port call}}
        \end{subfigure}
    \end{minipage}
    \hfill
    \begin{minipage}{0.45\textwidth}
        \centering
        \begin{subfigure}[b]{\textwidth}
            \centering
            \xygraph{  
            !{<0cm,0cm>;<1cm,0cm>:<0cm,1cm>::}  
            !{(-2,2) }*+[black]{\bullet_{X}}="X"   
            !{(0,4)}*+[black]{\bullet_{Y}}="Y"     
            !{(2,2) }*+[black]{\bullet_{Z}}="Z"    
            "X" *+!<5pt,-15pt>{\text{\scriptsize load}}    
            "Y" *+!<10pt,-15pt>{\text{\scriptsize discharge}}   
            "Z" *+!<-30pt,-15pt>{\text{\scriptsize discharge}} 
            "X":"Z" 
            "X":"Y"  
            "Z" :@{--} "Y"   
            }
            \caption{\textbf{Directed graph of trade flow}}
        \end{subfigure}
    \end{minipage}
    \caption{\textbf{Comparison of port call and trade flow.} In row (1), graph (a) presents ship movements when half of the cargo is loaded in port A, with the ship then traveling to B to load the other half. Finally, the ship travels to port C to discharge the entire cargo. Graph (b) shows how the port calls are recorded in terms of trade flows, i.e. where a cargo is loaded and discharged. In this case, a ship loads cargo in port A and delivers it to port B. It also loads in port B and discharges in port C. Hence there are two laden legs, both directing to the destination C. However, the journey between A and B is not recorded in trade flow data, resulting in an omission from the network's constitution. In row (2), Graph (c) illustrates the ship's movements from port X to discharge half of the cargo in port Y. It then travels to port Z to discharge the rest of the cargo. This graph shows a complete voyage which captures data as this ship travels from X to Y to Z. However, from a trade flow data angle, the movements are recorded in the form of graph (d) where a ship loads in port X, and then goes to port Y for the first discharge. The second laden leg is from X to Z. To distinguish between the fixture and the trade flow, Oceanbolt gives one unique voyage ID for the multiple load case and gives two unique flow IDs for the two laden legs in both (b) and (d) situations.}
    \label{fig:multiloadsdischarges}
\end{figure}

Finally, where there are journeys with multiple loads and discharges, the use of trade flow data significantly impacts the resulting connectivity and transitivity of the networks. This is shown in Figure \ref{fig:multiloadsdischarges}. The figure presents four directed graphs: the top row shows networks with multiple loads, while the bottom row shows networks with multiple discharges. The graphs on the left represent networks based on the sequence of ship movements (port calls), and the ones on the right are based on trade flow data.

Although the number of nodes and edges is the same across all the graphs, the impact on the network structure, particularly in terms of centrality measures, is different. In the case of multiple loads, in graph (a), port B has an indegree of 1 and an outdegree of 1, while port C only has an indegree of 1. However, in graph (b), port B has an indegree of 0 and an outdegree of 1, while port C has an indegree of 2. This observation highlights how trade flow data provides a more accurate representation of the import and export volumes, which is crucial to our objective of highlighting the importance of the underlying structural dynamics of the shipping network.

Another important effect of using trade flow data instead of port call data is the difference in transitivity within the network. In graphs (a) and (c), the dashed line between ports A and C indicates that if there is a trade flow between them, a complete clique is formed, resulting in a transitivity of 1 for each node. In graphs (b) and (d), the dashed lines suggest alternative edges that can form cliques, depending on whether there are additional trade flows between these pairs of nodes.

Finally, the trade flow dataset provides a more accurate representation of the shortest path length between importing and exporting regions. For instance, although a ship may travel from Port A to Port B before reaching Port C, maximising cargo usage along the way, the trade flow data reflects a direct trading relationship between Ports A and C as a single-step path. In contrast, port call data, which only tracks each stop, would count this as two steps. By prioritising a direct path measurement, we shift attention to understanding actual trade relationships.

\section{Methods}\label{sec:meth}
Networks can be presented mathematically through various frameworks, with the two predominant formats being the edge list and the adjacency matrix. In the context of maritime shipping networks, an edge list provides a suitable representation, detailing each journey of a ship from one port to another. An edge list could encompass both port calls and trade flow data. However, for analytical rigour, we employ an adjacency matrix, wherein the connectivity between nodes (ports) is captured in a square matrix format. In this matrix, the elements $A_{ij}$ are designated as 1 if an edge exists between nodes ( i ) and ( j ), and 0 otherwise. 

In scenarios where the graph is simple and weighted, wherein each edge $A_{i,j}$ is assigned a weight $w(A_{i,j})=w_{i,j}$, the elements of the adjacency matrix reflect the weights corresponding to these connections. In our study, the weights signify the frequency of trade journeys, the volumetric cargo capacity, and the deadweight tonnage, per trade flow.

To construct a static network, we utilise a complete dataset spanning nine years, allowing us to analyse network trends in a comprehensive way. To observe temporal changes, we dis-aggregate the network of ship movements into consecutive quarterly periods, thus forming temporal networks. However, as discussed by \cite{guinand2015time}, such constructed networks may be regarded as static, as individual ship movements occurring within a given quarter are considered in uniform manner. Nevertheless, we assert that this quarterly granularity is sufficient to address our research questions regarding shifts in dry bulk networks during and after the COVID-19 pandemic and the continuing war in Ukraine.

\subsection{Network properties}
At the node level, we focus on centrality measures, including degree centrality, strength centrality, average path length, and clustering coefficient. Degree centrality represents the number of connections a node has, reflecting the existing trading routes. In a weighted directed graph, strength centrality generalises this measure by incorporating the weights of connections, indicating the trading volume a port engages in.

The degree distribution reveals the connectivity distribution across nodes. Here, a power-law distribution equates to a scale-free network, where a few nodes exhibit very high connectivity while the majority have low connectivity. This pattern is commonly observed in real-world networks \cite{barabasi1999emergence}. 

The average path length measures the mean of the shortest paths (geodesics) between all pairs of nodes and provides insight into the particularities of network efficiency. A short average path length suggests that nodes can be reached with relatively few connections. The clustering coefficient measures the likelihood that a node's neighbors are interconnected. The clustering coefficient is commonly used for undirected graphs. However, we treat directed graphs as undirected to analyse clustering, acknowledging the interconnectedness of nodes regardless of edge direction.

Understanding the average path length is crucial for examining the small-world property of our networks. In small-world networks, the average path length between any two nodes is relatively short, even in larger structures, meaning that most nodes can be accessed through only a few connections. High clustering, along with a short average path length, characterises small-world networks, which can be navigated with relatively few steps.

Degree assortativity measures the tendency of nodes to connect with others of similar degree, with an assortativity coefficient ranging from -1 (disassortative) to 1 (assortative), akin to a correlation coefficient. Understanding these properties will enhance our comprehension of the structural dynamics within dry bulk shipping networks and inform subsequent analyses.

Finally, an important topological property is community structure \cite{pan2019connectivity}. In the context of trade flow networks, community structure analysis can show historical trade relations between countries \cite{yin2024temporal}. Community detection as a tool essentially groups ports based on edges between them. If ports form dense connections within a group and fewer connections between groups, the network naturally divides into groups as communities. This analysis is relevant for global seaborne trade, which operates within physical constraints and is influenced by geopolitical factors as well as geography. The first two factors often change over time. Hence, it's particular interesting to combine the community structure detection in combination with temporal network analysis.

In addition, identifying and understanding a network’s large-scale patterns of connection can be challenging, especially those that might not be easily visualised \cite{Newman2010networks}. To detect the dry bulk trade community structures, we apply the Louvain algorithm, developed by \cite{blondel2008fast}. (For a detailed explanation of the algorithm, please see the Supplementary Material B).   

\subsection{Null model and simulation}
To better understand patterns in the centralities of the dry bulk trade flow within the shipping network, we introduce a null model. This model assesses whether our observed measurements reflect typical network structures or indicate unusual behavior. In graph theory, a null model sets expectations based on the assumption that no specific patterns exist in the data. By comparing the data against a real-world network, we can quantify how much a network property deviates from what might be expected, enabling us to draw meaningful conclusions about the network's organisation \citep{fornito2016fundamentals}.

We utilise a calibrated random model that maintains the same degree distribution as the observed network. For benchmarking, we focus on the clustering coefficient and average path length of the simulated random null model. By aligning the null model's properties with those of the observed network, we isolate features of interest, facilitating our understanding of the mechanisms that drive the observed data. If the model replicates the observed network structure successfully, it suggests that the rules used to create the null model are significant in shaping the dynamics of the network.

In our study, we apply the simulated random network's clustering coefficient and average path length to benchmark our actual network properties concerning small-world phenomena. We utilise methods developed by \cite{humphries2008network} and \cite{telesford2011ubiquity} to derive small-world coefficients, integrating both the clustering coefficient and the average path length, as defined below:
\begin{equation}\label{eqn:sigma_test}
    \sigma=\frac{C/C_{\text{rand}}} {L/L_\text{rand}},
\end{equation}
\begin{equation}\label{eqn:w_test}
    \omega=\frac{L_{\text{rand}}}{L} - \frac{C}{C_\text{latt}},
\end{equation}
where $L$ is the path length of the observed network, $C$ is the clustering coefficient of the network, $L_{\text{rand}}$ is the path length of an equivalent random network, $C_{\text{rand}}$ is the clustering coefficient of that random network, and $C_\text{latt}$ is the clustering coefficient of an equivalent regular network.

Building on the work of \cite{humphries2008network}, \cite{telesford2011ubiquity} argue that it is insufficient to determine the small-world properties of a network by merely comparing the clustering coefficient and average path length to their random equivalents. The presence of $C\_{\text{rand}}$ in the denominator makes the ratio sensitive to small clustering values. Furthermore, the authors emphasise the importance of identifying whether networks exhibit specific behaviours, such as specialisation (lattices) or effective information transmission (random networks), which are critical for understanding the operational dynamics of the network.

\subsection{Distance matrix}
To measure dissimilarity between quarterly networks, we constructed a distance matrix based on the pairwise comparison of adjacency matrices for each quarter, as showcased by \citep{sugishita2021recurrence}. Denoted as $\mathbf{D}$, this ( $n \times n$ ) matrix represents the total number of quarters analysed, with each element $d\_{ij}$ indicating the distance between two networks, $G_i$ and $G_j$, corresponding to quarters  i and j.

The distance between two directed, unweighted networks $G_i$ and $G_j$ is computed using the following normalised distance formula:
\begin{equation}
    D(G_i, G_j) = 1 - \frac{M(G_i \cap G_j)}{\sqrt{M(G_i)M(G_j)}},
\end{equation}
where $M(G_i)$ and $M(G_j)$ represent the total number of edges in networks $G_i$ and $G_j$, respectively. The term $M(G_i \cap G_j)$ denotes the number of edges that are common between the two networks. The normalization by $\sqrt{M(G_i) M(G_j)}$ ensures that the distance lies between 0 and 1, where 0 indicates identical networks and 1 indicates no overlap in edges between the networks. The resulting distance matrix provides a comprehensive view of the dissimilarities between networks across different quarters. 

After constructing the distance matrix, we seek to deepen our understanding of the network dynamics by focusing on hierarchical clustering in the dry bulk trade patterns. Hierarchical clustering can group together quarters based on the similarity of their network structures, revealing underlying patterns and trends in the evolution of the network. We utilise the Ward linkage method, which minimises the variance within clusters during the merging process. This clustering approach enables us to group together similar quarters and provides insight into periods of stability in trade connectivity or disruptions caused by external factors within the network.

The identified clusters are then analysed to interpret key periods where significant structural changes occurred. This methodological approach not only facilitates a thorough examination of the changes in the dynamics of the network over time, but also aligns with our research objective to address how major disruptions like the COVID-19 pandemic and the Ukrainian war influence trade patterns and structural dynamics. Further details regarding the clustering algorithm can be found in Section \ref{cluster_algo} of the Supplemental Materials, which provides essential context for the analysis.

\section{Results}\label{sec:emp}
In this section, we present the empirical findings from our analysis of the network structures for various dry bulk commodities, specifically focusing on coal, grain, and iron ore. 
\subsection{Network properties}
\paragraph{Overall network observations}
The complete dry bulk network consists of 2,748 ports, 1,671,034 trade flows, and 171,631 unique routes, structured as a directed weighted graph utilising Python's Networkx. The adjacency matrix, with a zero diagonal, confirms the absence of self-loops, consistent with the expectation that ships do not return to the same port within a single trip. Our analysis focuses on sub-networks based on specific commodities, wherein ports play multifaceted roles due to the variety of cargo handled.

\paragraph{Sub-network comparisons}
We identify both similarities and significant differences between the three sub-networks, as well as with the full dry bulk network. As Table \ref{tbl:glo_centrality} shows, each cargo specific network is smaller in scale, indicating their specialised roles, as not all ports support every trade. Among them, the coal sub-network is the largest by port number but has fewer edges and a lower network density compared to grain. The grain network shows the highest connectivity with density of 0.011, compared to 0.008 for coal and 0.007 for iron ore, due to a greater number of edges and nodes. Iron ore is the smallest and least connected sub-network with just one third of the dry bulk ports active. The aggregate dry bulk network is the most interconnected overall, representing a densely connected structure due to overlapping multi-layered cargo networks. 

\begin{table}[ht!]
\tiny
\centering
\begin{tabular}{@{}rlllllllllllll@{}}
\toprule
\textbf{} &
  \textbf{n} &
  \textbf{e} &
  \textbf{k} &
  $\phi$ &
  \textbf{n\_w} &
  \textbf{p\_w} &
  \textbf{d\_w} &
  \textbf{n\_s} &
  \multicolumn{1}{c}{\textbf{p\_s}} &
  \multicolumn{1}{c}{\textbf{d\_s}} &
  \multicolumn{1}{c}{\textbf{l}} &
  \multicolumn{1}{c}{\textbf{c}} &
  \multicolumn{1}{c}{\textbf{a}} \\ \midrule
\textbf{Full Dry Bulk} & 2748 & 171631 & 62.46 & 0.023 & 1 & 100\%   & 7 & 685 & 75\% & 5 & 2.50 & 0.23 & -0.09 \\
\textbf{Coal}          & 1525 & 19206  & 12.59 & 0.008 & 1 & 100\%   & 6 & 985 & 35\% & 5 & 3.65 & 0.05 & -0.11 \\
\textbf{Grains}        & 1459 & 22966  & 15.74 & 0.011 & 1 & 100\%   & 7 & 971 & 34\% & 4 & 3.15 & 0.04 & -0.07 \\
\textbf{Iron Ore}      & 902  & 5970   & 6.62  & 0.007 & 1 & 100\%   & 7 & 738 & 18\% & 5 & 4.12 & 0.05 & -0.10 \\
\bottomrule
\end{tabular}
\caption{\textbf{Global network centralities of the full dry bulk network and sub-networks per commodity group.} The centrality measures of each network include: number of ports n; number of unique edges e; average in- or out-degree k; network density $\phi$; number of weakly-connected components (WCCs) n\_w; percentage of giant weakly-connected component of the full network p\_w, diameter of the largest weakly-connected component d\_w; number of strongly-connected components (SCCs) n\_s, percentage of giant strongly-connected component (GSCC) of the full network p\_s, diameter of the strongly-connected component d\_s; clustering coefficient c; the average shortest path length l, and degree assortativity a.}
\label{tbl:glo_centrality}
\end{table}

\begin{figure}[ht!]
    \centering \includegraphics[width=0.8\linewidth, clip,trim={0 0.5cm 0 2cm}]{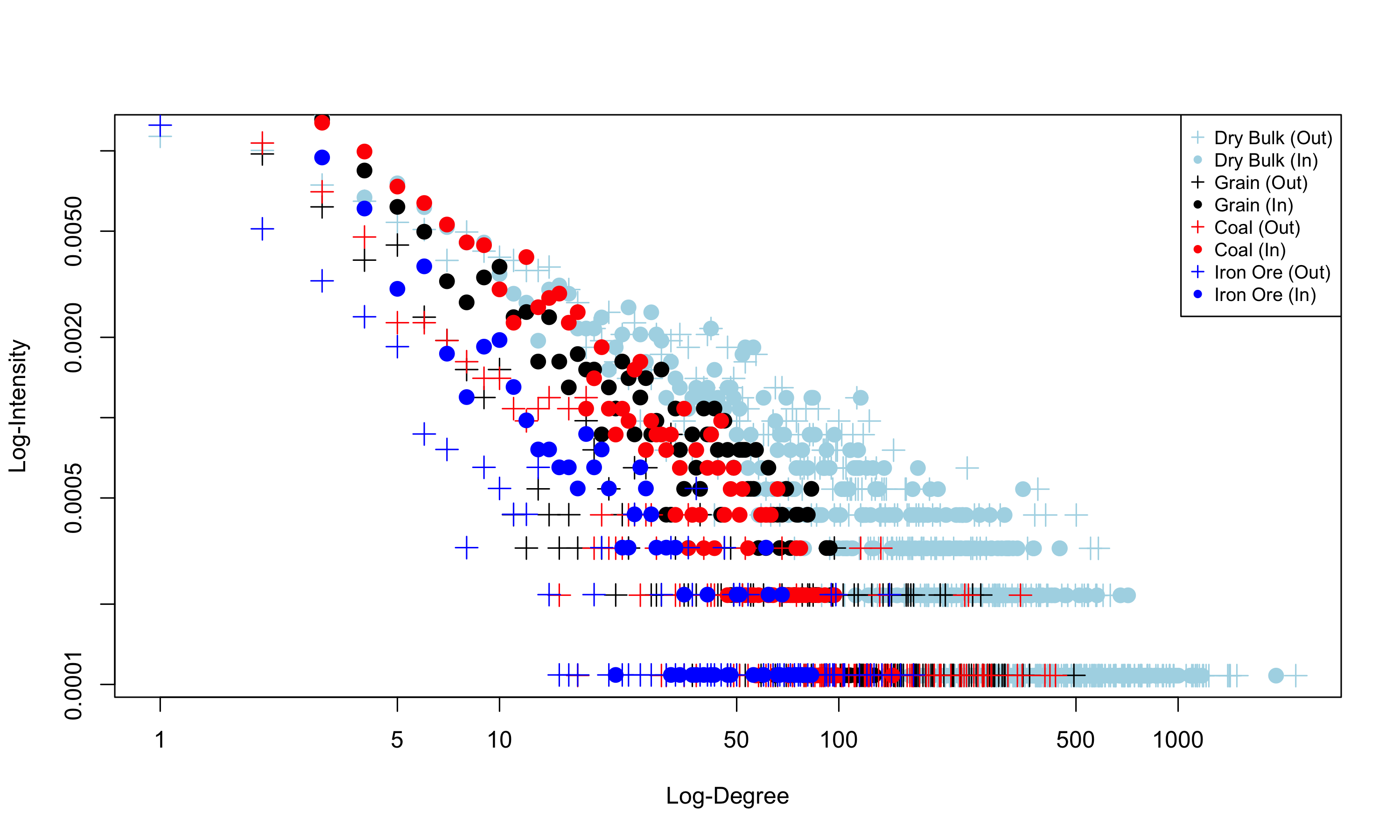} 
    \caption{\textbf{Log-logdegree distributions for all sub-networks.} The $x$-axis shows $\log(\text{degree})$, and the $y$-axis shows $\log(\text{frequency})$ of nodes with that degree. A linear trend in log--log plot indicates a power-law relationship of the form $f(k) = C\,k^{-\gamma}$. The slope in these log--log plots is $-\gamma$. The plus $(+)$ marker in the plot represents the log of out-degrees, whereas the round ($\circ$) marker represents the log of in-degrees. A colour scheme is used to distinguish different sub-networks: light blue denotes the full dry bulk network, black represents grain, red indicates coal and dark blue signifies the iron ore sub-network. }\label{log_log_degree}
\end{figure}

\paragraph{Degree distribution} 
We observe scale-free characteristics in the degree distributions of the full dry bulk network and sub-networks, indicating that all networks follow power-law behaviour. Figure~\ref{log_log_degree} illustrates the linear relationship between log-frequency and log-degree, which is consistent with the regression in Equation~\eqref{eqn:reg}, the log-transformed form of the power-law model $f(k) = C \, k^{-\gamma}$:

\begin{equation}\label{eqn:reg}
\log\bigl(f(k)\bigr) = \log(C) \;-\; \gamma\,\log(k).
\end{equation}

Here, $\gamma$ denotes the power-law exponent, and $C$ is the proportionality constant. Table~\ref{tbl:powerlaw} summarises the estimated exponents for the in-degree, out-degree, and total-degree distributions across the grain, coal, and iron-ore sub-networks, as well as the full dry bulk network. Notably, in the coal, grain, and iron ore sub-networks, the in-degree exponents are consistently larger than 1, whereas the out-degree exponents are all around 0.7. Because a larger exponent corresponds to a faster-decaying tail, the in-degree distributions thus exhibit fewer extremely high-degree nodes, implying greater heterogeneity in discharging activities. Conversely, the smaller out-degree exponents indicate heavier-tailed distributions, suggesting the presence of hubs responsible for extensive exporting connectivity.

In comparison, the full dry bulk network exhibits a notably low exponent for its total-degree distribution, indicating a strong concentration of connections among a relatively small number of nodes. However, its in-degree and out-degree exponents are both around 0.8, suggesting a more symmetric division between inbound and outbound connections. 

Across all networks, the lower out-degree exponents consistently point to a small set of key exporters (or load ports) that link to many discharge ports. Nevertheless, when combining all commodities, the dry bulk shipping network as a whole appears more balanced than the individual commodity-specific networks. Taken together, these findings show the complexity and heterogeneity of the sub-networks, as well as the inherent trade imbalances and specialised roles that shape global shipping.

\begin{table}[ht!]
\tiny
\centering
\begin{tabular}{@{}p{3cm}p{3cm}p{3cm}p{3cm}@{}}
\toprule
\textbf{} & \textbf{$\gamma\_all$} & \textbf{$\gamma\_indegree$} & \textbf{$\gamma\_outdegree$} \\
\midrule
Dry Bulk               & 0.704               & 0.834              & 0.811               \\
Coal                   & 1.075               & 1.331              & 0.749               \\
Grain                  & 1.010               & 1.207              & 0.710               \\
Iron Ore               & 1.069               & 1.328              & 0.761   \\
\bottomrule
\end{tabular}
\caption{\textbf{Estimated parameters of the power-law distributions across different sub-networks.} Estimated power-law exponents ($\gamma$) for different sub-networks, including: $\gamma\_{all}$ (total degree, combining in-degree and out-degree), $\gamma\_{indegree}$ (inbound connections), and $\gamma\_{outdegree}$ (outbound connections). Each row corresponds to a specific commodity sub-network as well as the full dry bulk network.}
\label{tbl:powerlaw}
\end{table}

While our findings focus on trade flow, traditional transport network studies tend to highlight a highly hub-spoke transportation system structure. These studies further infer that the reliance on central hubs can lead to vulnerability within shipping networks \citep{ducruet2010ports,ducruet2010centrality, kaluza2010complex, ducruet2012worldwide, ducruet2013network,williams2014degree,liu2018spatial,pan2019connectivity}. Consequently, the concentration of traffic through a few major ports raises concerns about resilience, as disruptions at these key ports can have disproportionate effects on the entire network. For example, targeted disruptions or failures at these hubs could severely impact maritime operations and, furthermore, significantly restrict global trade and supply chain operations from the transportation perspective. 

In contrast, our study draws on granular, micro-level maritime shipping trade-flow data from sub-networks to underscore the asymmetric demand and supply structures that underlie the hub-and-spoke system. Rather than viewing hubs solely as nodes with high traffic, we emphasise their role as critical import, export, or processing positions that consequently attract heavier traffic volumes, consistent with the traditional conception of hubs. One direct implication of this result is that policymakers and manufacturing stakeholders can use these structural insights when planning port expansions or identifying new hubs, thereby avoiding over-reliance on existing major centres.

\paragraph{Correlation matrix and assortativity} In addition to the trade imbalance shown by the power-law, we also examine the phenomena using the correlation matrices of weighted centrality measures. Table \ref{tab:corr} provides a quantitative overview of how each centrality measure is correlated. There are positive correlations within the same directional centralities across all networks. However, a notable absence of significant correlations is observed between the in- and out-centralities for the dry bulk and coal networks. In contrast, for the grain and iron ore networks, negative correlations are observed between in-strength and out-strength centralities. This suggests that ports with higher volumes of imports tend to export less and vice versa. The prominent asymmetric distribution of grain and iron ore globally show how resource rich regions only engage in exporting whereas resource scarce regions have high consumption demand.  

Together with the power-law exponents, this provides insight into worldwide trade imbalances. Imports tend to be more distributed across nodes whereas exports are more concentrated, and in addition exporting nodes import less and vice versa. Our results are consistent with \cite{williams2014degree}'s findings in the degree correlations of directed scale-free networks, and reveal that directed scale-free networks are largely uncorrelated concerning in-out degree correlations. 

In addition, \cite{williams2014degree} identify a notable disassortativity in the out-in correlation, where nodes with high out-degrees preferentially connect to those with low in-degrees. This further proves the trade imbalance where large exporting (or import) nations do not necessarily dominate the importing activities. In our study, as shown in Table \ref{tbl:glo_centrality}, we also observe negative assortativity across all sub-networks, though with values that are near zero. This is an indicator of the hub-spoke structure \citep{kaluza2010complex}. However, examining more granular commodity levels can reveal pronounced assortative patterns. For instance, we provide an example of detailed centrality measures for grain varieties in the Supplementary Material. This further proves that the dry bulk shipping network is a complex multi-layered network, with a wide variation in materials, and different forms of processed products being transported between resource rich and deficit regions.
\begin{table}[ht!]
\tiny
\centering
\begin{minipage}{0.45\textwidth}
    \centering
    \begin{tabular}{@{}cllllllll@{}}
    \toprule
    \multicolumn{1}{l}{} &
      \multicolumn{1}{c}{\textbf{k\_i}} &
      \multicolumn{1}{c}{\textbf{s\_i\_f}} &
      \multicolumn{1}{c}{\textbf{s\_i\_d}} &
      \multicolumn{1}{c}{\textbf{s\_i\_t}} &
      \multicolumn{1}{c}{\textbf{k\_o}} &
      \multicolumn{1}{c}{\textbf{s\_o\_f}} &
      \multicolumn{1}{c}{\textbf{s\_o\_d}} &
      \multicolumn{1}{c}{\textbf{s\_o\_t}} \\ \midrule
    \textbf{k\_i}    & 1    & 0.89 & 0.72 & 0.66 & 0.69 & 0.42 & 0.16 & 0.12 \\
    \textbf{s\_i\_f} & 0.89 & 1    & 0.85 & 0.80 & 0.58 & 0.39 & 0.13 & 0.10 \\
    \textbf{s\_i\_d} & 0.72 & 0.85 & 1    & 0.98 & 0.50 & 0.35 & 0.14 & 0.11 \\
    \textbf{s\_i\_t} & 0.66 & 0.80 & 0.98 & 1    & 0.47 & 0.34 & 0.15 & 0.11 \\
    \textbf{k\_o}    & 0.69 & 0.58 & 0.50 & 0.47 & 1    & 0.72 & 0.40 & 0.35 \\
    \textbf{s\_o\_f} & 0.42 & 0.39 & 0.35 & 0.34 & 0.72 & 1    & 0.73 & 0.69 \\
    \textbf{s\_o\_d} & 0.16 & 0.13 & 0.14 & 0.15 & 0.40 & 0.73 & 1    & 0.99 \\
    \textbf{s\_o\_t} & 0.12 & 0.10 & 0.11 & 0.11 & 0.35 & 0.69 & 0.99 & 1    \\ \bottomrule
    \end{tabular}
    \caption*{(a) Dry bulk}
    \label{tab:table_a}
\end{minipage}\hfill
\begin{minipage}{0.45\textwidth}
    \centering
    \begin{tabular}{@{}cllllllll@{}}
\toprule
\multicolumn{1}{l}{} &
  \multicolumn{1}{c}{\textbf{k\_i}} &
  \multicolumn{1}{c}{\textbf{s\_i\_f}} &
  \multicolumn{1}{c}{\textbf{s\_i\_d}} &
  \multicolumn{1}{c}{\textbf{s\_i\_t}} &
  \multicolumn{1}{c}{\textbf{k\_o}} &
  \multicolumn{1}{c}{\textbf{s\_o\_f}} &
  \multicolumn{1}{c}{\textbf{s\_o\_d}} &
  \multicolumn{1}{c}{\textbf{s\_o\_t}} \\ \midrule
\textbf{k\_i}    & 1    & 0.81 & 0.80 & 0.80 & 0.12 & 0.11 & 0.08 & 0.09 \\
\textbf{s\_i\_f} & 0.81 & 1    & 0.96 & 0.96 & 0.03 & 0.01 & 0.00 & 0.01 \\
\textbf{s\_i\_d} & 0.80 & 0.96 & 1    & 0.98 & 0.03 & 0.01 & 0.00 & 0.01 \\
\textbf{s\_i\_t} & 0.80 & 0.96 & 0.98 & 1    & 0.02 & 0.01 & 0.00 & 0.00 \\
\textbf{k\_o}    & 0.12 & 0.03 & 0.03 & 0.02 & 1    & 0.75 & 0.71 & 0.72 \\
\textbf{s\_o\_f} & 0.11 & 0.01 & 0.01 & 0.01 & 0.75 & 1    & 0.95 & 0.98 \\
\textbf{s\_o\_d} & 0.08 & 0.00 & 0.00 & 0.00 & 0.71 & 0.95 & 1    & 0.99 \\
\textbf{s\_o\_t} & 0.09 & 0.01 & 0.01 & 0.00 & 0.72 & 0.98 & 0.99 & 1    \\ \bottomrule
\end{tabular}
    \caption*{(b) Coal}
    \label{tab:table_b}
\end{minipage}

\vspace{1cm} 

\begin{minipage}{0.45\textwidth}
    \centering
\begin{tabular}{@{}cllllllll@{}}
\toprule
\multicolumn{1}{l}{} &
  \multicolumn{1}{c}{\textbf{k\_i}} &
  \multicolumn{1}{c}{\textbf{s\_i\_f}} &
  \multicolumn{1}{c}{\textbf{s\_i\_d}} &
  \multicolumn{1}{c}{\textbf{s\_i\_t}} &
  \multicolumn{1}{c}{\textbf{k\_o}} &
  \multicolumn{1}{c}{\textbf{s\_o\_f}} &
  \multicolumn{1}{c}{\textbf{s\_o\_d}} &
  \multicolumn{1}{c}{\textbf{s\_o\_t}} \\ \midrule
\textbf{k\_i}    & 1    & 0.83  & 0.77  & 0.77  & 0.05  & 0.02  & 0.03  & 0.04  \\
\textbf{s\_i\_f} & 0.83 & 1     & 0.97  & 0.94  & 0.01  & -0.01 & 0.00  & 0.00  \\
\textbf{s\_i\_d} & 0.77 & 0.97  & 1     & 0.96  & -0.01 & -0.02 & -0.02 & -0.02 \\
\textbf{s\_i\_t} & 0.77 & 0.94  & 0.96  & 1     & 0.00  & -0.01 & -0.01 & -0.01 \\
\textbf{k\_o}    & 0.05 & 0.01  & -0.01 & 0.00  & 1     & 0.84  & 0.80  & 0.81  \\
\textbf{s\_o\_f} & 0.02 & -0.01 & -0.02 & -0.01 & 0.84  & 1     & 0.98  & 0.97  \\
\textbf{s\_o\_d} & 0.03 & 0.00  & -0.02 & -0.01 & 0.80  & 0.98  & 1     & 0.98  \\
\textbf{s\_o\_t} & 0.04 & 0.00  & -0.02 & -0.01 & 0.81  & 0.97  & 0.98  & 1     \\ \bottomrule
\end{tabular}
    \caption*{(a) Grain}
    \label{tab:table_c}
\end{minipage}\hfill
\begin{minipage}{0.45\textwidth}
    \centering
\begin{tabular}{@{}lllllllll@{}}
\toprule
 & \textbf{k\_i} & \textbf{s\_i\_f} & \textbf{s\_i\_d} & \textbf{s\_i\_t} & \textbf{k\_o} & \textbf{s\_o\_f} & \textbf{s\_o\_d} & \textbf{s\_o\_t} \\ \midrule
\textbf{k\_i}    & 1     & 0.86  & 0.83  & 0.79  & 0.07 & -0.01 & -0.02 & -0.02 \\
\textbf{s\_i\_f} & 0.86  & 1     & 0.99  & 0.97  & 0.05 & 0.00  & -0.01 & -0.01 \\
\textbf{s\_i\_d} & 0.83  & 0.99  & 1     & 0.99  & 0.06 & 0.00  & -0.01 & -0.01 \\
\textbf{s\_i\_t} & 0.79  & 0.97  & 0.99  & 1     & 0.07 & 0.00  & -0.01 & -0.01 \\
\textbf{k\_o}    & 0.07  & 0.05  & 0.06  & 0.07  & 1    & 0.51  & 0.46  & 0.46  \\
\textbf{s\_o\_f} & -0.01 & 0.00  & 0.00  & 0.00  & 0.51 & 1     & 0.99  & 0.99  \\
\textbf{s\_o\_d} & -0.02 & -0.01 & -0.01 & -0.01 & 0.46 & 0.99  & 1     & 1.00  \\
\textbf{s\_o\_t} & -0.02 & -0.01 & -0.01 & -0.01 & 0.46 & 0.99  & 1.00  & 1     \\ \bottomrule
\end{tabular}
    \caption*{(d) Iron ore}
    \label{tab:table_d}
\end{minipage}
\caption{\textbf{Correlation of the weighted network centrality measures.} This matrix compares eight different centrality metrics. In order: (1) in-degree, (2) in-strength weighted by node frequency, , (3) in-strength weighted by deadweight ton (DWT), (4) in-strength weighted by cargo volume (metric tons), (5) out-degree, (6) out-strength weighted by node frequency, (7) out-strength weighted by DWT, (8) out-strength weighted by cargo volume. A higher positive correlation (closer to 1) indicates that the two measures tend to rank ports similarly in terms of importance, while a negative correlation suggests that the measures capture opposite facets of a port’s network role. Examining these correlations helps clarify whether, for instance, ports with high in-degree, meaning high connectivity with other ports, also handle large inbound volumes with frequent voyages, and whether they engage in export activities as well, or vice versa. Subtable (a) shows the correlations for the Dry bulk sub-network; (b) for Coal; (c) for Grain; and (d) for Iron Ore.}
\label{tab:corr}
\end{table}
\clearpage

Finally, in addition to centrality measures, we use maps to illustrate the spatial features of maritime trade flows which are vital characteristics of shipping networks. Map (a) displays the complete dry bulk trade flows, demonstrating the significance of maritime shipping trade worldwide, as nearly every country along the coastline is involved in some form of import or export activities.

Map (b) illustrates the trade flow of coal, Map (c) depicts grain, and Map (d) shows iron ore. These maps clearly demonstrate the geographic locations of demand and supply, with blue indicating importing regions (commodity deficit) and red indicating exporting regions (commodity surplus). When ports engage in both load and discharge activities, the blue and red overlap with the highest centrality dominating. Ports in red are larger on average than ports in blue, consistent with the power-law exponent estimate that the out-degree distribution contains more large hubs than the in-degree distributions, especially for the cargo-specific sub-networks.

Notably, for coal, major exporting countries include Australia, Indonesia, Russia, South Africa, Mozambique and the North America. For grain, Ukraine in Europe, Brazil, Argentina, and North America, along with Australia, play pivotal roles as significant exporters. Iron ore exports are more scattered globally, including Ukraine as a major resource-rich country, with China being the largest importer. In contrast, the importing countries for grain cover a vast area, especially in Africa, and Middle East.

Additionally, we observe that Australia plays a major role in these trade flows, possessing abundant resources such as coal, grain, and iron ore. Notably, the trading routes for coal are primarily concentrated in the eastern part of Australia, while iron ore routes are focused in the west, and grain routes are predominantly in South Australia.

This geographic information confirms that the well-connected dry bulk network results from complex, multi-layered trade sub-networks. For commodity traders and ship operators, it is crucial to understand the particularities of the dynamics of global trade flows and their intricate network structures (including country- and node-specific patterns) to make strategic decisions that maximise their operational efficiency.

\begin{figure}[htbp]
    \centering
    \begin{subfigure}{\linewidth}
        \includegraphics[width=\linewidth]{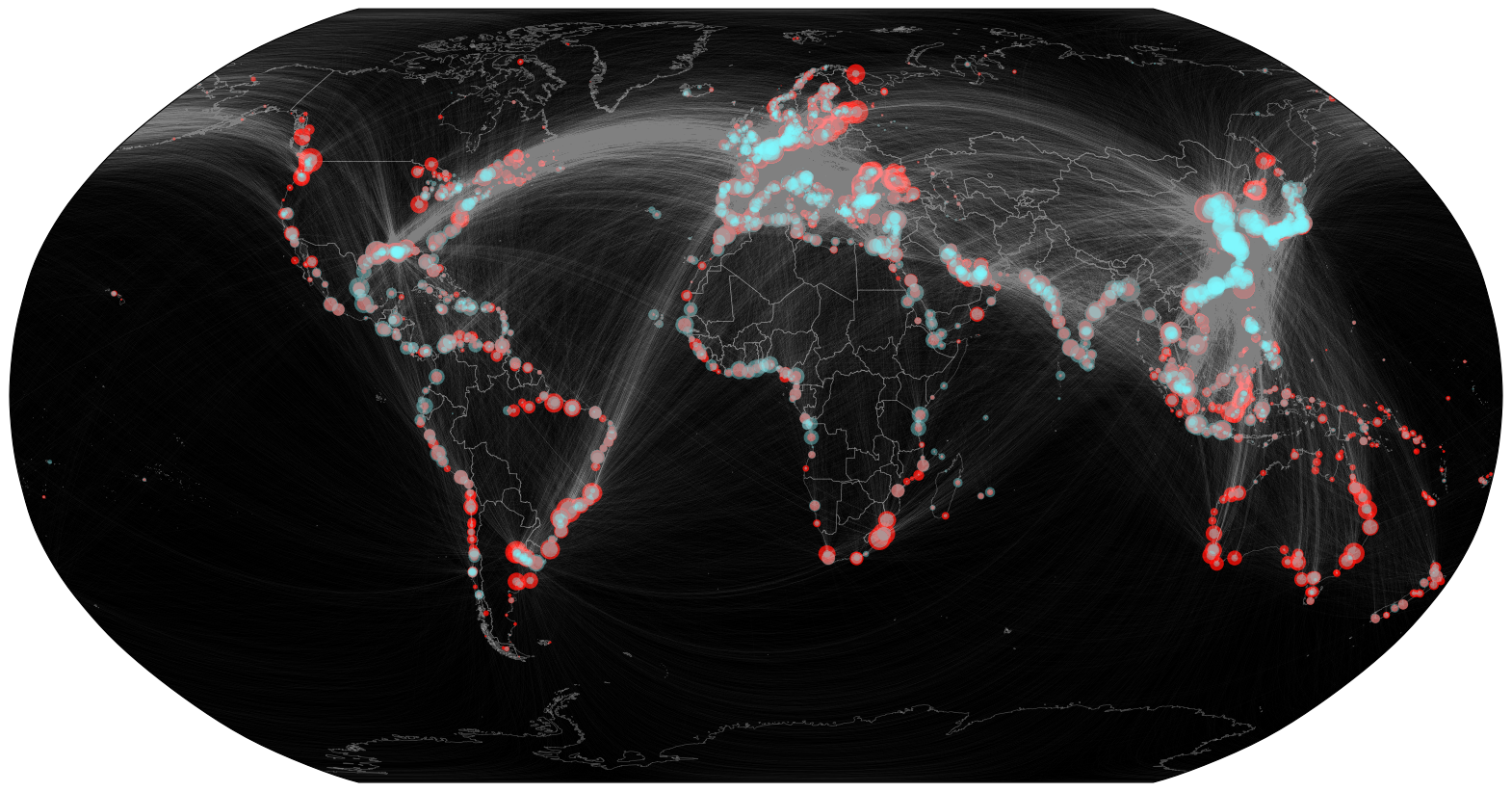}
        \caption*{(a) Full dry bulk trade flow network}
    \end{subfigure}
    
    \vspace{0.5cm}

    \begin{subfigure}{\linewidth}
        \includegraphics[width=\linewidth]{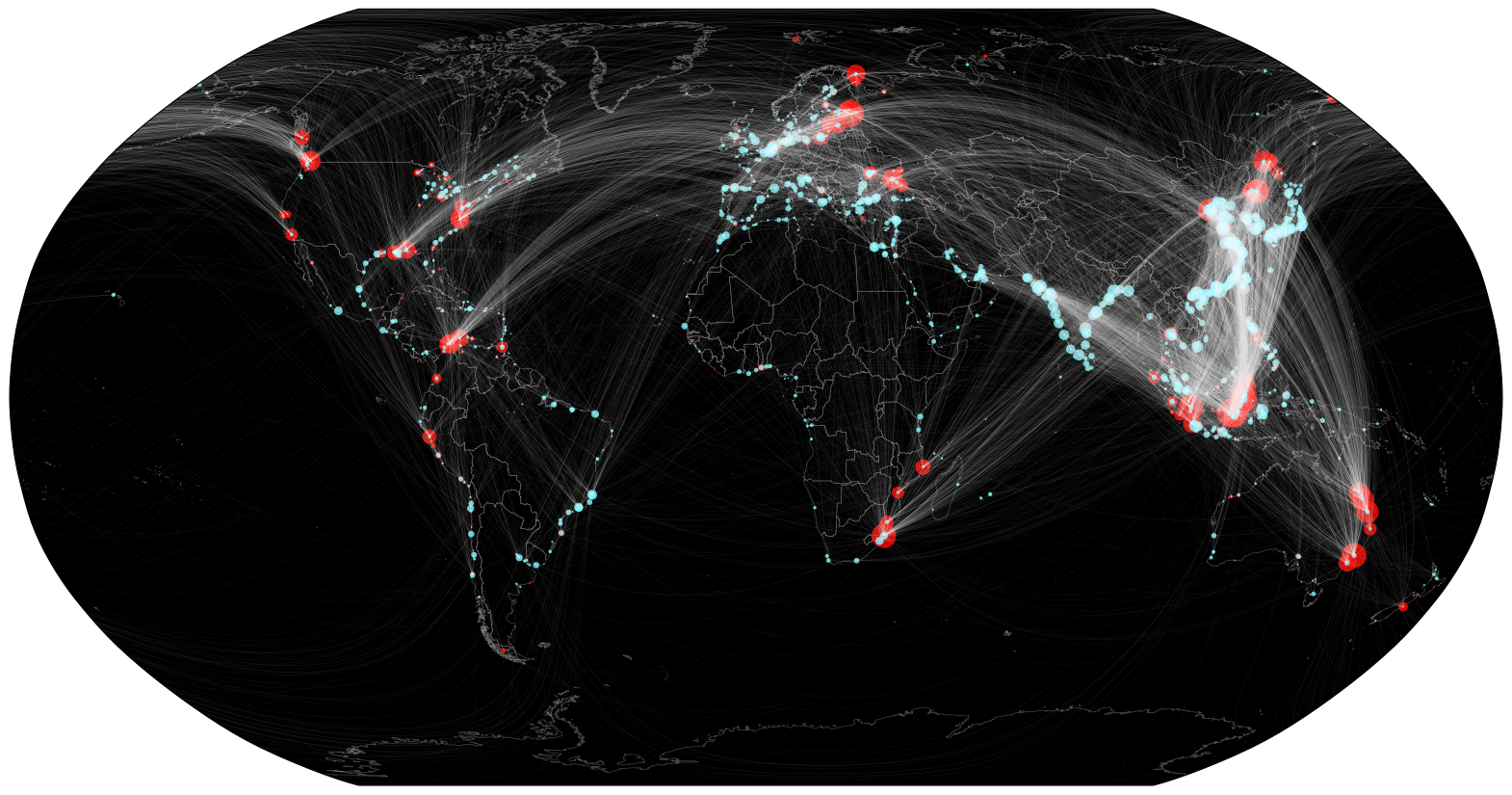}
        \caption*{(b) Coal cargo trade flow sub-network}
    \end{subfigure}
\end{figure}

\begin{figure}[htbp]\ContinuedFloat
    \centering
    \begin{subfigure}{\linewidth}
        \includegraphics[width=\linewidth]{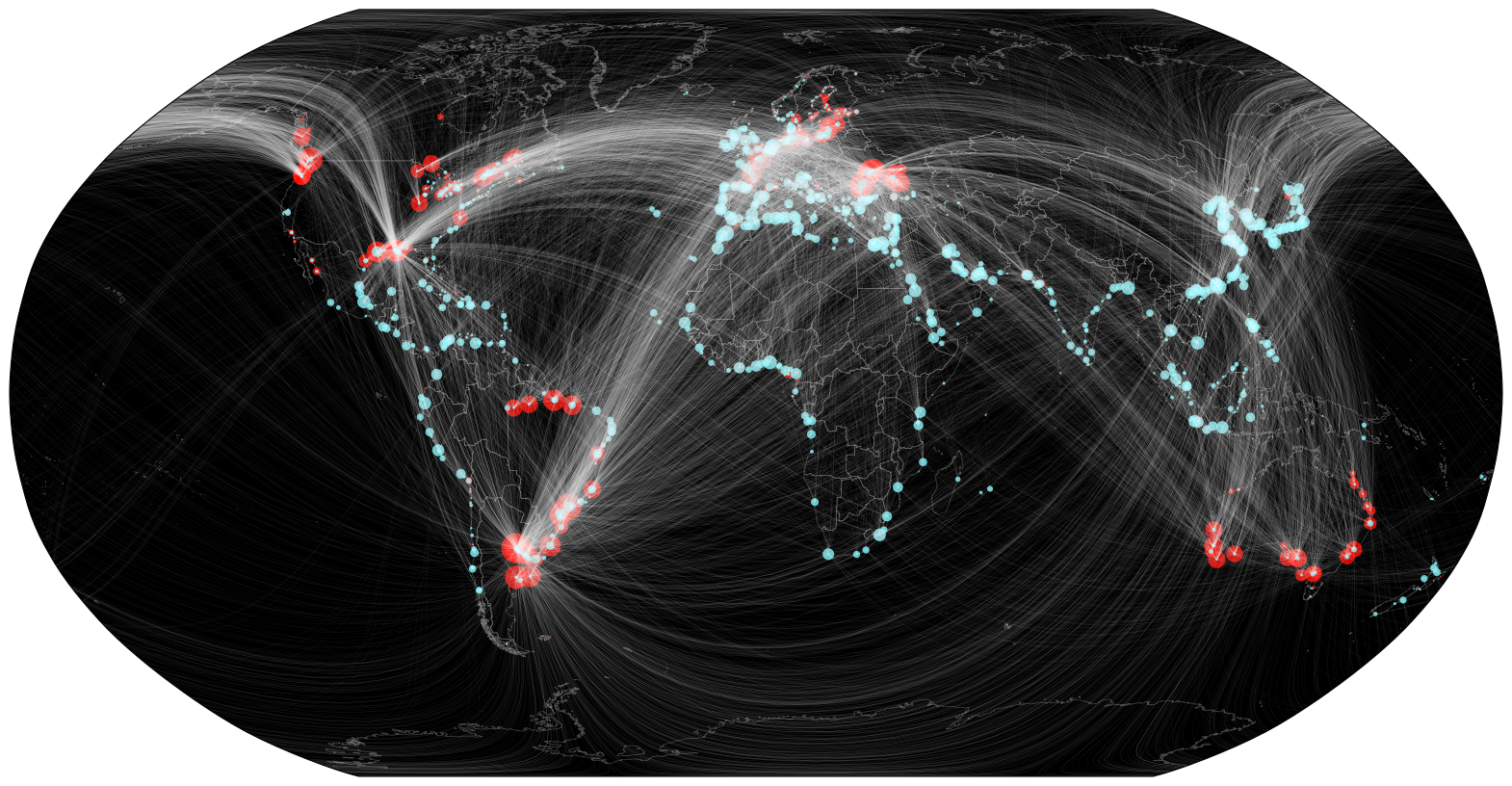}
        \caption*{(c) Grain trade flow sub-network}
    \end{subfigure}

    \vspace{0.5cm}

    \begin{subfigure}{\linewidth}
        \includegraphics[width=\linewidth]{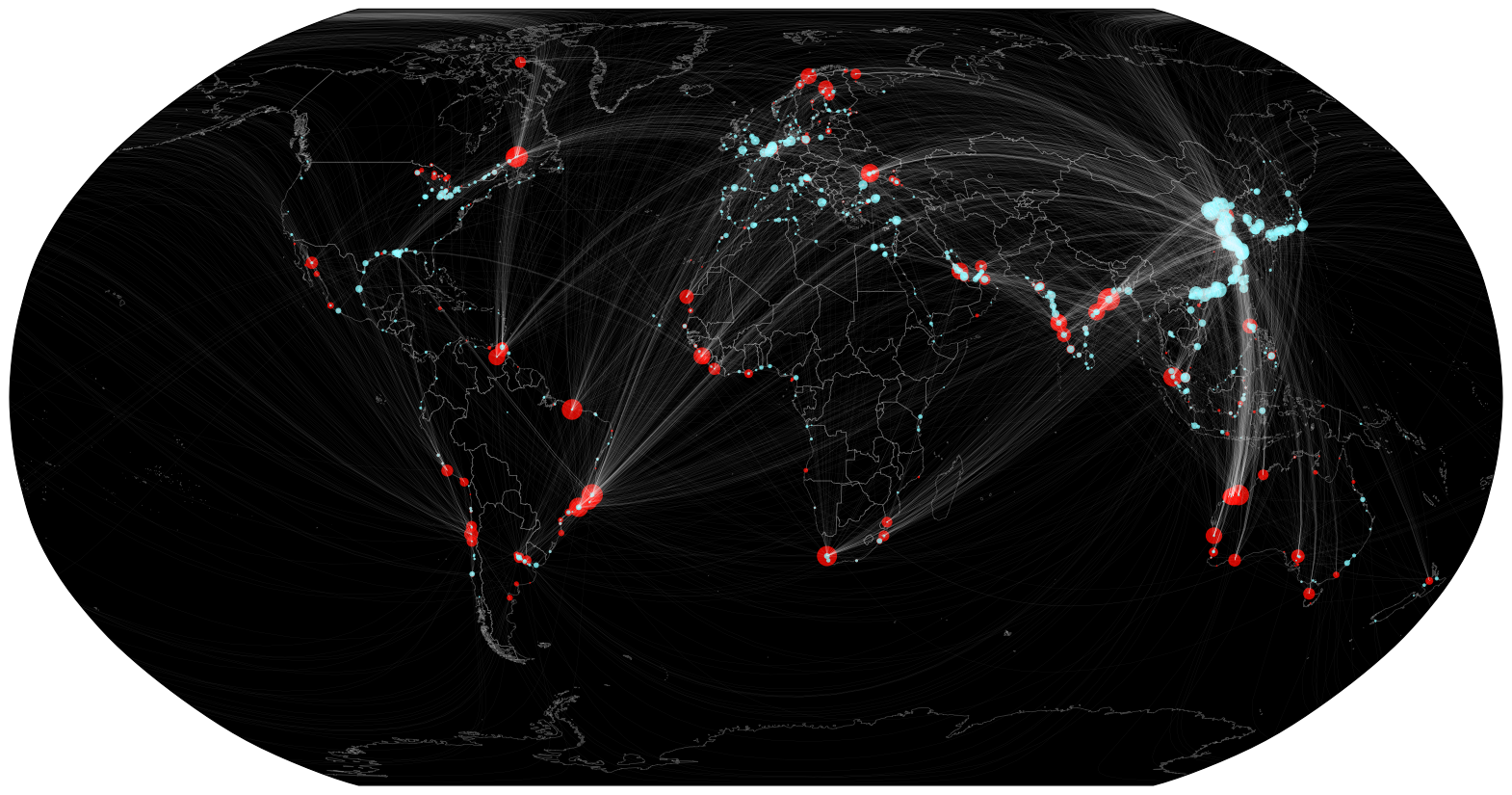}
        \caption*{(d) Iron ore trade flow sub-network}
    \end{subfigure}

    \caption{\textbf{Global dry bulk trade flow networks.}
        The maps illustrate the trade flow networks for (a) the full dry bulk network, and its sub-networks: (b) coal, (c) grain, and (d) iron ore. The size of each node reflects the highest of its normalised in-degree or out-degree centralities. The nodes in red are discharge ports reflecting the out-degrees and nodes in blue are load ports reflecting in-degrees. When ports have both load and discharge activities, the two colours overlap, with the highest centrality dominating. For Map (a), we reduce line width and opacity to improve comprehension.}
    \label{fig:all_maps}
\end{figure}
\clearpage

\paragraph{Component analysis} Having examined the size and the degree distribution of the sub-networks, we now focus on component analysis, which is a crucial factor in understanding connectivity within a directed network \cite{Kiss2006}. Our analysis considers weakly connected components (WCCs), strongly connected components (SCCs), and the Giant Strongly Connected Component (GSCC).

Weakly Connected Components (WCCs) are sets where nodes remain connected without considering the directions of the edges. As shown in Table \ref{tbl:glo_centrality}, having a single WCC for both the entire network and the main sub-networks indicates complete connectivity and reduces logistical barriers in international dry bulk trade.

In contrast, SCCs are subsets where directed paths connect nodes within the sub-graphs, whereas GSCC represents the largest of these components. In the context of the trade network, sub-graphs outside the GSCC reveal asymmetric trading patterns, where many ports primarily engage in either importing or exporting with either zero out-degree or in-degrees, but not both simultaneously. Nodes within the GSCC, however, often indicate bilateral trade flows between regions. These flows encompass the exchange of both raw materials and processed products, supporting \cite{rodrigue2006challenging}'s argument about the importance of intermediate locations in supply chains. In summary, the component analysis illustrates the complexity and interconnectedness of global intermediate commodities markets as well as highlights global trade imbalances. 

This phenomenon is illustrated by the roles of specific ports within global trade networks. Here, some serve primarily as major importers of raw materials due to their large economic demand as well as local processing capabilities, while others balance imports and exports by refining raw materials into goods for re-export of the excess of materials. These dynamics underscore the GSCCs as indicators of trade efficiency and of strategic hubs within the network. Ports within GSCC play central roles in the flow of goods, enhancing network efficiency by minimising steps needed to convert raw materials into exportable products. Moreover, they significantly influence trade routes by reducing logistical costs and transit times as they are able to leverage their dual role in processing and redistributing goods. This central positioning is key to optimising trade dynamics and enhancing resilience in the global dry bulk networks.

The diameter of a WCC represents the shortest path between the most distant ports in a weakly connected network, shown in Table \ref{tbl:glo_centrality}, which is notably small for the dry bulk, coal, and grain networks relative to their individual sizes. This observation suggests that the longest paths within sub-networks closely approximate those in the full network. Additionally, the diameter of each GSCC is approximately two steps shorter than that of its corresponding WCC, highlighting greater connectivity efficiency within the GSCCs. Calculating the average path length (a metric based on GSCCs due to their bi-directional edges) reveals that the aggregate dry bulk network is more dense than individual commodity networks. Thus, it exhibits a smaller average path length, consistent with findings by \cite{kaluza2010complex}, which further emphasises robust internal connectivity.

Similarly, the transitivity, also referred to as the clustering coefficient, of the full dry bulk network stands at 0.23, in contrast to the 0.43 noted by \cite{kaluza2010complex}. This discrepancy likely results from differing data sources; our reliance on laden-only trade flows omits ballast journeys, thereby reducing clique formation. This is illustrated in Figure \ref{fig:multiloadsdischarges}, where differences between port calls and trade flow datasets affect cluster formation, particularly regarding multiple loads and discharges. In scenarios where ballast legs are recorded, cliques form as the journey returns to the initial loading port. However, such cliques do not form in trade flow (laden legs only) datasets. Hence, at the sub-network level, low transitivity indicates smaller clusters. However, when integrating these into a comprehensive network, it suggests the potential for more cliques, reflecting robust inter-port trading connection.

In summary, analysing global centralities across networks provides an overview of size, connectivity, and clustering characteristics. The distinct patterns of supply and demand within each sub-network contribute to unique structural features, which offer valuable insights into the network's trade dynamics and operational efficiencies. The dry bulk network, integrating all sub-networks, is thus more condensed, featuring higher average degrees, larger GSCCs, reduced path lengths, and more balanced assortativity. This configuration promotes higher ship utilisation when chartering vessels for multiple commodities across the network.

Through component analysis, the GSCC topology critically impacts trade costs by enabling more direct, reciprocal routes, minimising empty return trips (ballast journeys) and thus reducing overall trade costs. Larger GSCC sizes correspond with fewer ballast legs in international trade. As noted by \cite{brancaccio2020geography}, while ballast journeys often raise costs, assessing GSCC structures more clearly distinguishes effective bi-lateral trades, emphasising segments prone to inefficiencies.

\paragraph{Small world of sub-networks}
Understanding small-world properties provides critical insights into trade network efficiency. Using the density of real networks, we simulate rewired random networks and employ the average degree to model a regular lattice network for each type of cargo network. Subsequently, we calculate the clustering coefficients ($C_{\text{rand}}, C_{\text{latt}}$) for both simulated networks, as well as the average path length ($L_{\text{rand}}$) for the simulated random network. We then compare these simulated statistics with the observed values. 

Finally, we apply the metrics $\sigma$ and $\omega$ as defined by \cite{telesford2011ubiquity} and \cite{humphries2008network} to quantify whether the observed networks exhibit small-world characteristics. A $\sigma > 1$ value supports the classification of a small-world network, as it indicates $C \gg C_{\text{rand}}$ and $L \approx L_{\text{rand}}$. Similarly, when $\omega$ is close to zero, the network is classified as small-world because $L \approx L_{\text{rand}}$ and $C \approx C_{\text{latt}}$. In contrast, a positive $\omega$ value ($\frac{L_{\text{rand}}}{L} > \frac{C}{C_{\text{latt}}}$) suggests that the network exhibits more random characteristics, while a negative $\omega$ value indicates features typical of regular networks.

The results indicate that the observed average path length is larger than, but still close to, that of the simulated random network ($L > L_{\text{rand}}$) across all sub-networks, with the exception of grain. In contrast, the clustering coefficients reveal that $C_{\text{rand}} \ll C_{\text{latt}}$, consistent with the expectation that the clustering coefficient of a random network is smaller than that of a lattice network. However, for coal, grain, and iron ore, we observe $C \ll C_{\text{rand}} \ll C_{\text{latt}}$, suggesting that these networks exhibit random, rather than small-world, characteristics. The full dry bulk network, however, demonstrates some small-world features, as $L \approx L_{\text{rand}}$ and $C \gg C_{\text{rand}}$, with $\sigma$ close to 1 and $\omega$ near 0. These findings present an intriguing distinction that warrants further investigation.

Although the $\omega$ measurements for coal, grain, and iron ore, as shown in Table \ref{tab:smallworld}, suggest small-world tendencies with more random network features, closer analysis indicates that these findings may misrepresent the overall connectivity. The notably low clustering coefficients in the simulated random networks suggest that the average path length disproportionately influences the results, overshadowing the role of clustering.

\begin{table}[ht!]
\tiny
\centering
\begin{tabular}{@{}p{0.5cm}p{1.5cm}p{1.5cm}p{1.5cm}p{1.5cm}p{1.5cm}p{1.5cm}p{1.5cm}p{1.5cm}@{}}
\toprule
\multicolumn{1}{l}{} &
  \multicolumn{1}{l}{\textbf{Network}} &
  \multicolumn{1}{l}{\textbf{L}} &
  \multicolumn{1}{l}{\textbf{L\_rand}} &
  \multicolumn{1}{l}{\textbf{C*}} &
  \multicolumn{1}{l}{\textbf{C\_rand}} &
  \multicolumn{1}{l}{\textbf{C\_latt}} &
  \multicolumn{1}{l}{\textbf{$\omega$}} &
  \multicolumn{1}{l}{\textbf{$\sigma$}} \\ \midrule
\textbf{1} & Dry bulk & 2.504 & 2.355 & 0.231 & 0.223 & 0.744 & 0.631 & 0.971 \\
\textbf{2} & Grain    & 3.146 & 3.298 & 0.040 & 0.146 & 0.724 & 0.993 & 0.287 \\
\textbf{3} & Coal     & 3.654 & 3.241 & 0.049 & 0.153 & 0.717 & 0.819 & 0.282 \\
\textbf{4} & Iron Ore & 4.120 & 3.581 & 0.049 & 0.150 & 0.682 & 0.797 & 0.285 \\ \bottomrule
\end{tabular}
\caption{\textbf{Network statistics of dry bulk sub-networks} L denotes the observed network average path length, $L_\text{rand}$ is the average path length of the rewired random graph, C is the clustering coefficient of the observed network, $C_\text{rand}$ and $C_\text{latt}$ represent the clustering coefficient of the rewired random graph and simulated lattice network respectively, $\omega$ and $\sigma$ are the small-world measures defined by \cite{telesford2011ubiquity,humphries2008network}.}\label{tab:smallworld}
\end{table}

We argue that the lack of small-world properties in the larger networks can be identified from the weakly connected nature of the directed graphs. As shown in Table \ref{tbl:glo_centrality}, the GSCC sub-graphs for coal, grain, and iron ore networks encompass less than 50\% of their original network sizes, which contributes to fewer connections and lower clustering coefficients across the broader network. To investigate this further, we conducted additional simulations focusing on the GSCCs' structure, particularly examining average path lengths and clustering coefficients. These simulations revealed that the GSCCs have higher densities and clustering coefficients, indicating that while the overall network does not exhibit small-world characteristics, individual GSCCs do demonstrate these properties. 

The results in Table \ref{tab:gscc} illustrate that the newly simulated graphs maintain the same number of nodes and edges as the original GSCC components. Each GSCC is strongly connected, demonstrated by a single component structure, and its density surpasses that of the corresponding full network. This is particularly true for the grain, coal, and iron ore sub-networks, where the densities are significantly higher. Such patterns are highlighted by the enhanced connectivity within the GSCC sub-graphs. This results in greater clustering coefficients across these sub-networks. Notably, when comparing the simulated rewired random networks of each GSCC with the original ones, we observe that $C_{\text{rand}} << C$, contrasting with the results of the full networks for coal, grain, and iron ore. In addition, we get the average path length being close to the observed values, $L \approx L_\text{rand}$. This comparison suggests that while the broader networks do not exhibit small-world properties, the GSCCs are small-world in nature, differentiating them from equivalent random networks. 

\begin{table}[ht!]
\centering
\tiny
\begin{tabular}{@{}l*{8}{p{1.25cm}}@{}}
\toprule
\multicolumn{1}{l}{\textbf{}} &
  \multicolumn{2}{c}{\textbf{Dry bulk}} &
  \multicolumn{2}{c}{\textbf{Grain}} &
  \multicolumn{2}{c}{\textbf{Coal}} &
  \multicolumn{2}{c}{\textbf{Iron Ore}} 
  \\ \specialrule{0.001pt}{0pt}{0pt} 
\textbf{GSCC} &
  \multicolumn{1}{c}{\textbf{obs}} &
  \multicolumn{1}{c}{\textbf{sim.}} &
  \multicolumn{1}{c}{\textbf{obs}} &
  \multicolumn{1}{c}{\textbf{sim.}} &
  \multicolumn{1}{c}{\textbf{obs}} &
  \multicolumn{1}{c}{\textbf{sim.}} &
  \multicolumn{1}{c}{\textbf{obs}} &
  \multicolumn{1}{c}{\textbf{sim.}} \\
  \midrule
\textbf{$\phi$}                & 0.04    & 0.04    & 0.05   & 0.05   & 0.04   & 0.04   & 0.06  & 0.06  \\
\textbf{$e$}              & 164,110 & 164,110 & 10,985 & 10,985 & 10,968 & 10,968 & 1,712 & 1,712 \\
\textbf{$n$}       & 2,062   & 2,062   & 489    & 489    & 541    & 541    & 164   & 164   \\
\textbf{$n_s$}               & 1       & 1       & 1      & 1      & 1      & 1      & 1     & 1     \\
\textbf{$a$}          & -0.09   & -0.19   & -0.07  & -0.31  & -0.11  & -0.27  & -0.16 & -0.21 \\
\textbf{$l$}        & 2.50    & 2.36    & 3.15   & 3.32   & 3.65   & 3.33   & 4.12  & 4.54  \\
\textbf{$c$} & 0.50    & 0.36    & 0.36   & 0.31   & 0.47   & 0.32   & 0.50  & 0.32  \\ \bottomrule
\end{tabular}
\caption{\textbf{GSCC network statistics of real dry bulk sub-networks and simulated networks.} The centrality measures of each network include: network density $\phi$; number of unique edges e; number of ports n; the
number of strongly-connected components (SCCs) $n_s$; degree assortativity a; the average shortest path length l, and clustering coefficient c.} 
\label{tab:gscc}
\end{table}

The small-world nature of the GSCCs implies that ports within these networks can be accessed quickly through a few intermediary connections, significantly reducing transport times and costs associated with trade. The high clustering observed indicates that ports with frequent trade interactions tend to form tightly knit networks, fostering enhanced logistical coordination and trade efficiency.

The entire dry bulk network does exhibit small-world properties, which are augmented by the characteristics of its GSCCs, consistent with \cite{kaluza2010complex}'s findings. This small-world nature indicates efficient connectivity and rapid navigability within the network, allowing ports to be reached through relatively few intermediary connections, often facilitated by a combination of cargo varieties. 

Understanding the small-world properties of the dry bulk shipping network is crucial for stakeholders in the shipping industry. Insights from this analysis can guide the optimisation of trade routes, thereby reducing transit times and trade costs by minimizing instances of ballast journeys. Stakeholders can thereby develop more effective operational strategies that ensure responsiveness to market demands and disruptions while improving their overall trade efficiency.

\subsection{Temporal network}
\subsubsection{Network structural change}
\begin{figure}[ht!]
    \centering
    \begin{minipage}{0.5\textwidth}
        \centering
        \includegraphics[width=\textwidth]{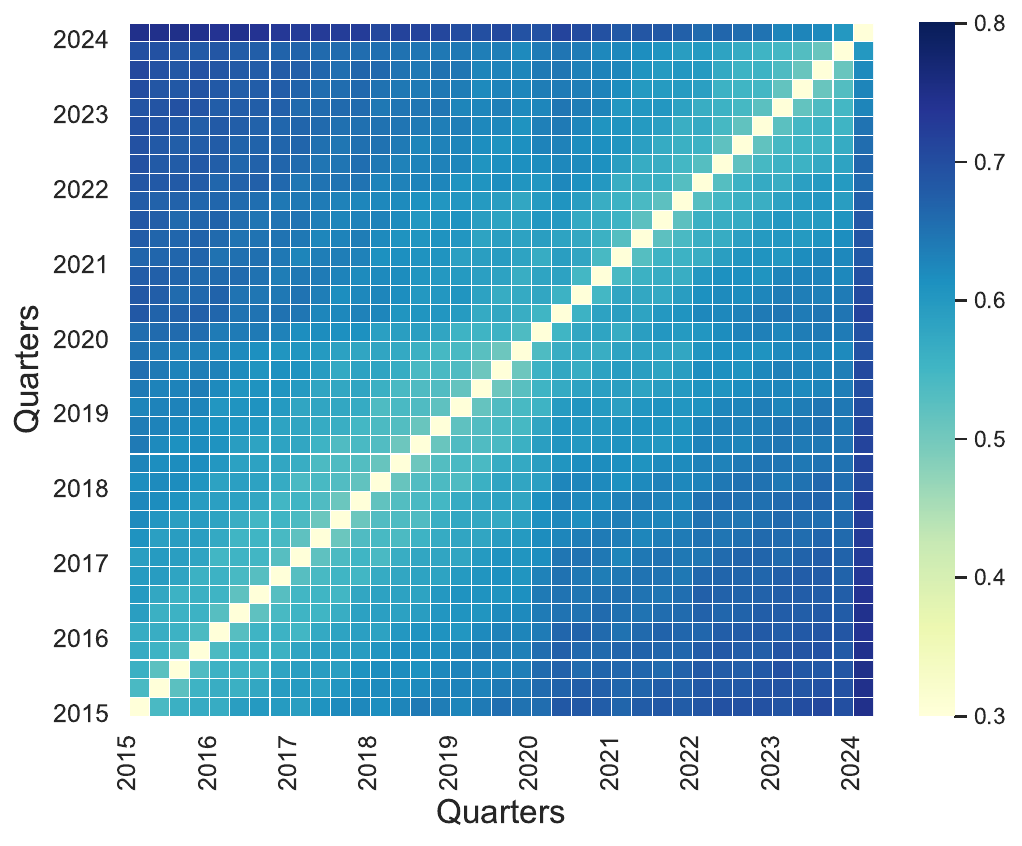}
        \caption*{(a) Dry bulk}
        \label{fig:distance_dry}
    \end{minipage}\hfill
    \begin{minipage}{0.5\textwidth}
        \centering
        \includegraphics[width=\textwidth]{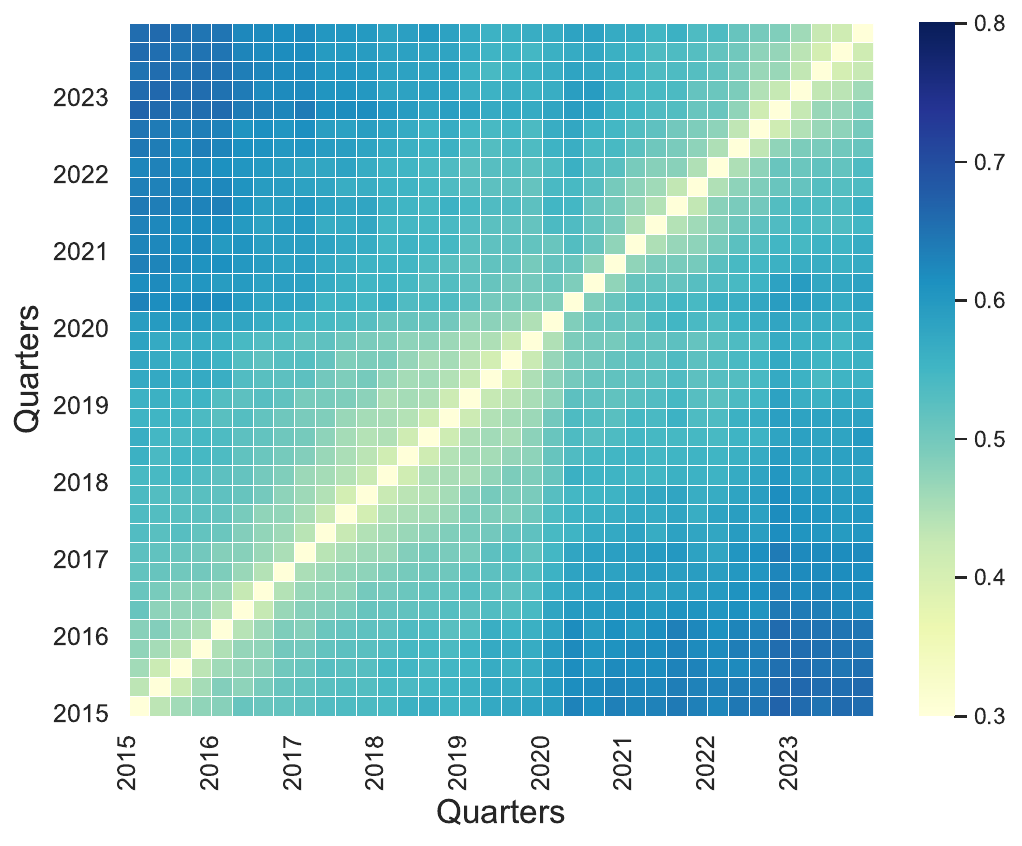}
        \caption*{(b) Coal}
        \label{fig:distance_coal}
    \end{minipage}
    \vspace{0.5cm}
    \begin{minipage}{0.5\textwidth}
        \centering
        \includegraphics[width=\textwidth]{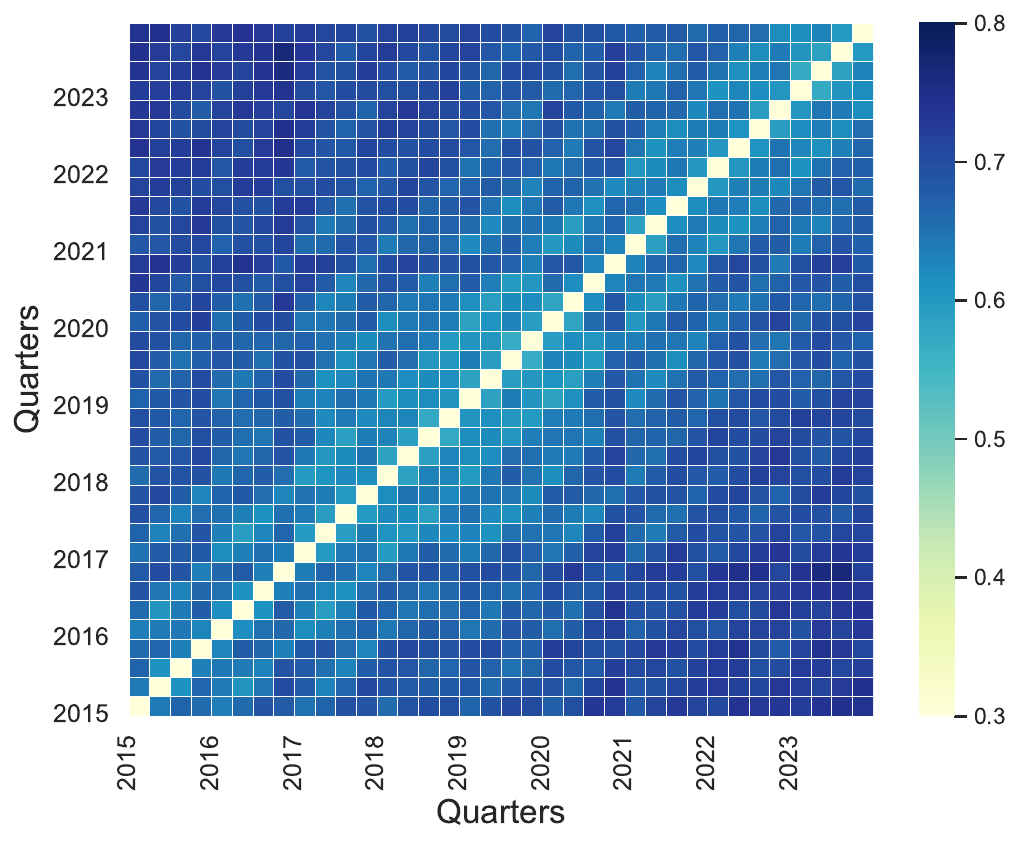}
        \caption*{(c) Grain}
        \label{fig:distance_grain}
    \end{minipage}\hfill
    \begin{minipage}{0.5\textwidth}
        \centering
        \includegraphics[width=\textwidth]{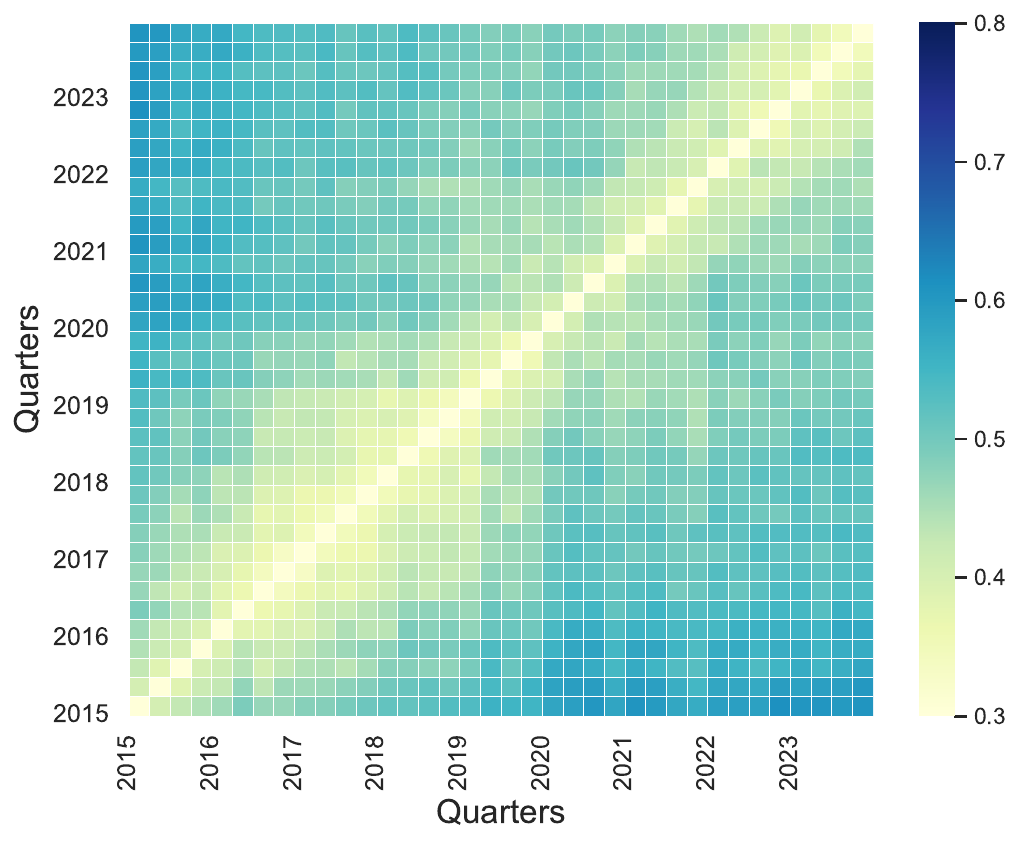}
        \caption*{(d) Iron ore}
        \label{fig:distance_iron}
    \end{minipage}
    \caption{\textbf{Network-distance matrices for all sub-networks.} Each heatmap visualises the trade distance among the respective commodity networks, including (a) dry bulk, (b) coal, (c) grain, and (d) iron ore, illustrating the temporal changes in trade connectivity in the global dry bulk market from 2015 to 2023. The colour scale represents dissimilarity values, with darker shades indicating higher divergence and lighter shades indicating greater similarity between the networks.}
    \label{fig:distancematrix}
\end{figure}

Using the distance matrices constructed in the method section, we explore the dynamic evolution of these networks over time. Figure \ref{fig:distancematrix} presents the heatmaps of the distance matrices, capturing the dissimilarities between the quarterly network structures for coal, grain, iron ore, and the overall dry bulk network. In these heatmaps, the colour scale represents dissimilarity values, with darker shades indicating higher divergence and lighter shades indicating greater similarity between the networks.

First, we observe clear differences between the four distance matrices. For instance, in Figure \ref{fig:distancematrix}(b), the heatmap for coal shows a significant structural shift in 2020, where the color transitions smoothly from light to dark, indicating increasing dissimilarity over time. This suggests that the coal network underwent notable structural changes during the COVID-19 pandemic period, with the granular sub-network analysis revealing the complexity of trade patterns disrupted by external factors. 

In contrast, the heatmaps in Figure \ref{fig:distancematrix}(c) and Figure \ref{fig:distancematrix}(d) (representing grain and iron ore, respectively) exhibit more concentrated patterns of dissimilarity. The grain network, as shown in Figure \ref{fig:distancematrix}(c), reveals a generally darker color pattern, suggesting more volatility and frequent changes in its network structure from quarter to quarter. Conversely, Figure \ref{fig:distancematrix}(d) displays the iron ore network, which shows lighter, more consistent colors, indicating a much more stable network structure over time compared to both grain and coal. This indicates that the iron ore sub-network, as explained above, responded better to the challenges arising from COVID-19. Through analysing these three sub-networks, a more detailed, granular picture of the dry bulk network emerges, one which emphasises that trade patterns vary by commodity and temporarily.   

The full dry bulk network distance matrix plot shows the combined results of the individual sub-networks (coal, grain, and iron ore). By applying hierarchical clustering to the distance matrices for each commodity, we identified clusters of quarters that exhibit similar network structures. Figure \ref{fig:Dencombined} (in Supplementary Materials) illustrates the hierarchical clustering results for each distance matrix. Upon reviewing these clusters, we observe that the clustering patterns vary depending on the type of network. When selecting two clusters for analysis, we are able to pinpoint the exact quarters in which significant structural changes occurred within each sub-network.

Specifically, for the dry bulk network, the key structural shift is observed in Q1 2020, while for iron ore, the change occurs in Q2 2019. For the grain network, the major structural transition is identified in Q1 2021, and for coal, it also occurs in Q1 2020. These findings suggest that the COVID-19 pandemic led to significant structural changes in both the coal and dry bulk networks. In contrast, the iron ore network experienced a notable shift prior to the pandemic, in 2019. Meanwhile, the grain network exhibited a structural change coinciding with the onset of the war in Ukraine, highlighting the impact of geopolitical events on its transportation network. 

If we examine further clustering into smaller groups, Figure \ref{fig:distancematrix}(b) indicates a third cluster emerging for coal after 2022, and Figure \ref{fig:distancematrix}(d) shows a third cluster post-2021 for iron ore. According to \href{https://www.reuters.com/world/asia-pacific/china-lifts-ban-five-australian-beef-exporters-2024-05-29/#:~:text=China%20imposed%20the%20bans%20between,the%20origin%20of%20COVID%2D19.}{Reuters}, between 2020 and 2022, China imposed bans on many export items from Australia due to its independent investigation into COVID-19. As reported by \href{https://www.drewry.co.uk/maritime-research-opinion-browser/maritime-research-opinions/chinas-shift-towards-overland-metallurgical-coal-imports-to-dampen-shipping-demand#:~:text=Furthermore%2C%20China's%20seaborne%20imports%20from,on%20Australia's%20coal%20in%20China.}{Drewry} , China has also increased its coal imports from Russia since 2023. Additionally, Figure \ref{fig:all_maps}(d) reveals that Ukraine serves as a major hub for iron ore exports. Accordingly, it is not surprising that the war in Ukraine appears to have caused a structural break in the iron ore trade routes.

In addition to the observed changes in network structure, Figure \ref{fig:distancematrix}(c) illustrates the distance matrix for grain, which reveals a distinct periodic pattern. Specifically, the matrix displays alternating strips of colour along its diagonal, indicating periodic fluctuations in distance measures. This periodic feature is not observed in the other networks analysed, suggesting that agricultural productivity exhibits unique cyclical patterns in trading activities compared to other sectors. Such data provide insights into the underlying temporal patterns through which agricultural productivity, and the grain sub-network, operate in distinct ways from the other dry bulk networks, including iron and ore. They thus provide insight for stakeholders in planning logistics and optimising shipping patterns to account for these cycles. The distance matrix addresses the challenges posed by small sample sizes, often found in maritime shipping due to limited data that make standard econometric techniques insufficient for detecting or adjusting seasonality.  

\subsubsection{Spatial distribution and evolution of community}
Community structure is a key property of shipping networks, offering insights into their efficiency and connectivity. Previous studies, such as \cite{kaluza2010complex} and \cite{pan2019connectivity}, emphasize that ports within the same community tend to be densely connected, while inter-community links are sparse. These sparse connections often act as bottlenecks, influencing trade flows and the overall resilience of the network. Understanding these community structures can support strategic planning for shipping companies and port operators by revealing critical trade routes and identifying potential vulnerabilities.

In the context of dry bulk trade flows, our static analysis confirms that dry bulk shipping network structures reflect underlying commodity demand and supply dynamics worldwide. However, these static patterns are not immutable; geopolitical events and market disruptions can alter trade relationships, leading to shifts in community groupings. For example, changes in international trade policies or conflicts can create new inter-community links or dissolve existing ones. This highlights the importance of integrating temporal analysis with community studies to capture the evolving nature of maritime trade networks.

Building on the structural changes identified in each sub-network, we now examine how community structures evolve by employing complementary visualizations. We use Sankey diagrams to illustrate the relative sizes of communities and how they flow, merge, and transition into new structures. However, maritime trade operates within a physical framework defined by oceans, seas, and geographic distances \cite{maritimeeconomics}. Therefore, we also display the network communities on world maps to align community structures with physical geography. Our analysis focuses on the grain and coal networks, where these changes are most prominent.

\begin{figure}
    \centering
    \includegraphics[width=1\linewidth]{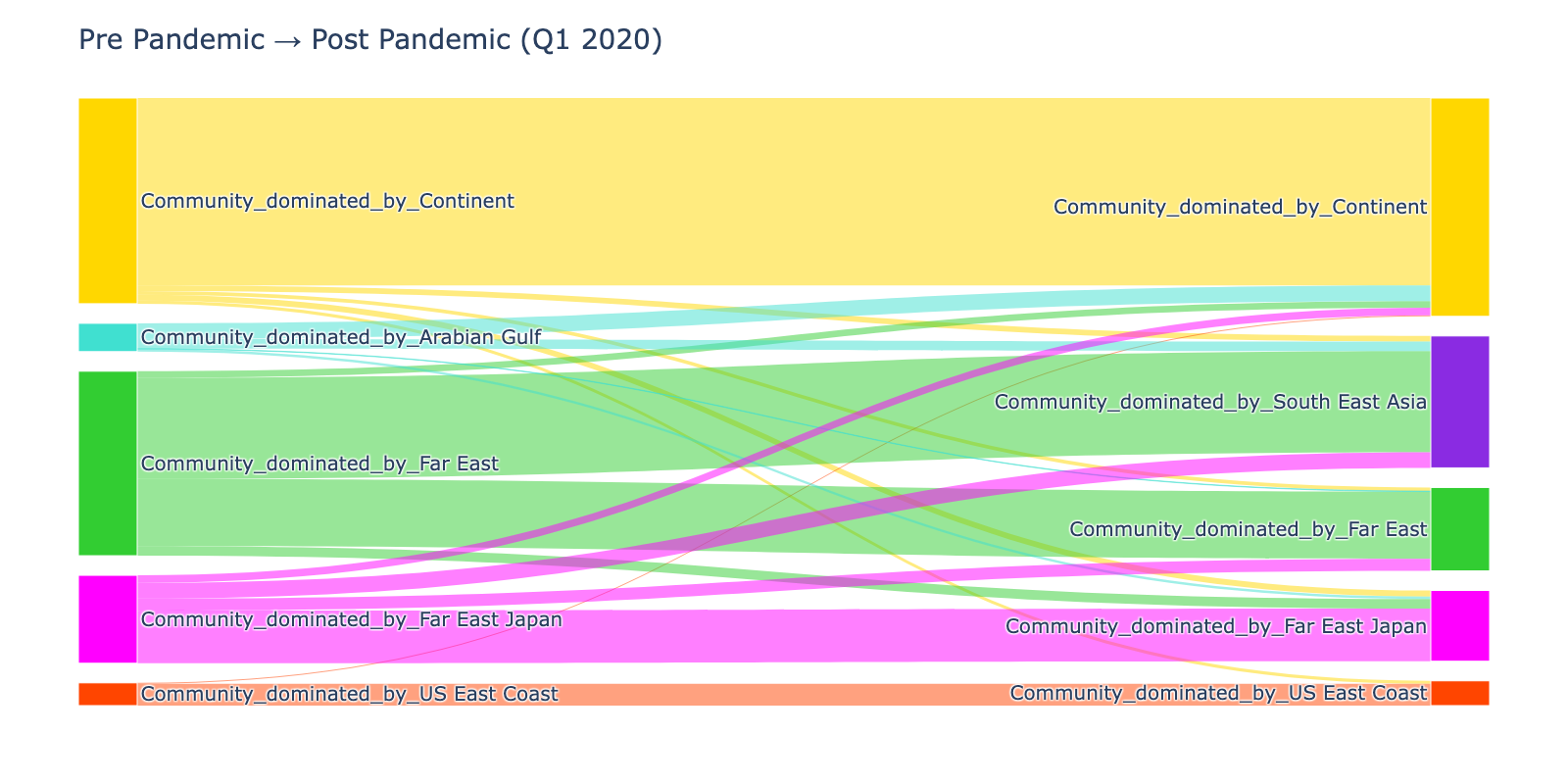}
    \caption{\textbf{Transitions in Coal Trading Communities Before and After the Pandemic:} Each node (represented as rectangular blocks on either side of the diagram) corresponds to a community, with its height proportional to the community size (measured by the number of ports). Nodes are colour-coded to reflect their dominant geographical region, which is determined by counting the number of ports from each region within the community and selecting the region with the highest representation. The colour scheme matches the world maps shown below. Links between nodes represent the transitions of ports from one community to another, with the width of each link indicating the number of ports transitioning. The network prior to Q1 2020 consisted of six communities, while the post-Q1 2020 network comprises five communities. Significant changes in community structure include: (1) Communities dominated by Continent, Japan, and US East Coast remain largely unchanged. (2) Smaller communities, such as those dominated by Arabian Gulf ports and Far East Russian ports, merged with larger groups after Q1 2020. (3) The largest pre-Q1 2020 community, dominated by Far East ports, splits into two distinct communities: one dominated by China and another by Southeast Asia.}
    \label{fig:sankey_coal}
\end{figure}

\begin{figure}[h!]
    \centering
    \begin{minipage}{\textwidth}
        \centering
        \includegraphics[width=\linewidth]{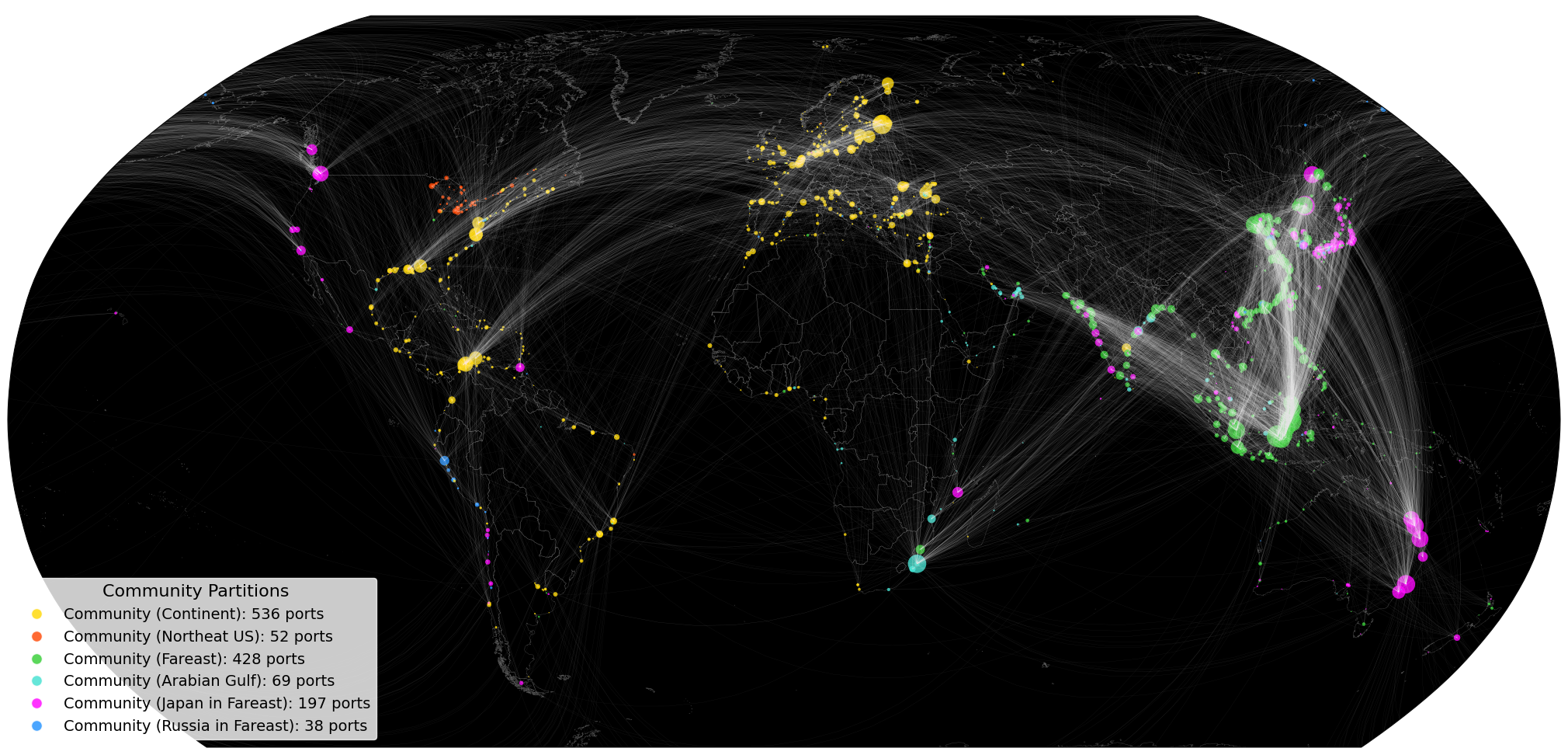} 
        \caption*{(a) Before Q1 2020}
        \label{fig:coal_before}
    \end{minipage}\\[1em] 
    \begin{minipage}{\textwidth}
        \centering
        \includegraphics[width=\linewidth]{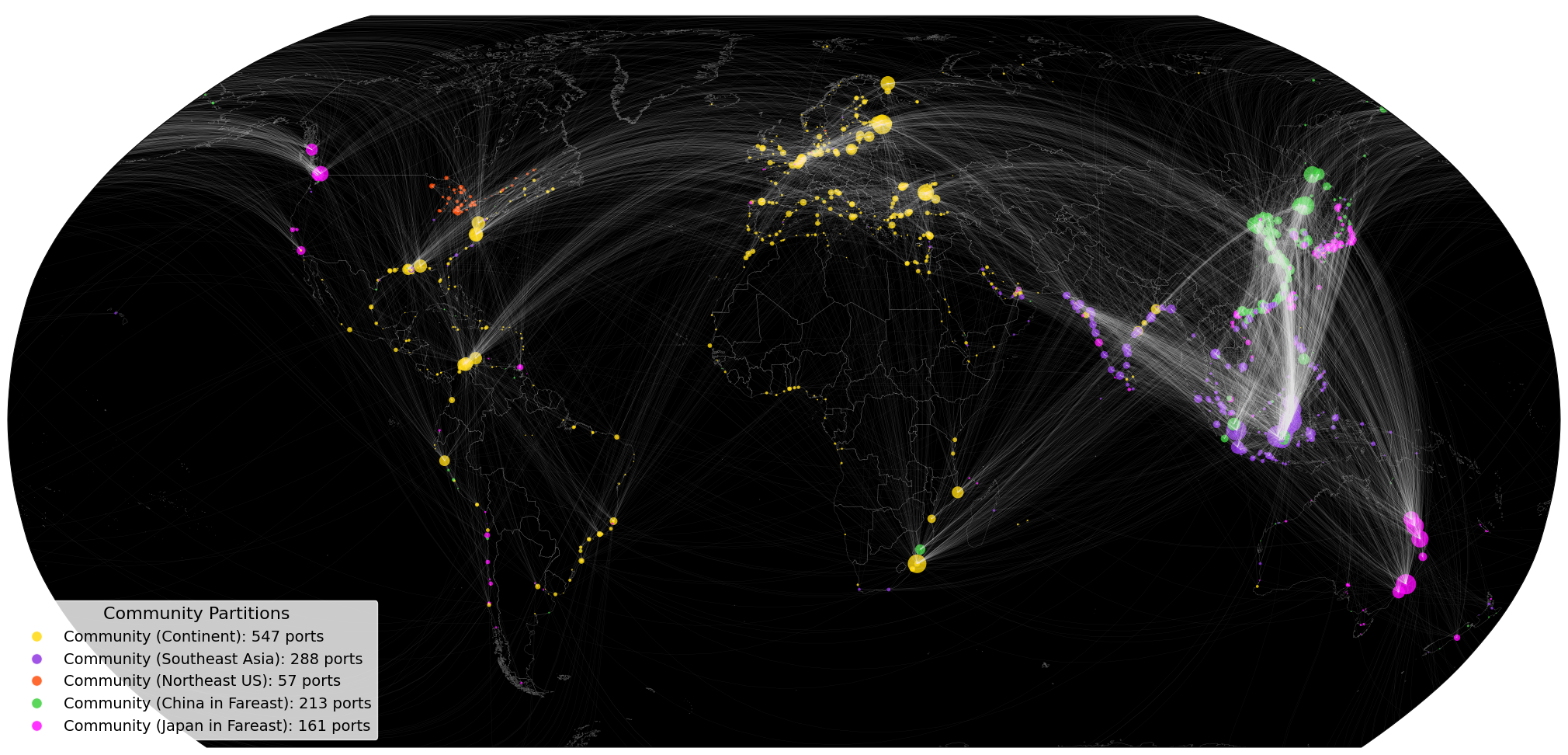}
        \caption*{(b) Q1 2020 and After}
        \label{fig:coal_after}
    \end{minipage}
    \caption{\textbf{Coal handling ports and shipping routes before and after Q1 2020.} Ports are grouped into communities, with distinct colours representing the dominant geographical region in each community. Dominance is determined by the region with the highest number of ports within the community. Changes in community structure are visualized through these region-specific colours, comparing the network before and after the COVID-19 pandemic (Q1 2020). The size of each node reflects the highest of its normalised in-degree or out-degree centrality. The primary structural change observable on the maps occurs in Asia (green), where ports previously grouped together split into two distinct communities: one dominated by China (green) and the other by Southeast Asia (purple).}
\label{fig:coal_mod}
\end{figure}

\paragraph{Coal sub-network}
In the coal sub-network, where a structural break was identified in Q1 2020, we use the Louvain method to detect communities in the networks before and after the break. As shown in Figures \ref{fig:sankey_coal} and \ref{fig:coal_mod}, the network consisted of six communities before Q1 2020, and five afterward. The six communities identified prior to Q1 2020 are as follows: (1) the largest community, consisting of 536 ports, includes ports in Europe, North Africa, and Central America, spanning the Atlantic, Mediterranean, and Black Sea regions; (2) a community dominated by Far East ports, comprising 428 ports in total; (3) a community of 197 ports centered around Japan, East Coast Australia, and West Coast North America across the Pacific; (4) a smaller community with 52 ports, primarily located along the US East Coast; (5) a community of 69 ports concentrated in the Arabian Gulf; and (6) the smallest community, consisting of 38 ports from the Russian Far East region, spanning 17 geographical regions.

As shown in Figure \ref{fig:coal_mod}, post Q1 2020, the new network comprises five communities. Three communities remain substantially the same as pre Q1 2020, dominated by ports in the same regions.  However, significant changes in routes across Far East Asia and the Indian Ocean after Q1 2020 lead to differences in community structure.

The most notable change occurs in Asia, where a previously tightly connected group splits into two. One community is now dominated by China, while another is centered on Southeast Asian ports. Additionally, ports in the Arabian Gulf, which were previously linked to Africa, now form stronger connections with Europe and the Southeast Asia community, leaving Africa more closely aligned with Europe. Furthermore, Russian ports in Northeast Asia, which previously formed a small, distinct community with other ports, are now integrated into the same community as China, reflecting strengthened trade ties since the COVID-19 pandemic and the onset of the war in Ukraine. (Source: \href{https://www.drewry.co.uk/maritime-research-opinion-browser/maritime-research-opinions/chinas-shift-towards-overland-metallurgical-coal-imports-to-dampen-shipping-demand#:~:text=Furthermore%2C%20China's%20seaborne%20imports%20from,on%20Australia's%20coal%20in%20China}{Drewry}). 

Despite the reduction in community count from six to five, modularity decreased from 0.34 to 0.31.\footnote{The global centralities are presented in a table in Supplementary Material.} This may be caused by the newly formed groups having fewer links within communities and more connections between communities. Modularity rewards communities with strong internal clustering and penalizes connections that blur the boundaries between groups. 

Finally it's worth noting the total loaded volume in 2023 exceeded the peak volume prior to the pandemic. Hence, despite environmental concerns over coal, consumption has not decreased, even though there was a dip during the COVID-19 lockdown period when industrial production was heavily affected worldwide.\footnote{Results are available on request.} These observations suggest that while geopolitical tensions and unexpected disruptions, such as pandemic, can directly affect community modularity in global trade networks, they may not necessarily alter overall trading volumes.

\begin{figure}
    \centering
    \includegraphics[width=1\linewidth]{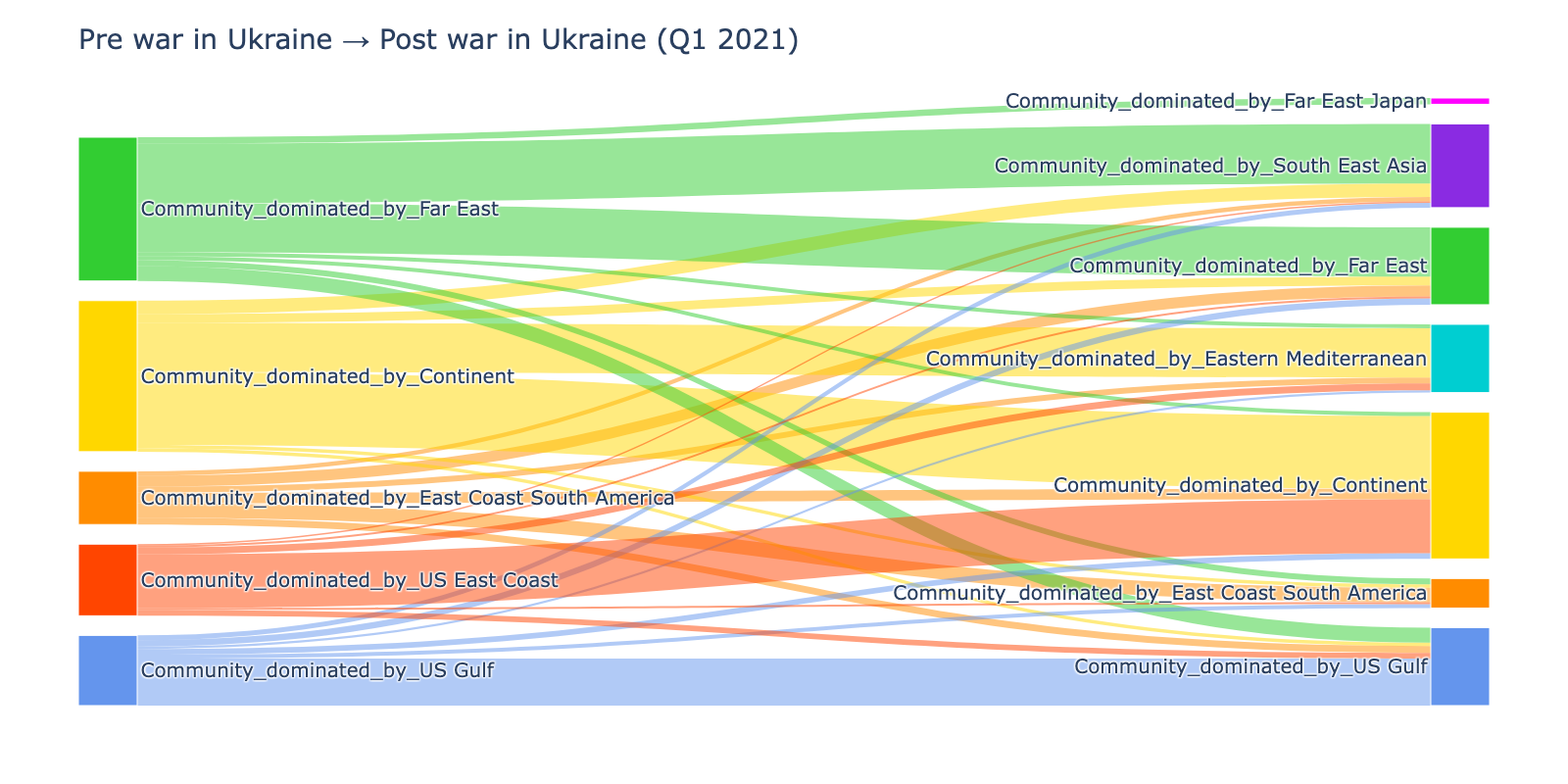}
    \caption{\textbf{Transitions in Grain Trading Communities Before and After the War in Ukraine:} Each node (represented as rectangular blocks on either side of the diagram) corresponds to a community, with its height proportional to the community size (measured by the number of ports). Nodes are color-coded to reflect their dominant geographical region, determined by counting the number of ports from each region within a community and selecting the region with the highest representation. The colours used match the world maps shown below. Links between nodes represent the transitions of ports from one community to another, with each link’s width indicating the number of ports undergoing that transition. The network prior to Q1 2021 consisted of five communities, while the post-Q1 2021 network comprises seven communities. Significant changes in community structure include: (1) ports in the Far East splitting into three distinct communities: one dominated by China, one by Japan, and another by Southeast Asia (2) a large community dominated by the European Continent splitting into an Eastern Mediterranean group and a remaining group still dominated by the European Continent (3) the community dominated by the US East Coast merging with the Continent-dominated community (4) most ports in the community dominated by East Coast South America transitioning to other communities, with a small proportion remaining unchanged.}
    \label{fig:sankey_grain}
\end{figure}

\paragraph{Grain sub-network}
In comparison to coal, the grain shipping network exhibits a more diverse evolution in community structure.\footnote{Using the Louvain community detection method, we identify six communities in the pre-Q1 2021 network and eight in the post-Q1 2021 network. However, in both networks, there is a small community of fewer than five nodes, each having degree 1. In addition to the small size, since most of these nodes within the community do not appear in the both networks, we exclude them from the community structure analysis in this section.} As shown in Figure \ref{fig:sankey_grain}, prior to the war in Ukraine, the Louvain community detection algorithm identifies five communities: (1) the largest community, dominated by continental Europe, with 418 ports;
(2) the second-largest community, dominated by the Far East, with 396 ports; (3) a trans-Atlantic community dominated by the U.S. East Coast, with 201 ports; (4) a community dominated by U.S. Gulf ports, with 182 ports; (5) a community dominated by East Coast South American ports, with 141 ports.

Following the outbreak of the war in Ukraine, these five communities merge, reorganise, and ultimately form seven groups, reflecting a new grain cargo network structure. The changes can be broadly grouped into two categories: those arising in Europe from transportation challenges stemming from the war in Ukraine, and those arising in Asia from rising tensions between China and Australia.

\begin{figure}[h!]
    \centering
    \begin{minipage}{\textwidth}
        \centering
        \includegraphics[width=\linewidth]{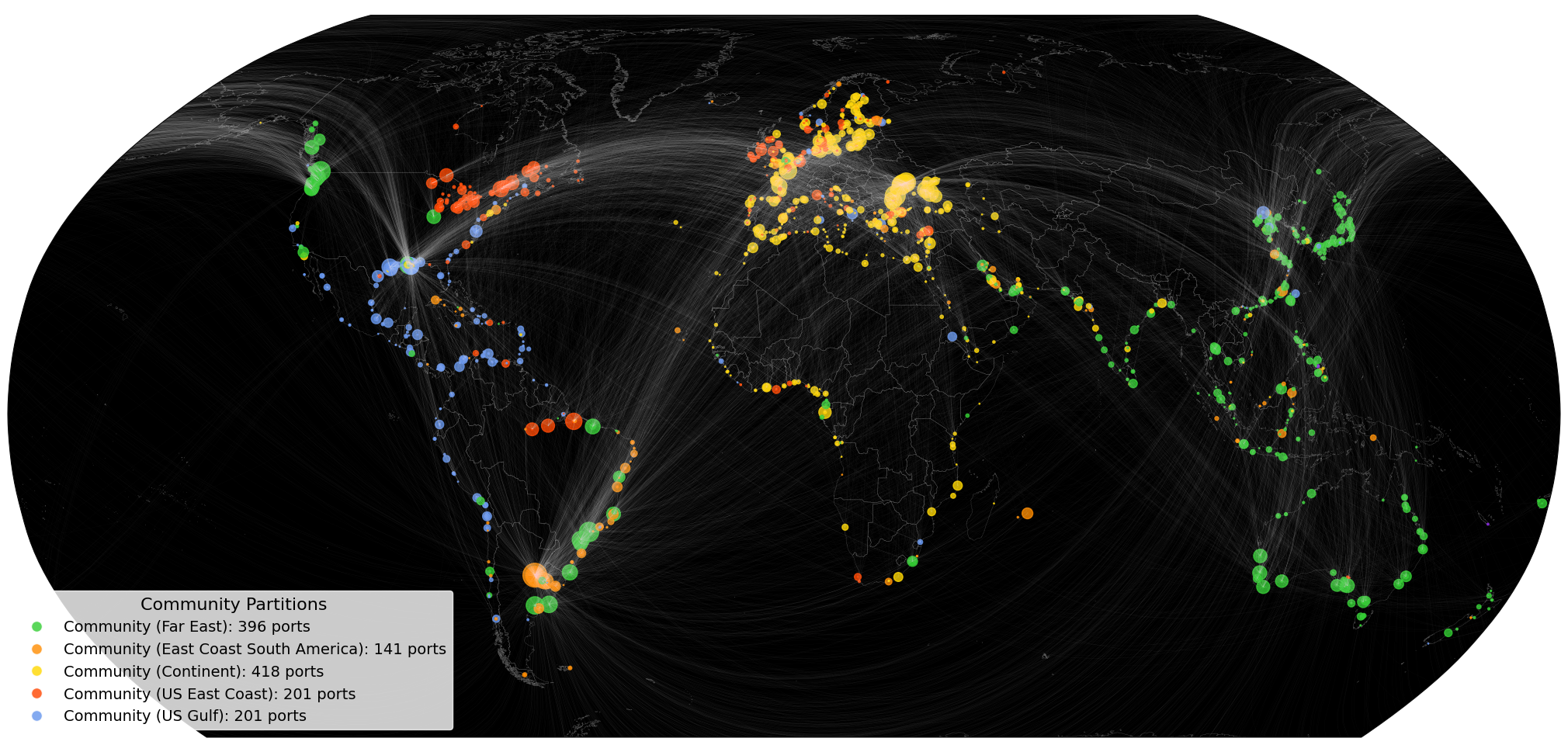} 
        \caption*{(a) Before Q1 2021}
        \label{fig:grain_before}
    \end{minipage}\\[1em] 
    \begin{minipage}{\textwidth}
        \centering
        \includegraphics[width=\linewidth]{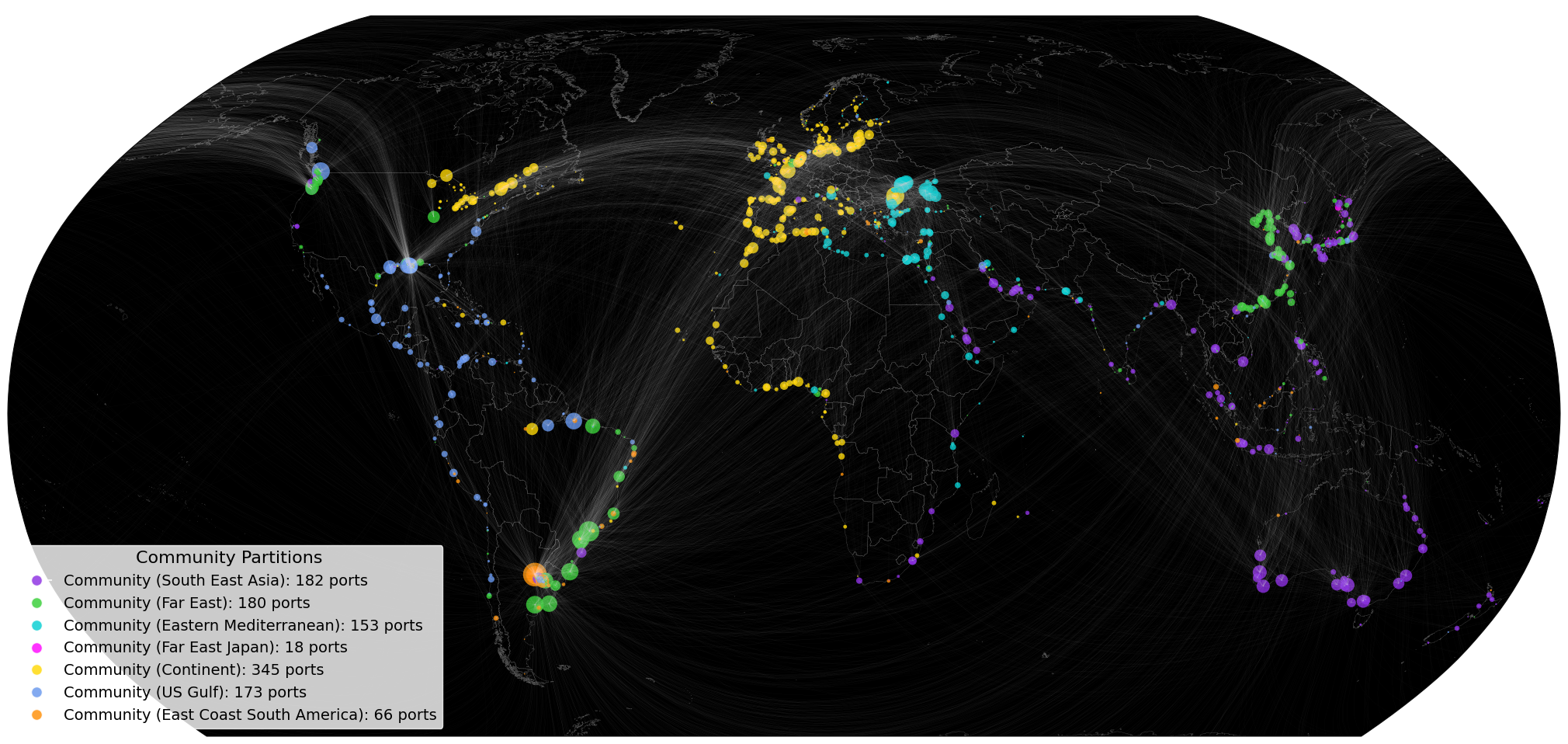}
        \caption*{(b) Q1 2021 and After}
        \label{fig:grain_after}
    \end{minipage}
    \caption{\textbf{Grain handling ports and shipping routes before and after Q1 2021.} Ports are grouped into communities, with distinct colours representing the dominant geographical region in each community. The community structure is visualised based on these region-specific colours before and after the war in Ukraine. The size of each node reflects the highest of its normalised in-degree or out-degree centrality. Significant changes in community structure include: (1) ports in the Far East (green) splitting into three distinct communities: one dominated by China (green), one by Japan (pink), and another by Southeast Asia (blue violet); (2) a large community dominated by the European Continent (yellow) splitting into an Eastern Mediterranean group (Dark Turquoise) and a remaining group dominated by Europe Continent (yellow) (3) the community dominated by the US East Coast (orange) merging with the Continent (yellow) dominated community. (4) the US Gulf dominated community (cornflower blue) remains largely unchanged while the East Coast South America  dominated community (orange red) become significantly smaller, as some ports transit to others. In total the number of communities increases from five to seven.}
    \label{fig:grain_mod}
\end{figure}

For instance, the community encompassing continental Europe and North Africa is split, with ports in the Eastern Mediterranean forming a distinct group, while the former U.S. East Coast community merges into the remaining continental Europe group. Meanwhile, the smaller East Coast South American community fragments into multiple new communities, leaving only a small proportion of ports intact. The Far East community splits into three, as illustrated in Figure \ref{fig:grain_mod}: There are increased trade flows between Chinese ports and East Coast South America, 18 Japanese ports form a small community on their own, whereas other Asian ports form a single community led by Southeast Asia.

After Q1 2021, the network’s modularity increases from 0.27 to 0.29. This is because  the network becomes smaller and less dense, with fewer nodes, edges, and a reduced average degree (as shown in Table \ref{tbl:grain_glo_centrality} in the Supplementary Materials). In a sparser network, community boundaries become more defined, as nodes cluster more distinctly. 

\begin{figure}[ht!]
    \centering
    \includegraphics[width=0.9\linewidth]{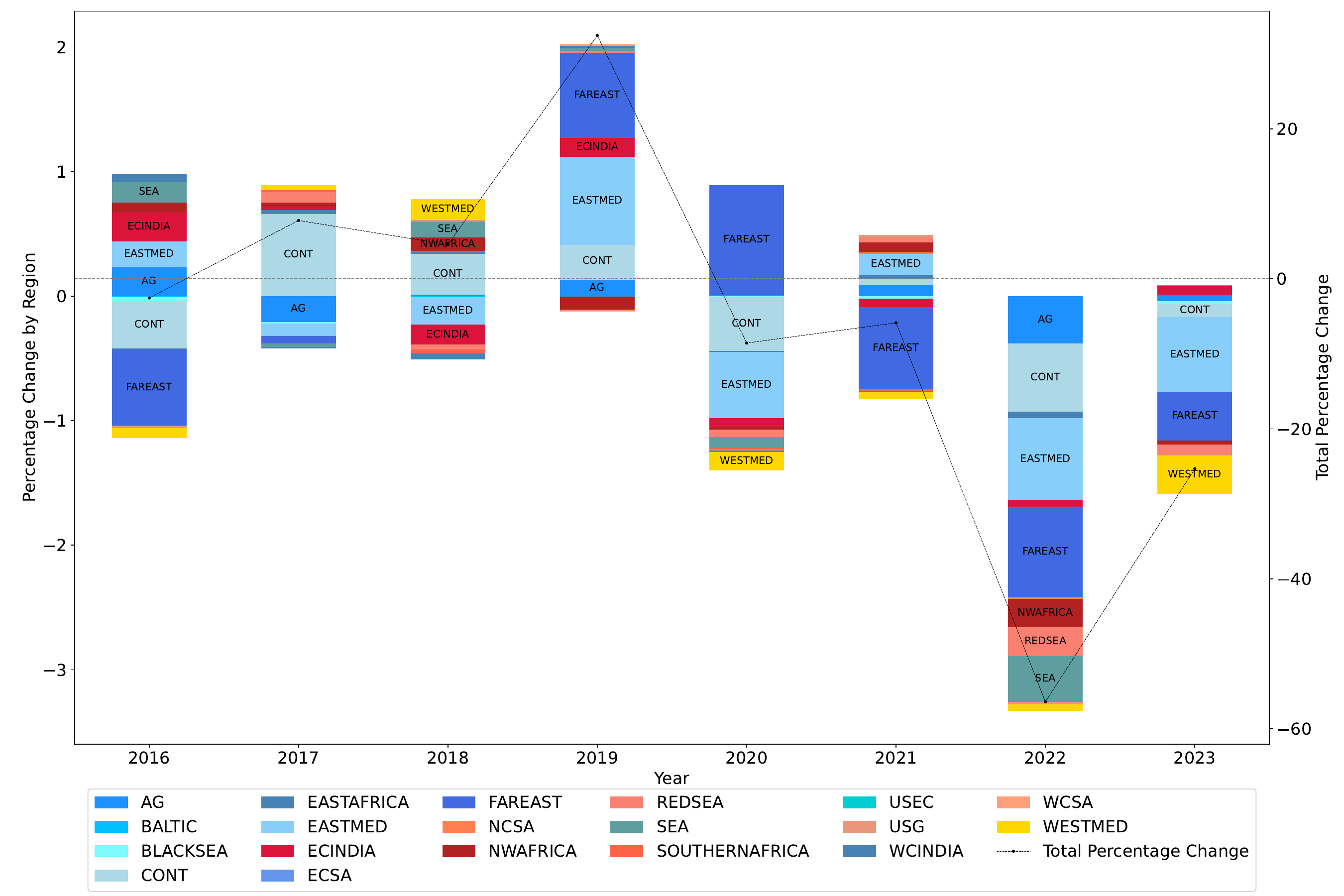}
    \caption{\textbf{Ukraine grain export volume change year on year.} This figure illustrates the annual changes in grain export volumes from Ukraine from 2016 to 2023 across regions. Regions are denoted by shorthand names: AG (Arabian Gulf), EASTAFRICA (East Africa), FAREAST (Far East), REDSEA (Red Sea), USEC (East Coast US), WCSA (West Coast South America), BALTIC (Baltic), EASTMED (East Mediterranean), NCSA (North Coast South America), SEA (Southeast Asia), USG (US Gulf), WESTMED (West Mediterranean), BLACKSEA (Black Sea), ECINDIA (East Coast India), NWAFRICA (Northwest Africa), SOUTHERNAFRICA (Southern Africa), WCINDIA (West Coast India), CONT (Continent), and ECSA (East Coast South America). The volume changes are expressed as a percent change compared to the previous year. The line chart represents the total percentage change in grain exports (right y-axis), showing the aggregated impact of all regions for each year. The contribution of each region's annual change to the overall trend can be calculated by dividing the change in export volume for each year by the total change over the entire period (left y-axis). Region short names are labeled on the bars when their annual contribution exceeds 10\% in absolute value, highlighting significant regional impacts on the overall export trend.}
    \label{fig:ukraine}
\end{figure}

In addition to the changes of structure, Figure \ref{fig:ukraine} shows year-on-year changes in Ukraine’s grain export volumes. From the onset of the war in 2021, there has been a pronounced reduction in exports to the Mediterranean, Black Sea, and Europe, followed by further decreases to Fareast in the subsequent year. By 2022, Ukraine’s total export volumes had fallen by nearly 60\% across all regions. This downward trend persisted in 2023, with an additional 30\% reduction in export volumes. Figure \ref{fig:q_metrics} in Supplementary Material illustrates the yearly changes in global grain cargo volumes, revealing a declining trend since 2021. 

In addition, in both the coal and grain sub-networks, the major importing and exporting nations, as identified by normalized degree centralities, remain largely unchanged after the structural breaks. Table \ref{tab:top8country} in the Supplementary Materials indicates that while the rankings of the largest trading countries may have shifted, the principal exporters and importers still come from a relatively small group. These rankings reflect connectivity, the number of links each port maintains with others, rather than the actual volume of goods transported.

\clearpage
\section{Conclusion}
\label{sec:conc}
The intervention of this study in relation to academic scholarship on shipping networks is based on employing micro-level trade flow data alongside an enhanced use of complex network analysis techniques to uncover disruptive patterns within dry bulk shipping networks. 

By exploring these dry bulk shipping sub-networks , our analysis reveals significant variability in sub-network sizes and characteristics. By capturing the direct network feature of tramp shipping, we focus on component analysis and conclude that while individual sub-networks generally lack small-world properties evidenced by low clustering coefficients, Giant Strongly Connected Components (GSCCs) in these networks exhibit these properties, suggesting the presence of efficient bilateral trade routes for commodities such as raw grain and processed goods. 

By aggregating layered sub-networks into a comprehensive dry bulk trade flow network, we uncover both small-world and scale-free characteristics, allowing ships to navigate a dense network that enhances trade efficiencies. The large proportion of GSCC within the full network reflects bilateral trade patterns across regions, incorporating all maritime-transported commodities. Conversely, the connections between nodes outside the GSCC indicate ballast journeys. Hence, a smaller GSCC in a dry bulk network may imply higher maritime shipping trade costs. 

In addition, a common degree distribution following a power-law pattern further underscores structural consistencies within the network. The correlation matrices with negative statistics further proves the scale-free and disassortativity features of the networks, implying the asymmetric distribution of the dry bulk commodities worldwide. 

While the static analysis using network properties depicts a vivid picture of the demand and supply of raw materials, it does not capture current trade affairs and conflicts, which significantly impact the network structure. Our temporal analysis utilising distance matrices identifies significant structural changes in the network which have been triggered by external factors. The COVID-19 pandemic had a pronounced impact on the coal network as well as the full dry bulk network, whereas the grain and iron ore networks demonstrated notable resilience to pandemic disruptions. In contrast, the war in Ukraine has caused substantial shifts in the grain network, with periodic changes indicating both distinct agricultural life cycles and geopolitical tensions that sets the grain sub-network apart from other networks. 

Without significant changes in demand and supply of the raw materials, the distribution, however, may shift due to the reshuffling of trade partners under geopolitical conditions. Network community analysis captures considerable changes in community structures, such as the clustering of China and Russia within the coal network and the post-Ukraine war separation of European and North African grain trade communities into distinct groups. These observations illustrate the profound influence of global conflicts on trade patterns, hence the transportation networks.

Recognising shifts in communities allows stakeholders to adapt trade routes in response to global events, thereby refining logistical strategies. Our hybrid approach, combining static and temporal network analyses of trade flow networks, not only sheds light on critical trade dynamics, but suggests the need for context-specific analysis of network formation. Moreover, our methodologies are applicable to other dry bulk networks, equipping stakeholders, including commodity traders, ship owners and logistics professionals, with strategies to enhance profitability and reduce operational risk to some extent.

Finally, building on a thorough understanding of laden networks across various cargo types, future research can incorporate analysis of freight rates to mitigate some of the unpredictability in the multi-layered dry bulk network. This approach would deepen insight into cost-related drivers of network dynamics and provide more robust predictive capabilities for industry stakeholders.

\clearpage

\bigskip
\begin{center}
{\large\bf SUPPLEMENTAL MATERIALS} \label{cluster_algo}
\end{center}

\appendix
\section{Data}
\section*{}\label{supplementA}

\subsection{Oceanbolt Trade Flow Database}
The table below is a sample of the full Oceanbolt trade flow database. 
\begin{longtable}{@{}lll@{}}

\toprule
   & Table Header                & Example (1)                          \\* \midrule
\endfirsthead
\multicolumn{3}{c}%
{{\bfseries Table \thetable\ continued from previous page}} \\
\toprule
   & Table Header                & Example (1)                          \\* \midrule
\endhead
\bottomrule
\endfoot
\endlastfoot
1  & voyage\_id                   & 6b4acf9b774dc311f39695791fe7b011 \\
2  & flow\_id                     & 4800fb9ba1158087e4ce7268cb746e8e \\
3  & imo                         & 9463449                          \\
4  & vessel\_name                 & EDENBORG                         \\
5  & segment                     & Handysize                        \\
6  & sub\_segment                 & Small Handysize (10-27k)         \\
7  & dwt                         & 11300                            \\
8  & commodity                   & Grains (unclassified)            \\
9  & commodity\_group             & Grains                           \\
10 & volume                      & 5500                             \\
11 & load\_port\_id                & 5319                             \\
12 & load\_port\_name              & Montreal                         \\
13 & load\_port\_unlocode          & CAMTR                            \\
14 & load\_berth\_id               & 12821                            \\
15 & load\_berth\_name             & Montreal Multibulk Berth         \\
16 & load\_country\_code           & CA                               \\
17 & load\_country                & Canada                           \\
18 & load\_region                 & USEC                             \\
19 & load\_port\_arrived\_at        & 2015-06-11 00:53:28+00:00        \\
20 & load\_port\_berthed\_at        & 2015-06-11T04:48:50Z             \\
21 & load\_port\_departed\_at       & 2015-06-11T12:47:07Z             \\
22 & load\_port\_days\_total        & 0.495                            \\
23 & load\_port\_days\_berthed      & 0.332                            \\
24 & load\_port\_days\_waiting      & 0.163                            \\
25 & discharge\_port\_id           & 694                              \\
26 & discharge\_port\_name         & Cadiz                            \\
27 & discharge\_port\_unlocode     & ESCAD                            \\
28 & discharge\_berth\_id          & 5051                             \\
29 & discharge\_berth\_name        & Cadiz Multibulk Berth 2          \\
30 & discharge\_country\_code      & ES                               \\
31 & discharge\_country           & Spain                            \\
32 & discharge\_region            & CONT                             \\
33 & discharge\_port\_arrived\_at   & 2015-06-23T00:58:00Z             \\
34 & discharge\_port\_berthed\_at   & 2015-06-23T03:58:20Z             \\
35 & discharge\_port\_departed\_at  & 2015-06-24T13:38:07Z             \\
36 & discharge\_port\_days\_total   & 1.527                            \\
37 & discharge\_port\_days\_berthed & 1.402                            \\
38 & discharge\_port\_days\_waiting & 0.125                            \\
39 & days\_steaming               & 11.507                           \\
40 & days\_total\_duration         & 13.531                           \\
41 & distance\_calculated         &                                  \\
42 & distance\_actual             & 3160.369669                      \\
43 & eta                         &                                  \\
44 & destination                 & CADIZ                            \\
45 & status                      & Complete                         \\
46 & parceling                   & TRUE                             \\
47 & ballast\_started\_at          & 2015-06-01T19:46:42Z             \\
48 & ballast\_port\_name           & Calumet Harbor                   \\
49 & ballast\_port\_id             & 3577                             \\
50 & ballast\_port\_unlocode       & USOUS                            \\
51 & ballast\_country             & United States                    \\
52 & ballast\_country\_code        & US                               \\
53 & ballast\_region              & USEC                               \\* \bottomrule
\caption{\textbf{An example of Oceanbolt tradeflow data sample} Fields description of main columns (from top to bottom): voyage\_id: unique ID per voyage assigned; flow\_id: unique ID per trip, i.e. one load to discharge journey. A full voyage may contain more than one flow due to multiple loads and discharges; imo: the unique vessel ID which never changes throughout its life time; vessel\_name: the name of a ship, which may change when a vessel is repurchased; segment: the vessel type based on its size; sub\_segment: a more granule vessel type; dwt: deadweight ton; commodity: a more granular commodity type; commodity\_group: main dry bulk commodity type; volume: cargo weight in metric ton; column (11) - (18) provide information about load port; load\_port\_id: port\_id is a unique port number given by Oceanbolt and port is specific for loading; load\_port\_name: the name of the port where cargo is loaded; load\_port\_unlocode: unlocode for the load port. UNLOCODE stands for the``United Nations Code for Trade and Transport Locations"; load\_berth\_id: a port where a vessel may be moored for loading; load\_berth\_name: the name for berth, load\_country\_code: the country code given by Oceanbolt where the load port is located; load\_country: the country name where the load port is located; load\_region: the region that the country belongs to; column (19) - (21) provide the timestamp for the arrival, berth, and departure time when loading; load\_port\_days\_total: is the sum of load\_port\_days\_berthed and load\_port\_days\_waiting; column (25) - (28) follows the same structure as column (11) - (24) with information for discharging; days\_steaming: calculated by using the total distance divided by the speed; days\_total\_duration: the sum of load\_port\_days\_total, discharge\_port\_days\_total and days\_steaming; distance\_calculated: the voyage distance calculated by Oceanbolt using their methodology; actual\_calculated: the geographical distance between ports; destination: the final port the vessel should travel to; status: voyage completion; parcelling; whether the vessel has to do multiple trips; column (47)- (53) the ballast leg information before this trade flow occurs.}   
\label{tab:tradeflow_ex}
\end{longtable}

The table below summarises the vessel type,size, and corresponding number of trade flows from 2015 t0 2023. We can see that the small-middle ranged vessels (Handysize, supramax and panamax) dominate the dry bulk fleet sector. The vessel type is commonly known as named after the route where the vessel fits. For example, panamax vessels can travel through panama canal whereas the capesize vessels are too large to go through. The capesize vessels are named after the cape town.  

\begin{table}[ht!]
\tiny
\centering
\setlength{\tabcolsep}{18pt}
\begin{tabular}{@{}llllll@{}}
\midrule
  \textbf{Segment} & \textbf{Size} & \multicolumn{4}{c}{\textbf{Number of vessels per   vessel type}}           \\ 
                   &               & \textbf{Dry bulk} & \textbf{Grain}  & \textbf{Coal}   & \textbf{Iron Ore} \\ 
\midrule
  Capesize  & 110-250 & 2,145 & 129   & 1,580 & 2,080 \\
  Handysize & 10-43   & 4,811 & 3,481 & 3,444 & 1,191 \\
  Panamax   & 68-110  & 3,331 & 2,646 & 3,176 & 2,500 \\
  Shortsea  & 0-10    & 1,742 & 711   & 793   & 151   \\
  Supramax  & 43-68   & 4,092 & 3,247 & 3,582 & 2,694 \\ 
\bottomrule
  \textbf{Total}   & \textbf{}     & \textbf{16,121}    & \textbf{10,214} & \textbf{12,575} & \textbf{8,616}   
\end{tabular}
\caption{\textbf{Number of ships in each ship type in dry bulk shipping data, 2015 - 2023} This table summarises the number of vessels categorised by ship type and their respective sizes (measured in thousand deadweight tonnage, DWT) in the dry bulk shipping industry over the period from 2015 to 2023. The data includes the total count of dry bulk, grain, coal, and iron ore vessels, providing insights into the composition of the fleet across different segments.}
\label{tab:vessel_type}
\end{table}

\begin{table}[ht!]
\tiny
\centering
\begin{tabular}{@{}cllll@{}}
\toprule
\multicolumn{1}{l}{} &
  \multicolumn{1}{c}{\textbf{commodity\_group}} &
  \multicolumn{1}{c}{\textbf{no.\_of\_voyages}} &
  \multicolumn{1}{c}{\textbf{no.\_of\_flows}} &
  \multicolumn{1}{c}{\textbf{volume}} \\ \midrule
\textbf{1}  & Coal*                  & 222945  & 274289  & 15.142 \\
\textbf{2}  & Iron Ore*              & 96523   & 127647  & 14.615 \\
\textbf{3}  & Unknown               & 261519  & 399230  & 6.104  \\
\textbf{4}  & Grains*                & 102734  & 157614  & 4.958  \\
\textbf{5}  & Cement and Clinker    & 151267  & 207142  & 1.851  \\
\textbf{6}  & Fertilizers           & 48859   & 78296   & 1.687  \\
\textbf{7}  & Aggregates            & 58783   & 77105   & 1.668  \\
\textbf{8}  & Ores and Concentrates & 36898   & 67443   & 1.639  \\
\textbf{9}  & Bauxite               & 16854   & 18853   & 1.467  \\
\textbf{10} & Steel                 & 58921   & 114777  & 1.432  \\
\textbf{11} & Forest Products       & 36976   & 75635   & 1.294  \\
\textbf{12} & Pet Coke              & 10858   & 16104   & 0.445  \\
\textbf{13} & Salt                  & 10926   & 13836   & 0.439  \\
\textbf{14} & Agribulk              & 10472   & 15091   & 0.431  \\
\textbf{15} & Alumina               & 9975    & 12120   & 0.364  \\
\textbf{16} & Scrap                 & 9640    & 13597   & 0.269  \\
\textbf{17} & Minerals              & 2326    & 4940    & 0.084  \\
\textbf{18} & Metals                & 423     & 740     & 0.012  \\
\textbf{19} & Break Bulk            & 648     & 996     & 0.009  \\
\textbf{20} & Caustic Soda          & 203     & 238     & 0.008  \\
\textbf{21} & Aluminum              & 150     & 366     & 0.005  \\
\textbf{22} & Other Minor Bulks     & 52      & 87      & 0.003  \\
\textbf{23} & \textbf{Total}        & \textbf{1147952} & \textbf{1676146} & \textbf{53.926} \\ \bottomrule
\end{tabular}
\caption{\textbf{Trade flow summary by commodity group, 2015 - 2023.} This table presents a summary of trade flows across various commodity groups from 2015 to 2023. It includes the total number of voyages, the number of trade flows, and the corresponding volume in billions of Gross Tonnage (GT) for each commodity. The data highlights the significant contributions of coal and iron ore to the overall trade volume, while also showcasing other commodities and their relative trade activities.}
\label{tab:commodity_sum}
\end{table}

\begin{tiny}
\begin{longtable}[c]{@{}lll@{}}
\toprule
\textbf{Commodity\_group}      & \textbf{Commodity}                                              & \text{No. of trade flows}   \\* \midrule
\endfirsthead
\endhead
\bottomrule
\endfoot
\endlastfoot
Aggregates            & Aggregates (unclassified)                              & 16,018  \\
Aggregates            & Bentonite                                              & 279     \\
Aggregates            & Clay                                                   & 81      \\
Aggregates            & Dolomite                                               & 93      \\
Aggregates            & Feldspar                                               & 871     \\
Aggregates            & Fly Ash                                                & 1,759   \\
Aggregates            & Granite                                                & 2,467   \\
Aggregates            & Gypsum                                                 & 8,991   \\
Aggregates            & Kaolin                                                 & 1,068   \\
Aggregates            & Limestone                                              & 27,784  \\
Aggregates            & Marble                                                 & 759     \\
Aggregates            & Perlite                                                & 11      \\
Aggregates            & Pumice                                                 & 389     \\
Aggregates            & Sand                                                   & 1,540   \\
Aggregates            & Slag                                                   & 14,901  \\
Agribulk              & Agribulk (unclassified)                                & 2,482   \\
Agribulk              & Animal Feed (unclassified)                             & 28      \\
Agribulk              & Bagged Sugar                                           & 323     \\
Agribulk              & DDGS (Animal Feed)                                     & 30      \\
Agribulk              & Palm Kernel Expellers                                  & 2,972   \\
Agribulk              & Raw Sugar                                              & 8,656   \\
Agribulk              & Rice                                                   & 520     \\
Agribulk              & Tapioca                                                & 3       \\
Alumina               & Alumina                                                & 12,089  \\
Aluminum              & Aluminum                                               & 366     \\
Bauxite               & Bauxite                                                & 18,732  \\
Break Bulk            & Break Bulk                                             & 460     \\
Break Bulk            & Project Cargo                                          & 534     \\
Caustic Soda          & Caustic Soda                                           & 236     \\
Cement and Clinker    & Cement                                                 & 180,588 \\
Cement and Clinker    & Cement and Clinker (Unclassified)                      & 5,573   \\
Cement and Clinker    & Clinker                                                & 20,824  \\
Coal                  & Anthracite                                             & 1,096   \\
Coal                  & Brown Coal                                             & 22,170  \\
Coal                  & Coal (unclassified)                                    & 115,052 \\
Coal                  & Coke                                                   & 8,086   \\
Coal                  & Coking Coal                                            & 37,437  \\
Coal                  & Thermal Coal                                           & 89,663  \\
Fertilizers           & Ammonium Nitrate                                       & 633     \\
Fertilizers           & Ammonium Sulphate                                      & 5,462   \\
Fertilizers           & Calcium Nitrate                                        & 4       \\
Fertilizers           & Calcium Phosphate                                      & 228     \\
Fertilizers           & Fertilizer Products (CAN)                              & 118     \\
Fertilizers           & Fertilizer Products (DAP)                              & 1,904   \\
Fertilizers           & Fertilizer Products (MAP)                              & 218     \\
Fertilizers           & Fertilizer Products (MOP)                              & 6,145   \\
Fertilizers           & Fertilizer Products (NPK)                              & 2,766   \\
Fertilizers           & Fertilizers (unclassified)                             & 20,031  \\
Fertilizers           & Phosphate Rock                                         & 7,202   \\
Fertilizers           & Potassium (Potash)                                     & 8,551   \\
Fertilizers           & Potassium Nitrate (Saltpeter)                          & 479     \\
Fertilizers           & Sulphur                                                & 7,524   \\
Fertilizers           & Urea                                                   & 16,817  \\
Forest Products       & Biomass                                                & 18      \\
Forest Products       & Forest Products (unclassified)                         & 4,836   \\
Forest Products       & Wood Chips                                             & 13,516  \\
Forest Products       & Wood Logs                                              & 37,336  \\
Forest Products       & Wood Pellets                                           & 3,191   \\
Forest Products       & Wood Pulp                                              & 16,458  \\
Grains                & Barley                                                 & 3,573   \\
Grains                & Beans                                                  & 199     \\
Grains                & Canola                                                 & 558     \\
Grains                & Corn                                                   & 20,824  \\
Grains                & Grains (Flour)                                         & 216     \\
Grains                & Grains (unclassified)                                  & 62,081  \\
Grains                & Heavy Soya Sorgum Grains                               & 1,062   \\
Grains                & Linseed (Flax)                                         & 1       \\
Grains                & Lupins                                                 & 1       \\
Grains                & Malt                                                   & 912     \\
Grains                & Oats                                                   & 14      \\
Grains                & Peas                                                   & 292     \\
Grains                & Rapeseeds                                              & 57      \\
Grains                & Sorghum                                                & 600     \\
Grains                & Soyabean Meal                                          & 3,038   \\
Grains                & Soybean Pellets                                        & 12      \\
Grains                & Soybeans                                               & 19,756  \\
Grains                & Sunflower Seed                                         & 698     \\
Grains                & Wheat                                                  & 43,083  \\
Iron Ore              & Iron Ore (Unclassified)                                & 101,092 \\
Iron Ore              & Iron Ore Fines                                         & 2,176   \\
Iron Ore              & Iron Ore Pellets                                       & 20,788  \\
Iron Ore              & Magnetite                                              & 3,108   \\
Metals                & Copper Anodes                                          & 68      \\
Metals                & Copper Refined                                         & 85      \\
Metals                & Ferrochrome                                            & 324     \\
Metals                & Ferromanganese                                         & 94      \\
Metals                & Ferronickel                                            & 44      \\
Metals                & Lead Refined                                           & 42      \\
Metals                & Manganese Refined                                      & 24      \\
Metals                & Metals (unclassified)                                  & 29      \\
Metals                & Nickel Refined                                         & 7       \\
Metals                & Silicon Manganese                                      & 15      \\
Metals                & Zinc Refined                                           & 7       \\
Minerals              & Barytes                                                & 130     \\
Minerals              & Borates                                                & 261     \\
Minerals              & Carbonates                                             & 482     \\
Minerals              & Ferrosulphate                                          & 5       \\
Minerals              & Fluorspar                                              & 212     \\
Minerals              & Hydroxides \& Peroxides (sodium, potassium)            & 17      \\
Minerals              & Iron Oxides                                            & 8       \\
Minerals              & Magnesite                                              & 160     \\
Minerals              & Magnesium                                              & 2       \\
Minerals              & Minerals (unclassified)                                & 759     \\
Minerals              & Nitrates                                               & 24      \\
Minerals              & Olivin Sand                                            & 71      \\
Minerals              & Phosphinates                                           & 54      \\
Minerals              & Rutile                                                 & 11      \\
Minerals              & Silicates (Garnets)                                    & 1       \\
Minerals              & Soda Ash                                               & 2,334   \\
Minerals              & Spodumene                                              & 79      \\
Minerals              & Sulphites \& Sulphates                                 & 323     \\
Ores and Concentrates & Chrome ore                                             & 5,615   \\
Ores and Concentrates & Copper Concentrate                                     & 20,114  \\
Ores and Concentrates & Ilmenite Ore                                           & 2,096   \\
Ores and Concentrates & Lead                                                   & 65      \\
Ores and Concentrates & Manganese Ore                                          & 9,489   \\
Ores and Concentrates & Nickel Matte                                           & 25      \\
Ores and Concentrates & Nickel Ore                                             & 14,963  \\
Ores and Concentrates & Ores \& Concentrates (unclassified)                    & 13,089  \\
Ores and Concentrates & Zinc                                                   & 1,813   \\
Other Minor Bulks     & Other Minor Bulks                                      & 87      \\
Pet Coke              & Calcined Pet Coke                                      & 2,234   \\
Pet Coke              & Green Pet Coke                                         & 3,765   \\
Pet Coke              & Pet Coke (unclassified)                                & 10,049  \\
Salt                  & Salt (Industrial)                                      & 2,553   \\
Salt                  & Salt (Rock)                                            & 569     \\
Salt                  & Salt (Solar)                                           & 282     \\
Salt                  & Salt (unclassified)                                    & 10,373  \\
Scrap                 & Mill Scale                                             & 120     \\
Scrap                 & Non-ferrious Metal Scrap                               & 1       \\
Scrap                 & Other Ferrious Metal Scrap                             & 28      \\
Scrap                 & Scrap (unclassified)                                   & 7,743   \\
Scrap                 & Steel Scrap                                            & 5,663   \\
Steel                 & Direct reduced iron (DRI)                              & 415     \\
Steel                 & Flat Products (Plates, Coils, Sheets)                  & 23,758  \\
Steel                 & Hot Briquette Iron (HBI)                               & 218     \\
Steel                 & Long Products (beams, angles)                          & 341     \\
Steel                 & Pig iron                                               & 2,556   \\
Steel                 & Semi-finished Products (Billets, slabs, wires, rebars) & 16,496  \\
Steel                 & Steel Products (unclassified)                          & 62,714  \\
Steel                 & Tubes \& Pipes                                         & 8,018   \\
Unknown               & Unknown                                                & 397,597 \\* \bottomrule
\caption{\textbf{Trade flow summary by commodity, 2015 - 2023} Summary of commodities at different levels of granularity. The table presents commodities at a high level (same as Table \ref{tab:commodity_sum}), and at a more granular level, such as specific types or classifications within each group. The higher level represents broader categories, while the lower level provides detailed classifications for each commodity group.}

\label{tab:full_commodity_list}\\
\end{longtable}
\end{tiny}

\begin{figure}[ht!]
    \centering  \includegraphics[width=0.75\linewidth]{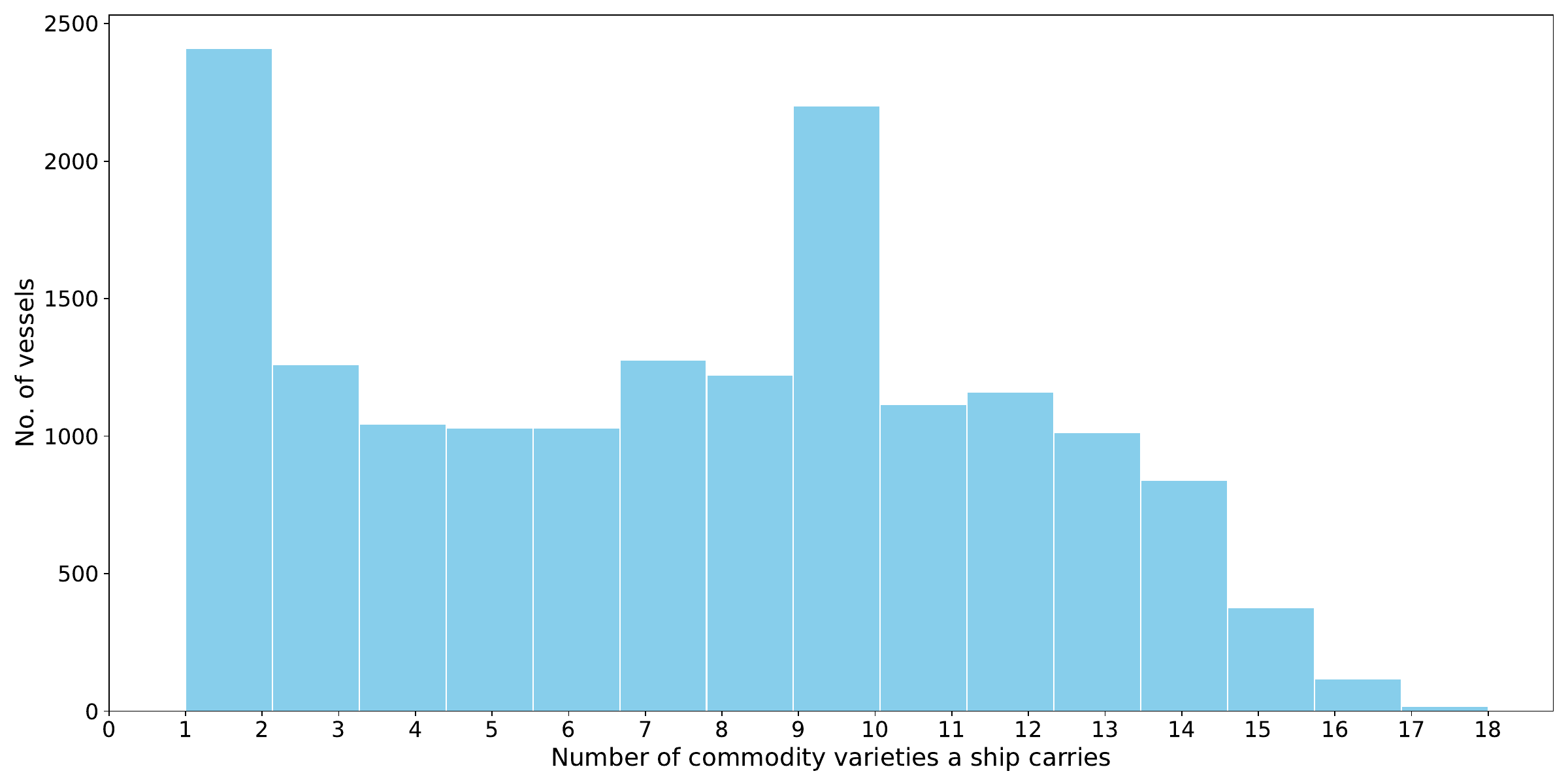}
    \caption{\textbf{Histogram of number of commodity varieties a ship carries.} This histogram illustrates the frequency distribution of the number of different commodity varieties carried by individual ships in the dry bulk trade from 2015 to 2023. The data highlights the diversity of cargo types transported by vessels.}
    \label{fig:his_vessel}
\end{figure}

\begin{table}[ht!]
\centering
\tiny
\begin{tabular}{@{}lllllllllllllll@{}}
\toprule
\textbf{} &
  Grain Product &
  \textbf{n} &
  \textbf{e} &
  \textbf{k} &
  \textbf{$\phi$} &
  \textbf{n\_w} &
  \textbf{p\_w} &
  \textbf{d\_w} &
  \textbf{n\_s} &
  \textbf{p\_s} &
  \textbf{d\_s} &
  \textbf{c} &
  \textbf{a} &
  \textbf{l} \\ \midrule
0   & \textbf{\textit{Entire Grain Network}}                   & 1459 & 22966 & 16   & 0.01 & 1 & 100\% & 7 & 971 & 33.52\% & 4 & 0.040 & -0.07 & 3.15 \\
1  & \textbf{Grains (unclassified)}    & 1326 & 12616 & 9.51 & 0.01       & 1 & 100\% & 7 & 957 & 27.90\% & 4 & 0.026 & -0.08 & 3.62 \\
2  & \textbf{Wheat*}                    & 979  & 9071  & 9.27 & 0.01       & 1 & 100\% & 6 & 814 & 16.96\% & 4 & 0.020 & -0.18 & 4.42 \\
3  & \textbf{Corn*}                     & 748  & 4625  & 6.18 & 0.01       & 1 & 100\% & 6 & 723 & 3.48\%  & 4 & 0.010 & -0.30 & 4.01 \\
4  & \textbf{Soybeans*}                 & 576  & 3068  & 5.33 & 0.01       & 2 & 53\%  & 6 & 542 & 5.90\%  & 4 & 0.013 & -0.33 & 3.75 \\
5  & \textbf{Barley}                   & 341  & 1075  & 3.15 & 0.01       & 1 & 100\% & 8 & 341 & 0.29\%  & 0 & 0.013 & -0.29 & 0.00 \\
6  & \textbf{Soyabean Meal}            & 329  & 916   & 2.78 & 0.01       & 1 & 100\% & 7 & 329 & 0.30\%  & 0 & 0.006 & -0.28 & 0.00 \\
7  & \textbf{Malt}                     & 159  & 281   & 1.77 & 0.01       & 1 & 100\% & 6 & 159 & 0.63\%  & 0 & 0.007 & -0.54 & 0.00 \\
8  & \textbf{Sunflower Seed}           & 150  & 259   & 1.73 & 0.01       & 1 & 75\%  & 6 & 150 & 0.67\%  & 0 & 0.002 & -0.45 & 0.00 \\
9  & \textbf{Heavy Soya Sorgum Grains} & 141  & 261   & 1.85 & 0.01       & 1 & 100\% & 7 & 141 & 0.71\%  & 0 & 0.001 & -0.42 & 0.00 \\
10 & \textbf{Sorghum}                  & 108  & 211   & 1.95 & 0.02       & 2 & 100\% & 6 & 108 & 0.93\%  & 0 & 0.005 & -0.37 & 0.00 \\
11 & \textbf{Peas}                     & 95   & 144   & 1.52 & 0.02       & 1 & 100\% & 6 & 95  & 1.05\%  & 0 & 0.000 & -0.63 & 0.00 \\
12 & \textbf{Grains (Flour)}           & 93   & 125   & 1.34 & 0.01       & 1 & 100\% & 6 & 93  & 1.08\%  & 0 & 0.003 & -0.45 & 0.00 \\
13 & \textbf{Canola}                   & 70   & 132   & 1.89 & 0.03       & 1 & 100\% & 7 & 70  & 1.43\%  & 0 & 0.011 & -0.36 & 0.00 \\
14 & \textbf{Beans}                    & 56   & 93    & 1.66 & 0.03       & 1 & 100\% & 4 & 56  & 1.79\%  & 0 & 0.000 & -0.43 & 0.00 \\
15 & \textbf{Rapeseeds}                & 19   & 21    & 1.11 & 0.06       & 1 & 100\% & 5 & 19  & 5.26\%  & 0 & 0.000 & -0.67 & 0.00 \\
16 & \textbf{Soybean Pellets}          & 15   & 11    & 0.73 & 0.05       & 4 & 100\% & 2 & 15  & 6.67\%  & 0 & 0.000 &       & 0.00 \\
17 & \textbf{Oats}                     & 8    & 7     & 0.88 & 0.13       & 2 & 96\%  & 3 & 8   & 12.50\% & 0 & 0.167 & -0.42 & 0.00 \\
18 & \textbf{Linseed (Flax)}           & 2    & 1     & 0.50 & 0.50       & 1 & 100\% & 1 & 2   & 50.00\% & 0 & 0.000 &       & 0.00 \\
19 & \textbf{Lupins}                   & 2    & 1     & 0.50 & 0.50       & 1 & 100\% & 1 & 2   & 50.00\% & 0 & 0.000 &       & 0.00 \\ \bottomrule
\end{tabular}
\caption{\textbf{Grain sub-networks global centrality measures.} The first row of the table is the global centrality measure of the full grain network, and from row 1 to row 19 are the specific grain products identified by the Oceanbolt trade flow data.}
\label{tbl:grainSubNetworkCentral}
\end{table}

\begin{figure}
    \centering
    \includegraphics[width=1\linewidth]{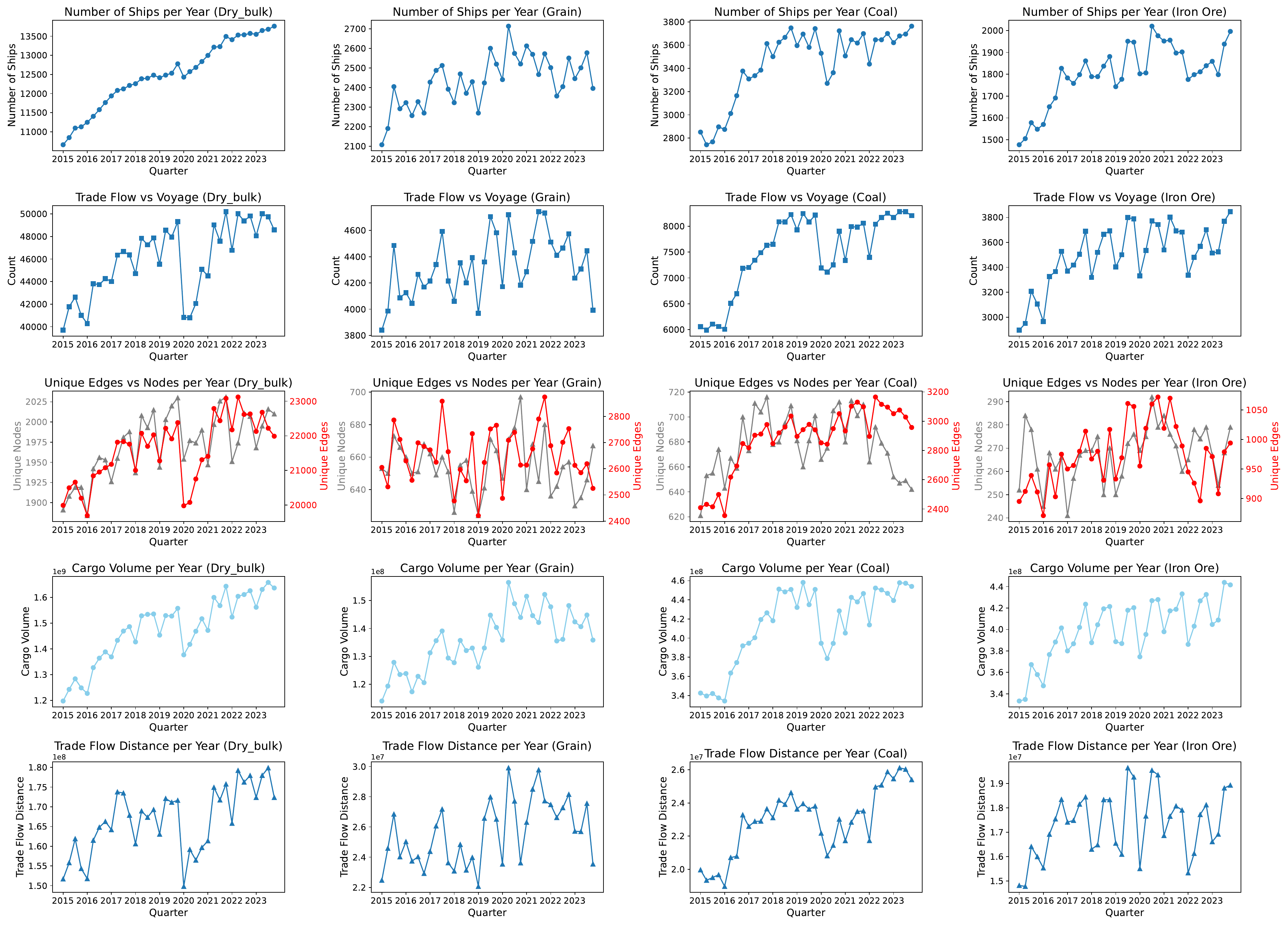}
    \caption{\textbf{Quarterly shipping and trade flow metrics across different sub-networks} Each column represents a different sub-network (namely dry bulk, grain, coal, and iron ore from left to right), showing the number of ships per year in row 1, trade flow versus voyages in row 2, unique edges and nodes in row 3, cargo volume in row 4, and trade flow distance in row 5. The x-axis displays time in quarters, while the y-axes capture the respective metrics. }
    \label{fig:q_metrics}
\end{figure}

\begin{table}[ht!]
\centering
\resizebox{\textwidth}{!}{ 
\begin{tabular}{l l l l l l l l l l l l}
\toprule
\multicolumn{1}{c}{\textbf{}} &
   &
  \multicolumn{2}{c}{\textbf{normalised out-degree}} &
  \multicolumn{1}{c}{\textbf{}} &
  \multicolumn{1}{c}{\textbf{}} &
  \multicolumn{2}{c}{\textbf{normalised in-degree}} &
   &
   &
  \multicolumn{2}{c}{\textbf{normalised\_betweenness}} \\ \midrule
GRAIN &
   &
  \textbf{prior} &
  \textbf{post} &
  \textbf{} &
  \textbf{} &
  \textbf{prior} &
  \textbf{post} &
   &
   &
  \textbf{prior} &
  \textbf{post} \\ \cmidrule(lr){3-4} \cmidrule(lr){7-8} \cmidrule(l){11-12} 
 &
  1 &
  United States &
  Argentina &
   &
  1 &
  China &
  Spain &
   &
  1 &
  United States &
  United States \\
 &
  2 &
  Argentina &
  United States &
   &
  2 &
  United States &
  China &
   &
  2 &
  Brazil &
  Brazil \\
 &
  3 &
  Russian Federation &
  Brazil &
   &
  3 &
  Turkey &
  Turkey &
   &
  3 &
  China &
  Turkey \\
 &
  4 &
  Brazil &
  Canada &
   &
  4 &
  Italy &
  United States &
   &
  4 &
  Turkey &
  China \\
 &
  5 &
  Canada &
  Russian Federation &
   &
  5 &
  Spain &
  Netherlands &
   &
  5 &
  Canada &
  France \\
 &
  6 &
  Ukraine &
  Ukraine &
   &
  6 &
  India &
  Brazil &
   &
  6 &
  Italy &
  Russian Federation \\
 &
  7 &
  France &
  Germany &
   &
  7 &
  Egypt &
  Italy &
   &
  7 &
  France &
  Spain \\
 &
  8 &
  Romania &
  Australia &
   &
  8 &
  United Kingdom &
  Egypt &
   &
  8 &
  Argentina &
  Netherlands \\
 & & & & & & & & & & &
   \\
COAL &
  \textbf{} &
  \multicolumn{2}{c}{\textbf{normalised out-degree}} &
  \textbf{} &
  \textbf{} &
  \multicolumn{2}{c}{\textbf{normalised in-degree}} &
   &
  \textbf{} &
  \multicolumn{2}{c}{\textbf{normalised\_betweenness}} \\ \cmidrule(lr){3-4} \cmidrule(lr){7-8} \cmidrule(l){11-12} 
 &
  \textbf{} &
  \textbf{prior} &
  \textbf{post} &
  \textbf{} &
  \textbf{} &
  \textbf{prior} &
  \textbf{post} &
   &
  \textbf{} &
  \textbf{prior} &
  \textbf{post} \\ \cmidrule(lr){3-4} \cmidrule(lr){7-8} \cmidrule(l){11-12} 
 &
  1 &
  Russian Federation &
  Russian Federation &
   &
  1 &
  China &
  China &
   &
  1 &
  United States &
  United States \\
 &
  2 &
  United States &
  Indonesia &
   &
  2 &
  India &
  India &
   &
  2 &
  China &
  Indonesia \\
 &
  3 &
  Indonesia &
  United States &
   &
  3 &
  Brazil &
  United States &
   &
  3 &
  Russian Federation &
  China \\
 &
  4 &
  South Africa &
  Colombia &
   &
  4 &
  United States &
  Netherlands &
   &
  4 &
  Colombia &
  South Africa \\
 &
  5 &
  Colombia &
  South Africa &
   &
  5 &
  Egypt &
  Turkey &
   &
  5 &
  Indonesia &
  Colombia \\
 &
  6 &
  China &
  Australia &
   &
  6 &
  Turkey &
  Italy &
   &
  6 &
  India &
  Russian Federation \\
 &
  7 &
  Australia &
  Mozambique &
   &
  7 &
  Netherlands &
  United Kingdom &
   &
  7 &
  South Africa &
  India \\
 &
  8 &
  Latvia &
  China &
   &
  8 &
  Italy &
  Spain &
   &
  8 &
  Brazil &
  United Kingdom \\
 &
   & & & & & & & & & &
   \\
IRON ORE &
   &
  \multicolumn{2}{c}{\textbf{normalised out-degree}} &
   &
   &
  \multicolumn{2}{c}{\textbf{normalised in-degree}} &
   &
   &
  \multicolumn{2}{c}{\textbf{normalised\_betweenness}} \\ \cmidrule(lr){3-4} \cmidrule(lr){7-8} \cmidrule(l){11-12} 
 &
  \textbf{} &
  \textbf{prior} &
  \textbf{post} &
  \textbf{} &
  \textbf{} &
  \textbf{prior} &
  \textbf{post} &
   &
  \textbf{} &
  \textbf{prior} &
  \textbf{post} \\ \cmidrule(lr){3-4} \cmidrule(lr){7-8} \cmidrule(l){11-12} 
 &
  1 &
  Brazil &
  Brazil &
   &
  1 &
  China &
  China &
   &
  1 &
  India &
  India \\
 &
  2 &
  India &
  India &
   &
  2 &
  United States &
  Netherlands &
   &
  2 &
  Brazil &
  China \\
 &
  3 &
  Canada &
  Canada &
   &
  3 &
  Netherlands &
  India &
   &
  3 &
  China &
  Brazil \\
 &
  4 &
  South Africa &
  Bahrain &
   &
  4 &
  Hong Kong &
  Hong Kong &
   &
  4 &
  Canada &
  Netherlands \\
 &
  5 &
  Sierra Leone &
  Sierra Leone &
   &
  5 &
  Germany &
  United States &
   &
  5 &
  Netherlands &
  South Africa \\
 &
  6 &
  Ukraine &
  South Africa &
   &
  6 &
  United Kingdom &
  Turkey &
   &
  6 &
  Malaysia &
  Mexico \\
 &
  7 &
  Venezuela &
  Ukraine &
   &
  7 &
  Japan &
  Japan &
   &
  7 &
  United States &
  Canada \\
 &
  8 &
  Norway &
  Norway &
   &
  8 &
  Turkey &
  Belgium &
   &
  8 &
  South Africa &
  Philippines \\ \bottomrule
\end{tabular}
}
\caption{\textbf{Top 8 countries with largest normalized degree and betweenness centralities before and after structural breaks}
Table showing the top 8 countries with the largest normalized out-degree, normalized in-degree, and betweenness centralities. The statistics are provided for the networks before and after the structural breaks in the global trade network. These values highlight the leading countries in terms of their centrality measures, both prior to and following the identified structural changes.}\label{tab:top8country}
\end{table}

\begin{table}[ht!]
\tiny
\centering
\begin{tabular}{@{}llllllllllllll@{}}
\toprule
 Coal &
  \textbf{n} &
  \textbf{e} &
  \textbf{k} &
  \textbf{$\phi$} &
  \textbf{n\_w} &
  \textbf{p\_w} &
  \textbf{d\_w} &
  \textbf{n\_s} &
  \textbf{p\_s} &
  \textbf{d\_s} &
  \textbf{l} &
  \textbf{c} &
    \textbf{a} \\ \midrule
\textbf{before Q1  2020} &
  1320 &
  12790 &
  19.38 &
  0.00735 &
  1 &
  100\% &
  6 &
  941 &
  29\% &
  6 &
  0.04 &
  -0.07 &
  4.16 \\
\textbf{after Q1   2020} &
  1266 &
  13048 &
  20.61 &
  0.00815 &
  1 &
  100\% &
  7 &
  903 &
  29\% &
  6 &
  0.04 &
  -0.08 &
  3.82 \\ \bottomrule
\end{tabular}
\caption{\textbf{Global coal network centralities.} The centrality measures of each network include: number of ports n; number of unique edges e; average in- or out-degree k; network density $\phi$; number of weakly-connected components (WCCs) n\_w; percentage of giant weakly-connected component of the full network p\_w, diameter of the largest weakly-connected component d\_w; number of strongly-connected components (SCCs) n\_s, percentage of giant strongly-connected component (GSCC) of the full network p\_s, diameter of the strongly-connected component d\_s; clustering coefficient c; the average shortest path length l, and degree assortativity a.}
\label{tbl:coal_glo_centrality}
\end{table}

\begin{table}[ht!]
\tiny
\centering
\begin{tabular}{@{}clllllllllllll@{}}
\toprule
\multicolumn{1}{l}{Grain} &
  \multicolumn{1}{c}{\textbf{n}} &
  \multicolumn{1}{c}{\textbf{e}} &
  \multicolumn{1}{c}{\textbf{k}} &
  \multicolumn{1}{c}{\textbf{$\phi$}} &
  \multicolumn{1}{c}{\textbf{n\_w}} &
  \multicolumn{1}{c}{\textbf{p\_w}} &
  \multicolumn{1}{c}{\textbf{d\_w}} &
  \multicolumn{1}{c}{\textbf{n\_s}} &
  \multicolumn{1}{c}{\textbf{p\_s}} &
  \multicolumn{1}{c}{\textbf{d\_s}} &
  \multicolumn{1}{c}{\textbf{l}} &
  \multicolumn{1}{c}{\textbf{c}} &
  \multicolumn{1}{c}{\textbf{a}} \\ \midrule
\textbf{before Q1   2021} &
  1341 &
  18228 &
  27.1857 &
  0.0101 &
  1 &
  100\% &
  7 &
  932 &
  31\% &
  5 &
  0.0321 &
  -0.07 &
  3.36 \\
\textbf{after Q1 2021} &
  1121 &
  12306 &
  21.9554 &
  0.0098 &
  1 &
  100\% &
  7 &
  838 &
  25\% &
  5 &
  0.0256 &
  -0.02 &
  3.45 \\ \bottomrule
\end{tabular}%
\caption{\textbf{Global grain network centralities.} The centrality measures of each network include: number of ports n; number of unique edges e; average in- or out-degree k; network density $\phi$; number of weakly-connected components (WCCs) n\_w; percentage of giant weakly-connected component of the full network p\_w, diameter of the largest weakly-connected component d\_w; number of strongly-connected components (SCCs) n\_s, percentage of giant strongly-connected component (GSCC) of the full network p\_s, diameter of the strongly-connected component d\_s; clustering coefficient c; the average shortest path length l, and degree assortativity a.}
\label{tbl:grain_glo_centrality}
\end{table}
\clearpage

\section{Python algorithm}
\section*{}
\label{supplementB}

\begin{description}

\item [Rewiring (edge-switching) Null Model] Python algorithm
To simulate a rewired random graph, we start from the original observed graph and randomly select a pair of edges, (v1,v2) and (v3,v4). We then keep the four nodes as they are, but remove the existing edges and then build two new connections (v1,v4) and (v2,v3), without adding or removing any of these vertices. We iterate this process multiple times to simulate the graph. If through this process, one or both of these new connections already exist in the network, we write the algorithms so that we can skip this round of iteration and randomly select another pair of connections in the next round. This rule ensures that new connections are not placed between the same pair of nodes. A repeated rewiring process randomizes the network topology. Since we do not add or remove nodes or edges, the rewired network will have the same size in terms of the number of nodes and network density as the original observed real world network. 

More importantly, this rewiring procedure ensures that the same number of links attached to each node is unchanged, thus preserving the degree distribution of the original network. This works particularly well for any directed networks. The rewiring algorithm ensures that each connection switch is such that an outgoing edge at one node is always replaced with an outgoing edge from another node, thus preserving both the in- and out-degrees. The output of the algorithm is thus a randomized network that is matched to our empirical network in terms of size, connection density, and degree distribution.

\item [Python Louvain:] community detection algorithm

The official Louvain community detection algorithm implementation is shown on the \href{https://github.com/taynaud/python-louvain/blob/master/community/community_louvain.py#L373}{GitHub website}, which illustrates how the function is written and all the steps that happen behind the scenes. Here is a brief summary of the key steps that occur in the algorithm.
\begin{verbatim}
import networkx as nx 
import community as community_louvain  
community = community_louvain.best_partition(G)
\end{verbatim}

When \verb|community.best_partition()| is called, the following steps occur to obtain the final communities:
\begin{enumerate}
  \item \textbf{Dendrogram Creation}: This entry function creates the hierarchical decomposition (dendrogram) by calling an internal function \verb|generate_dendrogram()|.
  \item \textbf{Calculations in} \verb|generate_dendrogram()|: 
    \begin{itemize}
      \item \emph{Initialisation}: A \verb|Status| object tracks the initial communities and degrees per node.
      \item \emph{Local Moving} (\verb|__one_level()|): Shuffles nodes around to maximise modularity at the node level.
      \item \emph{Check Improvement} (\verb|__modularity()|): Computes modularity.
      \item \emph{Aggregate Communities} (\verb|induced_graph()|): Merges each community into a single super-node and returns a new, smaller graph. The \verb|Status| is then re-initialised for the next iteration.
      \item \emph{Repeat} these three steps until no further improvement in modularity is possible.
    \end{itemize}
  \item \textbf{Final Partition}: The last level of the dendrogram (highest modularity) is returned.
\end{enumerate}

\item[Python Scipy:] hierarchical clustering algorithm 

The algorithm begins by treating each quarter as a separate cluster and iteratively merges the most similar clusters until all quarters are combined into a single cluster. This process is represented as a dendrogram, which visually shows the hierarchical structure of the clusters.

The algorithm was implemented in Python using the \texttt{SciPy} library. The linkage matrix was computed with the following code:

\begin{verbatim}
from scipy.cluster.hierarchy import linkage
linkage_matrix = linkage(distance_matrix, method='ward')
\end{verbatim}

The linkage matrix contains the information needed to determine how and when clusters are merged. To form a predetermined number of clusters, we applied the \texttt{fcluster} function, which cuts the dendrogram at a specific level based on the number of clusters we wish to obtain:

\begin{verbatim}
from scipy.cluster.hierarchy import fcluster
clusters = fcluster(linkage_matrix, num_clusters, criterion='maxclust')
\end{verbatim}

Here, \texttt{num\_clusters} represents the desired number of clusters. The \texttt{fcluster} function assigns each quarter to a specific cluster, allowing us to analyse the structural similarities between quarters.

By visualising the dendrogram and applying the clustering algorithm, we were able to detect significant changes in network structure over time. This clustering approach enabled us to group similar quarters together, providing insights into periods of stability or disruption within the network. The detected clusters were then analysed to interpret the key periods where major structural changes occurred, helping us understand the dynamics of the network over time.

\begin{figure}[ht!]
    \centering
    \begin{minipage}{0.5\textwidth}
        \centering
        \includegraphics[width=\textwidth]{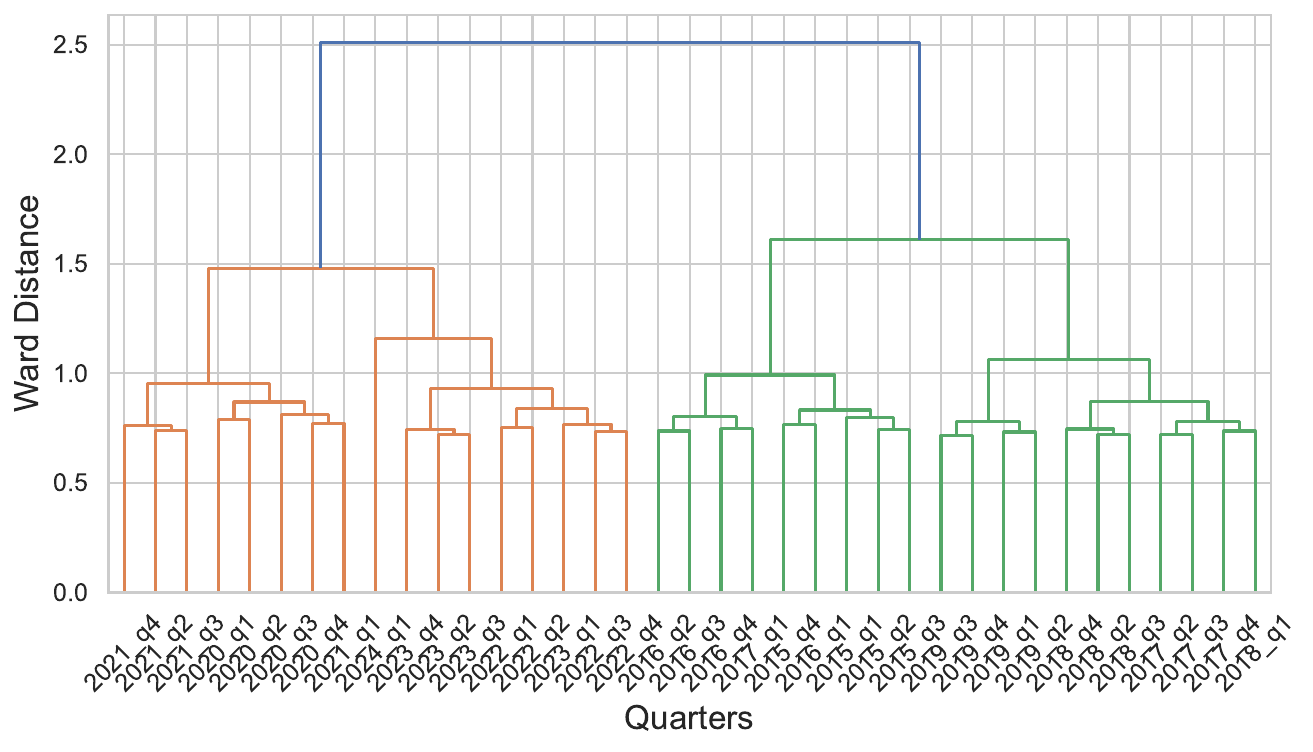}
        \caption*{(a) Dry bulk}
        \label{fig:image1}
    \end{minipage}\hfill
    \begin{minipage}{0.5\textwidth}
        \centering
        \includegraphics[width=\textwidth]{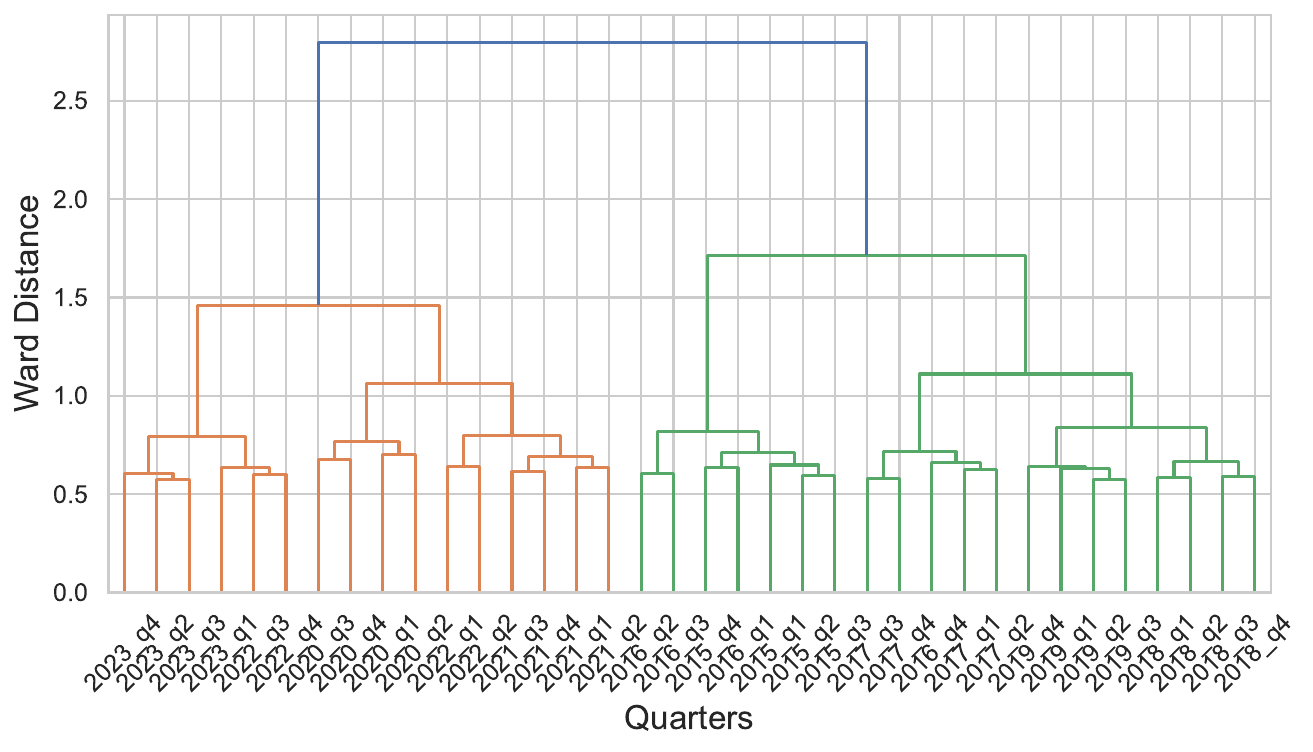}
        \caption*{(b) Coal}
        \label{fig:image2}
    \end{minipage}
    \vspace{0.5cm}
    \begin{minipage}{0.5\textwidth}
        \centering
        \includegraphics[width=\textwidth]{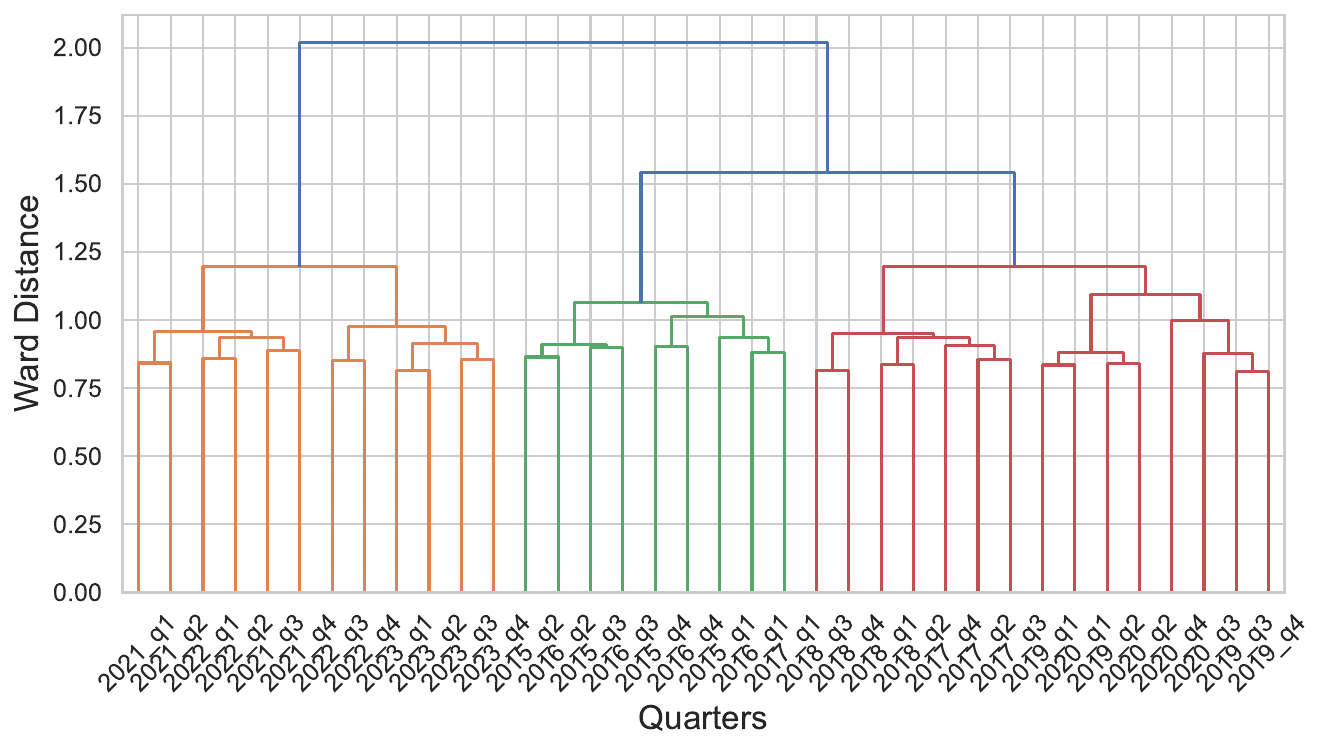}
        \caption*{(c) Grain}
        \label{fig:image3}
    \end{minipage}\hfill
    \begin{minipage}{0.5\textwidth}
        \centering
        \includegraphics[width=\textwidth]{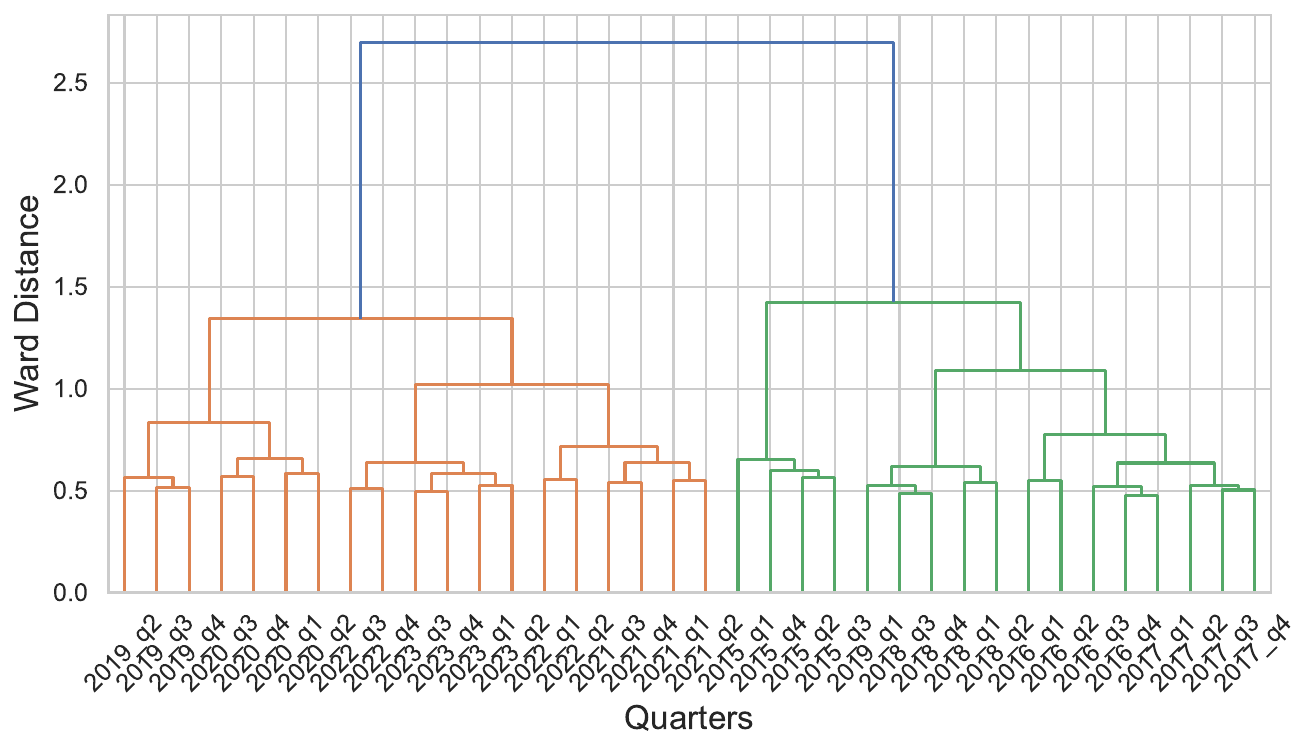}
        \caption*{(d) Iron ore}
        \label{fig:image4}
    \end{minipage}
    \caption{\textbf{Dendrogram for all sub-networks} Hierarchical clustering dendrograms illustrating the community structures for different dry bulk shipping sub-networks: (a) overall dry bulk, (b) coal, (c) grain, and (d) iron ore. Each dendrogram depicts clustering patterns within each commodity-specific network, highlighting similarities and differences in network structures, which reflect unique trade flows and connectivity properties across these commodities.}
    \label{fig:Dencombined}
\end{figure}

\end{description}

\clearpage
\phantomsection
\addcontentsline{toc}{chapter}{Bibliography}
\bibliography{bib}

\begin{thebibliography}{}

\bibitem[Abouarghoub et~al., 2018]{abouarghoub2018reconciling}
Abouarghoub, W., Nomikos, N.~K., and Petropoulos, F. (2018).
\newblock On reconciling macro and micro energy transport forecasts for strategic decision making in the tanker industry.
\newblock {\em Transportation Research Part E: Logistics and Transportation Review}, 113:225--238.

\bibitem[Bai et~al., 2023]{bai2023data}
Bai, X., Ma, Z., and Zhou, Y. (2023).
\newblock Data-driven static and dynamic resilience assessment of the global liner shipping network.
\newblock {\em Transportation Research Part E: Logistics and Transportation Review}, 170:103016.

\bibitem[Barab{\'a}si and Albert, 1999]{barabasi1999emergence}
Barab{\'a}si, A.-L. and Albert, R. (1999).
\newblock Emergence of scaling in random networks.
\newblock {\em Science}, 286(5439):509--512.

\bibitem[Blondel et~al., 2008]{blondel2008fast}
Blondel, V.~D., Guillaume, J.-L., Lambiotte, R., and Lefebvre, E. (2008).
\newblock Fast unfolding of communities in large networks.
\newblock {\em Journal of statistical mechanics: theory and experiment}, 2008(10):P10008.

\bibitem[Brancaccio et~al., 2020]{brancaccio2020geography}
Brancaccio, G., Kalouptsidi, M., and Papageorgiou, T. (2020).
\newblock Geography, transportation, and endogenous trade costs.
\newblock {\em Econometrica}, 88(2):657--691.

\bibitem[Calatayud et~al., 2017a]{calatayud2017connectivity}
Calatayud, A., Mangan, J., and Palacin, R. (2017a).
\newblock Connectivity to international markets: A multi-layered network approach.
\newblock {\em Journal of Transport Geography}, 61:61--71.

\bibitem[Calatayud et~al., 2017b]{calatayud2017vulnerability}
Calatayud, A., Mangan, J., and Palacin, R. (2017b).
\newblock Vulnerability of international freight flows to shipping network disruptions: A multiplex network perspective.
\newblock {\em Transportation Research Part E: Logistics and Transportation Review}, 108:195--208.

\bibitem[Cheung et~al., 2020]{cheung2020eigenvector}
Cheung, K.-F., Bell, M.~G., Pan, J.-J., and Perera, S. (2020).
\newblock An eigenvector centrality analysis of world container shipping network connectivity.
\newblock {\em Transportation Research Part E: Logistics and Transportation Review}, 140:101991.

\bibitem[Conway, 2023]{worldmaterial}
Conway, E. (2023).
\newblock {\em World Material}.
\newblock WH Allen, paperback 2024 edition.

\bibitem[Ducruet, 2013]{ducruet2013network}
Ducruet, C. (2013).
\newblock Network diversity and maritime flows.
\newblock {\em Journal of Transport Geography}, 30:77--88.

\bibitem[Ducruet et~al., 2010a]{ducruet2010centrality}
Ducruet, C., Lee, S.-W., and Ng, A.~K. (2010a).
\newblock Centrality and vulnerability in liner shipping networks: revisiting the northeast asian port hierarchy.
\newblock {\em Maritime Policy \& Management}, 37(1):17--36.

\bibitem[Ducruet and Notteboom, 2012]{ducruet2012worldwide}
Ducruet, C. and Notteboom, T. (2012).
\newblock The worldwide maritime network of container shipping: spatial structure and regional dynamics.
\newblock {\em Global networks}, 12(3):395--423.

\bibitem[Ducruet et~al., 2010b]{ducruet2010ports}
Ducruet, C., Rozenblat, C., and Zaidi, F. (2010b).
\newblock Ports in multi-level maritime networks: evidence from the atlantic (1996--2006).
\newblock {\em Journal of Transport Geography}, 18(4):508--518.

\bibitem[Fornito et~al., 2016]{fornito2016fundamentals}
Fornito, A., Zalesky, A., and Bullmore, E. (2016).
\newblock {\em Fundamentals of brain network analysis}.
\newblock Academic press.

\bibitem[Guinand and Pign{\'e}, 2015]{guinand2015time}
Guinand, F. and Pign{\'e}, Y. (2015).
\newblock Time considerations for the study of complex maritime networks.
\newblock In {\em Maritime Networks}, pages 187--213. Routledge.

\bibitem[Hook et~al., 2025]{HookEtAl2025}
Hook, L., Leahy, J., and Ding, W. (2025).
\newblock The china commodities supercycle is over. will there be another?
\newblock \url{https://www.ft.com/content/8ae50718-31ce-4e18-8491-33a600bbd3a5}.
\newblock Accessed: Jan 15, 2025.

\bibitem[Humphries and Gurney, 2008]{humphries2008network}
Humphries, M.~D. and Gurney, K. (2008).
\newblock Network ‘small-world-ness’: a quantitative method for determining canonical network equivalence.
\newblock {\em PloS one}, 3(4):e0002051.

\bibitem[Iapadre and Tajoli, 2014]{iapadre2014emerging}
Iapadre, P.~L. and Tajoli, L. (2014).
\newblock Emerging countries and trade regionalization. a network analysis.
\newblock {\em Journal of Policy Modeling}, 36:S89--S110.

\bibitem[Kaluza et~al., 2010]{kaluza2010complex}
Kaluza, P., K{\"o}lzsch, A., Gastner, M.~T., and Blasius, B. (2010).
\newblock The complex network of global cargo ship movements.
\newblock {\em Journal of the Royal Society Interface}, 7(48):1093--1103.

\bibitem[Kanrak et~al., 2019]{kanrak2019maritime}
Kanrak, M., Nguyen, H.~O., and Du, Y. (2019).
\newblock Maritime transport network analysis: A critical review of analytical methods and applications.
\newblock {\em Journal of International Logistics and Trade}, 17(4):113--122.

\bibitem[Kingsman, 2017]{kingsman2017commodity}
Kingsman, J. (2017).
\newblock {\em Commodity Conversations: An Introduction to Trading in Agricultural Commodities}.
\newblock CreateSpace Independent Publishing Platform.

\bibitem[Kiss et~al., 2006]{Kiss2006}
Kiss, I.~Z., Green, D.~M., and Kao, R.~R. (2006).
\newblock The network of sheep movements within great britain: network properties and their implications for infectious disease spread.
\newblock {\em Journal of The Royal Society Interface}, 3(10):669--677.

\bibitem[Laxe et~al., 2012]{laxe2012maritime}
Laxe, F.~G., Seoane, M. J.~F., and Montes, C.~P. (2012).
\newblock Maritime degree, centrality and vulnerability: port hierarchies and emerging areas in containerized transport (2008--2010).
\newblock {\em Journal of Transport Geography}, 24:33--44.

\bibitem[Liu et~al., 2018]{liu2018spatial}
Liu, C., Wang, J., and Zhang, H. (2018).
\newblock Spatial heterogeneity of ports in the global maritime network detected by weighted ego network analysis.
\newblock {\em Maritime Policy \& Management}, 45(1):89--104.

\bibitem[Newman, 2010]{Newman2010networks}
Newman, M. E.~J. (2010).
\newblock {\em Networks: an introduction}.
\newblock Oxford University Press.

\bibitem[Pan et~al., 2019]{pan2019connectivity}
Pan, J.-J., Bell, M.~G., Cheung, K.-F., Perera, S., and Yu, H. (2019).
\newblock Connectivity analysis of the global shipping network by eigenvalue decomposition.
\newblock {\em Maritime Policy \& Management}, 46(8):957--966.

\bibitem[Rodrigue, 2006]{rodrigue2006challenging}
Rodrigue, J.-P. (2006).
\newblock Challenging the derived transport-demand thesis: geographical issues in freight distribution.
\newblock {\em Environment and Planning A}, 38(8):1449--1462.

\bibitem[Stopford, 2008]{maritimeeconomics}
Stopford, M. (2008).
\newblock {\em Maritime Economics}.
\newblock Taylor \& Francis Ltd, 3rd edition edition.

\bibitem[Sugishita and Masuda, 2021]{sugishita2021recurrence}
Sugishita, K. and Masuda, N. (2021).
\newblock Recurrence in the evolution of air transport networks.
\newblock {\em Scientific reports}, 11(1):5514.

\bibitem[Telesford et~al., 2011]{telesford2011ubiquity}
Telesford, Q.~K., Joyce, K.~E., Hayasaka, S., Burdette, J.~H., and Laurienti, P.~J. (2011).
\newblock The ubiquity of small-world networks.
\newblock {\em Brain connectivity}, 1(5):367--375.

\bibitem[Tsiotas and Polyzos, 2015]{tsiotas2015analyzing}
Tsiotas, D. and Polyzos, S. (2015).
\newblock Analyzing the maritime transportation system in greece: a complex network approach.
\newblock {\em Networks and Spatial Economics}, 15(4):981--1010.

\bibitem[Wang et~al., 2016]{wang2016study}
Wang, N., Wu, N., Dong, L.-l., Yan, H.-k., and Wu, D. (2016).
\newblock A study of the temporal robustness of the growing global container-shipping network.
\newblock {\em Scientific reports}, 6(1):1--10.

\bibitem[Williams and Del~Genio, 2014]{williams2014degree}
Williams, O. and Del~Genio, C.~I. (2014).
\newblock Degree correlations in directed scale-free networks.
\newblock {\em PloS one}, 9(10):e110121.

\bibitem[Yin et~al., 2024]{yin2024temporal}
Yin, Z., Hu, J., Zhang, J., Zhou, X., Li, L., and Wu, J. (2024).
\newblock Temporal and spatial evolution of global major grain trade patterns.
\newblock {\em Journal of Integrative Agriculture}, 23(3):1075--1086.

\end{thebibliography}

\end{document}